\documentclass[review]{elsarticle}

\usepackage{amsmath}
\usepackage{amssymb}
\usepackage{amsfonts}
\usepackage{graphicx}
\usepackage{fullpage}
\usepackage{graphics}
\usepackage{multirow}
\usepackage{epsfig}
\usepackage{setspace} 
\usepackage{epstopdf}
\usepackage{newfloat}
\usepackage{subcaption}
\usepackage[table]{xcolor}
\usepackage{algorithm}
\usepackage{algorithmic}
\usepackage{varioref}
\usepackage{hyperref}
\usepackage{bm}
\usepackage{array}
\usepackage{tcolorbox}
\definecolor{forestgreen}{rgb}{0.13, 0.55, 0.13}
\usepackage[normalem]{ulem}

\usepackage{accents}

\usepackage{blkarray}

\usepackage{lineno,hyperref}

\modulolinenumbers[5]

\graphicspath{{./}{./images-suppl/}{./images/}}

\journal{Journal of \LaTeX\ Templates}

\begin{document}

\begin{frontmatter}

\title{Master Stability Functions in Complex Networks}

\author[mymainaddress1]{Suman Acharyya}
\ead{sumanacharyya2006@gmail.com}

\author[mymainaddress1,mymainaddress2,mymainaddress4]{Priodyuti Pradhan}
\corref{mycorrespondingauthor}
\ead{priodyutipradhan@gmail.com}

\author[mymainaddress3]{Chandrakala Meena}\ead{meenachandrakala@gmail.com}


\address[mymainaddress1]{Complex Network Dynamics Lab, Department of Mathematics, Bar-Ilan University, Ramat Gan - 5290002, Israel}
    
\address[mymainaddress2]{networks.ai Lab, Department of Computer Science \& Engineering, Indian Institute of Information Technology Raichur, Karnataka - 584135, India}

\address[mymainaddress4]{Department of Computer Science, University of Petroleum and Energy Studies, Dehradun, Uttarakhand - 248007, India}

\address[mymainaddress3]{School of Physics, Indian Institute of Science Education and Research Thiruvananthapuram, Kerala - 695551, India}

\begin{abstract}
Synchronization is an emergent and fundamental phenomenon in nature and engineered systems. Understanding the stability of a synchronized phenomenon is crucial for ensuring functionality in various complex systems. The stability of the synchronization phenomenon is extensively studied using the Master Stability Function (MSF). This powerful and elegant tool plays a pivotal role in determining the stability of synchronization states, providing deep insights into synchronization in coupled systems. However, a major challenge lies in determining the MSF for complex dynamical networks driven by nonlinear interaction mechanisms. These mechanisms introduce additional complexity due to the intricate connectivity of the interacting elements and the complex dynamics governed by nonlinear processes, diverse parameters, and the higher dimensionality of the system. Although MSF analysis has been used for 25 years to study the stability of synchronization states, a systematic investigation of MSF across various networked systems remains missing from the literature. In this article, we present a simplified and unified MSF analysis for diverse undirected and directed networked systems. We begin with the analytical MSF framework for pairwise-coupled identical systems with diffusive and natural coupling schemes and extend our analysis to directed networks and multilayer networks, considering both intra-layer and inter-layer interactions. Furthermore, we revisit the MSF framework to incorporate higher-order interactions alongside pairwise interactions. To enhance understanding, we also provide a numerical analysis of synchronization in coupled Rössler systems under pairwise diffusive coupling and propose algorithms for determining the MSF, identifying stability regimes, and classifying MSF functions. Overall, the primary goal of this review is to present a systematic study of MSF in coupled dynamical networks in a clear and structured manner, making this powerful tool more accessible. Furthermore, we highlight cases where the study of synchronization states using MSF remains underexplored. Additionally, we discuss recent research focusing on MSF analysis using time series data and machine learning approaches.
\end{abstract}

\begin{keyword}
\texttt{Nonlinear dynamics\sep Networks\sep Synchronization\sep Master stability function
}
\end{keyword}

\end{frontmatter} 

\vspace{1mm}
\begin{flushright}
\textit{Celebrating $25$ years of Master Stability Function}   
\end{flushright}


\section{Introduction}
Many complex systems can be modeled using dynamical network frameworks of interacting units. This framework consists of two essential ingredients to model the system: network structures and dynamics. The network structure consists of many interacting units, where elements are known as \emph{nodes} or \emph{vertices}, and interactions between nodes are known as \emph{links} or \emph{edges} \cite{BarabasiBook,RevModPhys.74.47}. 
Network science theory plays an important role in understanding the various types of interactions that occur between elements in complex systems~\cite{Newman2018, Winfree1980, Kuramoto1984, StrogatzSyncBook, PhysRep.469.93}. Dynamics capture the interaction mechanisms in the system and help in tracking the temporal activity of each node in the network. In a dynamical network framework, the evolution of the dynamical activity of nodes can be modeled as either a differential or a difference equation. For example, to study a network of population dynamics, we can take the dynamics of the Ricker map or logistic differential equation ~\cite{meena2017threshold,ducrot2022differential}. The dynamical network framework helps to understand many
problems in geology~\cite{TurcotteBook}, ecology~\cite{RMMayBook}, mathematical biology~\cite{Science.275.334}, neuroscience~\cite{NatRevNeuroSci.10.186}, and in  physics to understand the physics of neutrinos~\cite{PhysRevD.58.073002} and superconductors~\cite{PhysRevE.57.1563}.
Complex dynamical networks are capable of exhibiting rich dynamical behaviors ~\cite{BarratBook}, for example, synchronization~\cite{PikovskyBook, PhysRep.366.1,rungta2017network}, cascading failures~\cite{ValdezJCM2020}, information spreading~\cite{BarratBook} that cannot be observed in an isolated system. Synchronization is one of the most emergent phenomena observed in many real and man-made systems~\cite{PhysRep.424.175, ArenasPRL2006}.

The word \emph{synchronization} originated from a Greek word that means ``sharing common time". Two or more systems evolving with time will be synchronized if their time evolution is correlated and remains correlated in the future. For the first time, in the seventeenth century in $1673$, the Dutch scientist Christian Huygens documented the synchronization between two pendulums that hang from the same wooden beam~\cite{Huygens1665}. In $1680$, another Dutch physicist, Engelbert Kaempfer, observed the synchronization of the flashing of fireflies in southeast Asia~\cite{buck1968mechanism}. Later, people observed synchronization in many natural systems ranging from planetary motion, applauding audiences, chirping crickets, clocks, communication, navigation systems, and many more \cite{PikovskyBook,StrogatzSyncBook}.
Since synchronization is ubiquitous, we use a dynamic network framework that models various real and engineered systems~\cite{PikovskyBook, PhysRep.366.1,rungta2018,meena2020resilience} to understand this. The most important discovery in this topic is the study of synchronization in coupled chaotic dynamical systems. 
Chaotic dynamical systems have a sensitive dependence on the initial condition; an infinitesimal difference in the initial conditions will increase exponentially, and the two trajectories of the same chaotic dynamical system starting from two infinitesimally different initial conditions will become uncorrelated very quickly~\cite{DevaneyBook}. So chaotic systems, by definition, defy synchronization. 
Understanding the stability of a synchronized state of such coupled chaotic systems is crucial for ensuring functionality in various complex natural and engineered systems. For instance, in natural systems like, in neuroscience, neuronal synchronization enables coherent brain activity ~\cite{NatRevNeuroSci.10.186,PNAS.95.7092}; disruptions can lead to disorders like epilepsy and Parkinson's disease; in biological science, the synchronized beating of pacemaker cells in heart is vital for a stable heartbeat; desynchronization can cause arrhythmias~\cite{PikovskyBook, StrogatzSyncBook, PhysRep.366.1}, and biological clocks synchronize to maintain sleep-wake cycles across organisms, in ecological science: fireflies synchronize flashing patterns, and fish shoals exhibit synchronized swimming to improve their survival~\cite{PhysRevE.96.258102}.
In the same way, in engineered systems, synchronization of components in power grid networks is necessary to prevent blackouts. As we all know, the northeast blackout of $2003$ affected some parts of the United States and Canada, and the $2012$ India blackout affected north and east India \cite{BarabasiBook,AlbertPRE2004}. In telecommunications \& internet networks, synchronization is crucial for data transmission protocols and distributed computing~\cite{LamportACM1978}. In quantum systems, synchronization in qubits can enhance the performance of quantum algorithms~\cite{quantum_sync_2010}.
The study of synchronization in complex systems benefits from advances in understanding the structures and functions of complex networks~\cite{RevModPhys.74.47, PhysRep.469.93}.

Understanding the stability of a synchronized state is crucial for ensuring functionality in various natural and engineered complex systems. One of the most critical aspects of studying dynamical systems is understanding and predicting the trajectories' long-term or asymptotic behavior. The stability analysis of the trajectories of the dynamical systems answers the same~\cite{GlendinningBook}. Similarly, in the study of synchronization, the stability analysis of the synchronization state is essential to understand its long-term behaviors. In $1983$, Fujisaka and Yamada studied the stability of synchronous motion in coupled oscillator systems using the Lyapunov matrix~\cite{FujisakaYamada1983, YamadaFujisaka1983}. Later, in $1990$, Pecora and Carroll showed in their seminal paper that synchronization between two identical chaotic dynamical systems is possible under suitable coupling~\cite{PecoraCarrollPRL1990}, and the stability of the synchronization can be determined using the Lyapunov exponents~\cite{PecoraCarrollPRL1990,BarreiraBook}.
The Lyapunov exponents provide the exponential rate of convergence or divergence of nearby trajectories. Synchronization will be stable when all Lyapunov exponents that are transverse to the synchronization manifold are negative~\cite{PecoraCarrollPRL1990}. 
\begin{figure*}[tbh]
	\centering
\includegraphics[width=6.5in, height=1.8in]{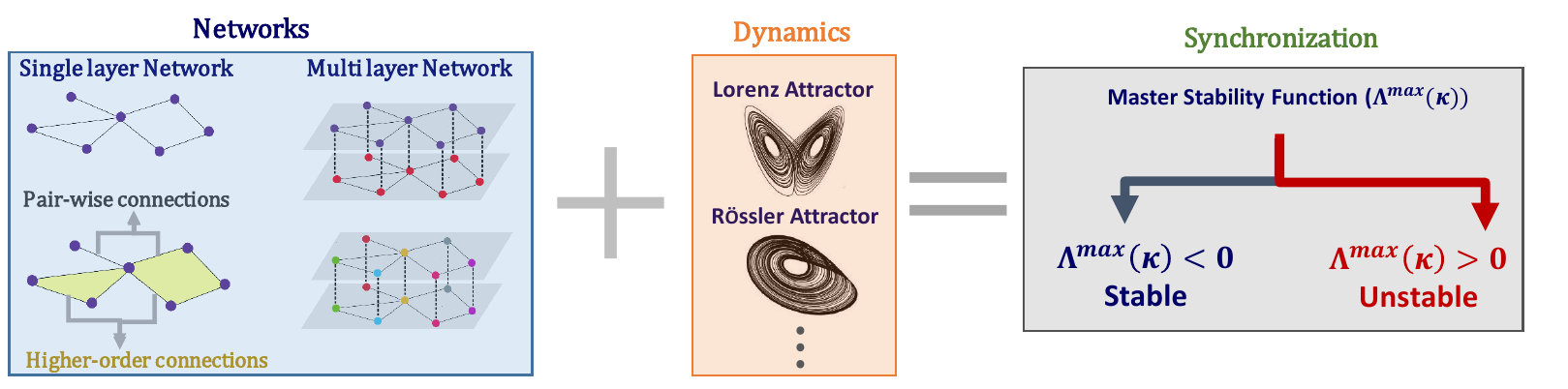}
\caption{\textbf{Portray the master stability function (MSF) in different networks and dynamical systems.} To predict the stability of synchronization, we analyze the MSF, which depends on the generic parameter $\kappa$. The parameter $\kappa$ (Eq.~(\ref{mse1})) encapsulates the influence of network topology through the Laplacian matrices. The functional form and values of MSF are determined by the dynamics of coupled oscillators, such as the Lorenz, R\"ossler, and other nonlinear dynamical systems (Table~\ref{table_MSF}).}
\label{schematic_MSF}
\end{figure*}

The Master Stability Function (MSF) provides a systematic way to assess the stability of synchronization across different networks.
The MSF was introduced by Pecora and Carroll in $1998$ to study the stability of synchronization for coupled identical systems in networks~\cite{PecoraCarrollPRL1998}. The MSF is calculated as the maximum Lyapunov exponent from a simple set of equations known as the master stability equations (MSE) as a function of the generic coupling parameter. The MSF separates the network information from the dynamical evolution of the coupled systems and can solve the stability problem for any network of a particular dynamical system and coupling function. The simple and elegant approach to characterize and compare the stability of synchronization has made MSF very popular and powerful~\cite{BarahonaPRL2002}. Further studies have shown that stable synchronization in many complex dynamical networks is constrained within a finite range of coupling parameters \cite{HuangPRE2009, BarahonaPRL2002, MotterPRE2005}. 
In Ref.~\cite{BarahonaPRL2002}, Barahona and Pecora derived the exact condition for stable synchronization using MSF and have shown that the small-world structures and clustering in networks can promote stable synchronization. 
In another study, Motter et al. have observed that degree heterogeneity can hinder synchronization in networks~\cite{MotterPRE2005}. In Ref.~\cite{motter2013spontaneous}, using the MSF approach, Motter et al. have shown that it is possible to identify the synchronization conditions for generators in power grid networks and identify the critical parameters of the generators for synchronization. Another study by Nishikawa and Motter identified the positive impact of negative interaction in synchronization using the MSF approach~\cite{NishikawaPNAS2010}. The MSF approach has recently been used to study pattern formation in ecology~\cite{PhysRevE.97.032307, PhysRevE.105.044310}. 
In the past $25$ years, several extensions of the MSF have been studied, such as in networks of coupled non-identical systems~\cite{SunEPL2009, SorrentinoEPL2011, AcharyyaEPL2012, AcharyyaPRE2015, NazerianEPL2023, NazerianCommPhys2023}, directed networks~\cite{NishikawaPRE2006,NishikawaPhysicaD2006}, multilayer networks \cite{BianconiBookML, PhysREvE.99.012304, SciAdv.2.e1601679},  hypergraphs \cite{MulasPRE2020} and in simplicial complexes \cite{NatComm.12.1255, Gambuzza2022}.

In this article, through an extensive literature survey, we provide a concise and clear derivation and algorithm of the MSF for networks of coupled oscillators.
In the illustrative Fig.~\ref{schematic_MSF}, we show that synchronization emerges as one of the collective phenomena when we consider dynamics on networks. The MSF ($\Lambda^{max}$) provides the criteria for stable synchronization. When the MSF $\Lambda^{max}<0$, the synchronization is stable; otherwise it is unstable. 
The MSF depends on the dynamics, the coupling function, and the generic parameter $\kappa$, which captures the information of the network structure. We begin MSF analysis by considering a network of coupled identical systems with diffusive coupling and then generalizing it to natural coupling schemes. We consider continuous-time dynamical systems that are modeled using differential equations, such as Lorenz and R\"ossler systems.

We also show systematic numerical and analytical MSF analysis for the coupled R\"ossler system under diffusive coupling function. We present the MSF framework in a simple and unified way to analyze the stability of synchronization for a broad class of dynamical networks, including undirected and directed networks with diffusive and natural coupling functions, multilayer networks with intralayer and interlayer couplings, and higher-order interactions networks. Furthermore, we highlight cases where the study of synchronization using MSF remains underexplored. Additionally, we discuss recent research focusing on MSF analysis using machine learning approaches.

\section{MSF in identical dynamical systems with diffusive coupling}\label{msf-iden-diff}

Consider an undirected graph or network, $\mathcal{G} =\{V, E\}$ where $V=\{v_1,\ldots,v_N\}$ is the set of vertices (nodes), $E = \{(v_i, v_j) | v_i, v_j \in V \}$ is the set of edges (connections) which contains the unordered pairs of vertices. We denote the adjacency matrix corresponding to $\mathcal{G}$ as ${\bm A} \in \mathbb{R}^{N \times N}$ which is defined as $a_{ij} = 1$ if there is an edge or link between the nodes $i$ and $j$, and $0$ otherwise. The cardinality of the set $|V|=N$ represents the number of nodes of the graph $\mathcal{G}$. Further, we consider $\bm{x}_{i}(t) \in \mathbb{R}^d$ as the dynamic state of the node $i$ at time $t$. The dynamics of the node $i$ can be written in a general form as,
\begin{equation}\label{1stdyneqn}
\begin{split}
    \dot{\bm{x}}_{i}(t) &= \bm {f}(\bm{x}_i(t)) + \sigma \sum_{j=1}^N a_{ij} \bm{h}(\bm{x}_j(t)-\bm{x}_i(t)),\;\; i = 1, \ldots, N 
\end{split}    
\end{equation}
where $\bm{x}_i(t)=(x^1_{i}(t),x^2_{i}(t),\ldots,x^d_{i}(t))^{T}$ is the state variables, $\bm {f}(\bm{x}_i):\mathbb{R}^d \mapsto \mathbb{R}^d$ is the self dynamics ($\bm {f}(\bm{x}_i)=(f^1(\bm{x}_i),f^2(\bm{x}_i),\ldots,f^d(\bm{x}_i)^{T}$) or the uncoupled dynamics of the node $i$, $\sigma$ is the scalar coupling parameter, $a_{ij}$ is the adjacency matrix entry, and $\bm {h}(\bm{x}_j): \mathbb{R}^d \mapsto \mathbb{R}^d$ ($\bm{h}(\bm{x}_j)=(h^1(\bm{x}_j),h^2(\bm{x}_j),\ldots,h^d(\bm{x}_j))^{T}$) is a linear diffusive coupling function. The dimension of the phase space is $N\times d$. Since, $\bm{h}(\bm{x})$ is a linear function, we can write Eq. (\ref{1stdyneqn}) in terms of the network Laplacian matrix as below
\begin{equation}\label{2stdyneqn}
\begin{split} 
    \dot{\bm{x}}_{i}(t) 
    &= \bm{f}(\bm{x}_i) + \sigma \sum_{j=1}^N a_{ij} [\bm{h}(\bm{x}_j)-\bm{h}(\bm{x}_i)]\\
    &= \bm{f}(\bm{x}_i) + \sigma \left[\sum_{j=1}^N a_{ij} \bm{h}(\bm{x}_j)-\bm{h}(\bm{x}_i)\sum_{j=1}^N a_{ij}\right]\\
    &= \bm{f}(\bm{x}_i) + \sigma \left[\sum_{j=1}^N a_{ij} \bm{h}(\bm{x}_j)-\bm{h}(\bm{x}_i) k_i\right]\\
    &= \bm{f}(\bm{x}_i) - \sigma \sum_{j=1}^N L_{ij} \bm{h}(\bm{x}_j)
\end{split}    
\end{equation}
where, 
\begin{equation}\label{laplacian_matrix}
L_{ij}=
   \begin{cases}
     - a_{ij}       & \quad \text{if} \ i \neq j \\
     k_i  & \quad i=j\\
   \end{cases}
\end{equation}
and in matrix form ${\bm L}={\bm D}-{\bm A}$ is the network Laplacian matrix where ${\bm D}=\text{diag}(k_1, k_2, \ldots, k_N)$ is a diagonal matrix with $i$th diagonal entry as the degree of node $i$ such that $k_i=\sum_{j=1}^{N} a_{ij}$. 

For a suitable coupling function $\bm{h}(\bm {x})$ and scalar coupling parameter $\sigma$, the coupled systems will synchronize. In the synchronized state, we can write $\bm{x}_1(t) = \bm{x}_2(t) = \cdots = \bm{x}_N(t) = \bm{s}(t)$, where $\bm{s}(t)\in \mathbb{R}^{d}$ is the solution of the uncoupled system
\begin{equation}
    \dot{\bm{s}}(t) = \bm{f}(\bm{s}(t)).
\label{syncdyn}
\end{equation}
In the synchronized state, the dynamic evolution of the coupled oscillators converges to the dynamical evolution of a single isolated oscillator in a $d$ dimensional hyperplane of the $d\times N$ dimensional phase space. This hyperplane is called the \emph{synchronization manifold}. The remaining direction in the phase space define the \emph{transverse manifold}.

We determine the stability of the synchronized state ($\bm{s}(t)$) using linear stability analysis by deriving the dynamics of small perturbations to the synchronization state.
Let us consider $\bm{\varepsilon}_i(t) 
 = \bm{x}_i(t)-\bm{s}(t)$ is the small perturbation to the synchronization state of the oscillator $i$. From Eq.~(\ref{1stdyneqn}) we can write
\begin{equation}
    \dot{\bm{s}}(t)+\dot{\bm{\varepsilon}}_{i}(t) = \bm{f}(\bm{s}(t) + \bm{\varepsilon}_i(t)) - \sigma\sum_{j=1}^N L_{ij} \bm{h}(\bm{s}(t) + \bm{\varepsilon}_j(t))
\label{pertdyn_1}
\end{equation}
Expanding $\bm{f}(\bm{s}(t) + \bm{\varepsilon}_i(t))$ and $\bm{h}(\bm{s}(t) + \bm{\varepsilon}_j(t))$ in Taylor's series and ignoring higher order terms we get,
\begin{subequations}
\label{tse_fh}
\begin{align}
    \bm{f}(\bm{s}(t) + \bm{\varepsilon}_i(t)) = \bm{f}(\bm{s}(t)) +  {\bm F}(\bm{s}(t)) \bm{\varepsilon}_i(t) \label{tse_f}
    \\
    \bm{h}(\bm{s}(t) + \bm{\varepsilon}_j(t)) = \bm{h}(\bm{s}(t)) +  {\bm H}(\bm{s}(t)) \bm{\varepsilon}_j(t)  \label{tse_h}
\end{align}
\end{subequations}
where, ${\bm F}(\bm{s}(t))=\left[\frac{\partial f^i}{\partial x^j}\right]$ and ${\bm H}(\bm{s}(t))=\left[\frac{\partial h^i}{\partial x^j}\right]$ are the $d\times d$ Jacobian matrices of $\bm{f}$ and $\bm{h}$ respectively at the synchronized solution $\bm{s}(t)$. Substituting Eqs.~(\ref{tse_f}) and ~(\ref{tse_h}) into Eq.~(\ref{pertdyn_1}) we get
\begin{equation}
    \dot{\bm{s}}(t)+\dot{\bm{\varepsilon}}_{i}(t) = \bm{f}(\bm{s}(t)) + {\bm F}(\bm{s}(t))\bm{\varepsilon}_i(t) -  \sigma\sum_{j=1}^N L_{ij}\left[ \bm{h}(\bm{s}(t)) + {\bm H}(\bm{s}(t)) \bm{\varepsilon}_j(t) \right]
    \label{perturb_Eqn_general}
\end{equation}
In the above equation, the first term, in the L.H.S. $\dot{\bm s}(t)$ is equal to the first term in the R.H.S. $\bm{F}(\bm{s}(t))$ from Eq. (\ref{syncdyn}). Hence, we can rewrite Eq. (\ref{perturb_Eqn_general}) as follow
\begin{equation}
    \dot{\bm{\varepsilon}}_{i}(t) = {\bm F}(\bm{s}(t))\bm{\varepsilon}_i(t) -  \sigma\sum_{j=1}^N L_{ij}\left[ \bm{h}(\bm{s}(t)) + {\bm H}(\bm{s}(t)) \bm{\varepsilon}_j(t) \right]
    \label{simplfy_perturb_Eqn_general}
\end{equation}
The summation term in the Eq. (\ref{simplfy_perturb_Eqn_general}) can be simplified further as follows: 
\begin{eqnarray}\label{laplacian_condition}
    \sigma\sum_{j=1}^N L_{ij}\left[ \bm{h}(\bm{s}(t)) + {\bm H}(\bm{s}(t))\bm{\varepsilon}_j(t)  \right] 
    & = & \sigma\sum_{j=1}^N L_{ij} \bm{h}(\bm{s}(t)) + \sigma\sum_{j=1}^N L_{ij}\left[ \bm{H}(\bm{s}(t))\bm{\varepsilon}_j(t) \right] \nonumber \\
    & = & \bm{h}(\bm{s}(t)) \sigma \sum_{j=1}^N L_{ij} + \sigma\sum_{j=1}^N L_{ij}{\bm H}(\bm{s}(t))\bm{\varepsilon}_j(t) \nonumber \\
    & = & \sigma\sum_{j=1}^N L_{ij}{\bm H}(\bm{s}(t))\bm{\varepsilon}_j(t)
\end{eqnarray}
since $\sum_{j=1}^{N} L_{ij} =0$. Thus, the linearized dynamics of the small perturbation can be written as
\begin{equation}\label{lindynpert}
    \dot{\bm{\varepsilon}}_{i}(t) = {\bm F}(\bm{s}(t)) \bm{\varepsilon}_i(t) -  \sigma\sum_{j=1}^N L_{ij} {\bm H}(\bm{s}(t))\bm{\varepsilon}_j(t),\;\; i=1,\ldots,N
\end{equation}
Here, Eq. (\ref{lindynpert}) captures the dynamics of all small perturbations to the synchronization state $\bm{s}(t)$. The synchronization state will be stable if all perturbations that are transverse to the synchronization manifold die with time. Note that the perturbations that are on the synchronization manifold or parallel to the synchronization manifold will not affect the stability of the synchronization manifold; only the perturbations that are transverse to the synchronization manifold will affect the stability of the synchronization manifold. In the following, we separate the perturbations which are transverse to the synchronization manifold from the perturbations which are parallel to the synchronization. We consider a $d\times N$ matrix as,
\begin{equation}
    \bm{E}(t) = \begin{bmatrix}
    \vert & \vert & \ldots &\vert&\ldots& \vert \\
    \bm{\varepsilon}_1(t) & \bm{\varepsilon}_2(t) & \ldots &\bm{\varepsilon}_i(t) & \ldots
    & \bm{\varepsilon}_N(t) \\
    \vert & \vert & \ldots& \vert& \ldots & \vert 
    \end{bmatrix}_{d\times N}
\end{equation}
where $i$th column of $\bm{E}(t)$ is the perturbation vector of $i$th oscillator. Using this form, we write the dynamical equations of small perturbation for all coupled oscillators in a single matrix equation (Eq. (\ref{lindynpert})) as 
\begin{equation}\label{lindynmat}
\begin{split}
\dot{\bm E}_{d\times N}&= {\bm F}_{d\times d}{\bm E}_{d\times N} - \sigma {\bm H}_{d\times d}{\bm E}_{d\times N}{\bm L}^{T}_{N\times N}\\ 
\end{split}
\end{equation}
where, ${\bm L}^T$ is the transpose of the coupling (Laplacian) matrix ${\bm L}$. Equation (\ref{lindynmat}) captures the dynamics of the perturbation in the full $d\times N$ phase space, including the perturbations transverse and parallel to the synchronization manifold. To separate the perturbations transverse to the synchronization manifold, we use a property of ${\bm L}$, the row sum is equal to zero ($\sum_j L_{ij}=0$). This will result in at least one zero eigenvalue of ${\bm L}$. If the network is connected, there will be exactly one zero eigenvalue of ${\bm L}$. We know the eigenvalues of a matrix and its transpose are the same. Thus, the transpose of the Laplacian ${\bm L}^T$ will have exactly one zero eigenvalue, let's say this eigenvalue is $\mu_1=0$, and the corresponding eigenvector is $\bm{e}_1= \bm{1}_n = (1,1,\ldots,1)^{T}$. Without loss of generality, we can arrange the eigenvalues of $\bm{L}^T$ in ascending order $0=\mu_1<\mu_2\leq\ldots\leq\mu_N$. The $k$th eigenvector $\bm{e}_k$ of matrix ${\bm L}^T$  and the corresponding eigenvalue ($\mu_k$) satisfy the eigenvalue equation,
\begin{equation}\label{eval_eq}
    {\bm L}^T \bm{e}_k = \mu_k \bm{e}_k
\end{equation}
Now we multiply Eq. (\ref{lindynmat}) by $\bm{e}_k$ from the right and using Eq. (\ref{eval_eq}) we get
\begin{eqnarray}
\label{intermediate_lindynmat}
\dot{\bm{E}}(t) \bm{e}_k & = & \bm{F}(\bm{s}(t)) \bm{E}(t) \bm{e}_k - \sigma \bm{H}(\bm{s}(t)) \bm{E}(t) {\bm L}^T \bm{e}_k 
 =  \bm{F}(\bm{s}(t)) \bm{E}(t) \bm{e}_k - \sigma \mu_k \bm{H}(\bm{s}(t)) \bm{E}(t) \bm{e}_k
\end{eqnarray}
There will be $N$ such equations corresponding to each eigenvalue of the Laplacian matrix. The above operation will separate the dynamics of small perturbations along the directions of the eigenvectors of the Laplacian matrix. We emphasize that we started with a $Nd$ dimension system, and using the spectral properties of ${\bm L}$ leads to a $N$ set of $d$ dimensional system. We denote this small perturbation along the direction of the $k$th eigenvector as $\bm{\psi}_k(t) = \bm{E}(t) \bm{e}_k$, Eq.~(\ref{intermediate_lindynmat}) is written as
\begin{eqnarray}
\dot{\bm{\psi}}_k(t) & = & \left[ \bm{F}(\bm{s}(t)) - \sigma \mu_k \bm{H}(\bm{s}(t)) \right] \bm{\psi}_k(t) \label{spectral_decom}
\end{eqnarray}
Here, we have $k=1,2,\ldots, N$ equations, which correspond to the respective eigen-mode of ${\bm L}$. The dynamical evolution of the small perturbations corresponding to $\mu_1 = 0$ lies within the synchronization manifold and does not affect its stability, and it is given by the below equation.
\begin{equation}
    \dot{\bm{\psi}}_1(t)  =  \bm{F}(\bm{s}(t)) \bm{\psi}_1(t),\;\text{for $\mu_1=0$}.
    \label{trlindyn}
\end{equation}
The remaining eigenvalues of the Laplacian matrix $(\mu_2, \ldots, \mu_N)$ will correspond to the transverse manifold. So, to determine the stability of the synchronization manifold, we only consider the dynamics of the perturbations corresponding to the nonzero eigenvalues of ${\bm L}$, and these dynamics are given by
\begin{equation}
    \dot{\bm{\psi}}_k(t) =  \left[\bm{F}(\bm{s}(t))  - \beta_k \bm{H}(\bm{s}(t)) \right] \bm{\psi}_k(t);\; k=2,\ldots, N. 
    \label{trlindyn}
\end{equation}
where, $\beta_k = \sigma \mu_k$. Equation  (\ref{trlindyn}) is same for all $k$, except $\beta_k$. At this point, to determine the stability of synchronization, we need to solve Eq.~(\ref{trlindyn}), which contains $N-1$ equations. It is a time varying equation, since matrix $\bm{F}(\bm{s}(t)) - \beta_k \bm{H}(\bm{s}(t))$ depends on the synchronized trajectory $\bm{s}(t)$, thus its eigenvalues are not constant. Therefore, to determine the stability of the synchronization manifold, we compute the maximum Lyapunov exponent ($\Lambda^{max}$) from Eq.~(\ref{trlindyn}). For time-varying systems, the Lyapunov exponent ($\Lambda = \lim_{t \to \infty} \frac{1}{t} \ln \frac{\|\bm{\psi}(t)\|}{\|\bm{\psi}(0)\|}$ \cite{lyapunov_1985}) measures the asymptotic growth rate of \(\|\bm{\psi}(t)\|\). We compute the Lyapunov exponent by employing a numerical approach \cite{lyapunov_1985}. The synchronized state will be stable if all (transverse) Lyapunov exponents are negative and solving Eq.~(\ref{trlindyn}) becomes tedious for systems with large $N$ (Algorithm \ref{stability_algo}). Instead we construct a single equation from Eq.~(\ref{trlindyn}) with the introduction of a generic coupling parameter $\kappa=\beta_k$,
\begin{equation}
    \dot{\bm{\psi}}(t)=\widetilde{{\bm F}}(s(t), \kappa) \bm{\psi}(t)
    \label{mse1}
\end{equation}
where $\widetilde{{\bm F}}(\bm{s}(t),\kappa) = \left[{\bm F}(\bm{s}(t))  - \kappa {\bm H}(\bm{s}(t)) \right]$. Equation (\ref{mse1}), is called the {\em Master Stability Equation} (MSE). When the system has a fixed point or time-independent solution, then the stability of the system can be determined from the eigenvalues of its Jacobian matrix $\widetilde{{\bm F}}(\bm{s}(t),\kappa)$ \cite{GlendinningBook}. However, when the system has a chaotic or time-dependent solution, the eigenvalues of the Jacobian matrix are no longer useful in determining its stability. Instead, we need to calculate the Lyapunov exponents \cite{BarreiraBook}. An important aspect of the MSE (Eq. (\ref{mse1})) is that it encapsulates the essence of the network structure within the generic coupling parameter $\kappa$, effectively decoupling the effects of the dynamics from the network structure (Algorithm \ref{MSF_algo}).

The maximum Lyapunov exponent ($\Lambda^{max}(\kappa)$) as a function of the generic coupling parameter ($\kappa$) is called the \textit{Master Stability Function} (MSF). The synchronization state is stable where $\Lambda^{max}(\kappa) < 0$. We denote a stable region as $\mathcal{R}_{\kappa} = \{ \kappa | \Lambda^{max}(\kappa) < 0 \}$. From the MSF, we get the stability criteria for all networks with the same self-dynamics and coupling functions (Algorithm \ref{Appl_MSF_algo}). The synchronization manifold is stable when the $\Lambda^{max}(\kappa)$ is negative. Note that the generic coupling parameter $\kappa$ is the product of the eigenvalues of the network Laplacian matrix $\mu_i$ and the coupling parameter $\sigma$. For any given network, we can easily determine its non-zero eigenvalues ($\mu_2,\ldots,\mu_N$) of the network Laplacian matrix ${\bm L}$. From the knowledge of $\mathcal{R}_{\kappa}$ and the non-zero eigenvalues of $\bm{L}$, we can determine the stability of synchronization in terms of the coupling parameter $\sigma$ for any given network. Thus, the MSF is a powerful tool and provides a unified solution to predict the stability of the synchronized state of any network for identical dynamics on nodes. 

\begin{algorithm}[h]
\caption{stability($\bm{f}$, $\bm{h}$, $\mathcal{G}(V,E)$)} \label{stability_algo}
\begin{algorithmic}
\STATE $N \leftarrow |V|$
\STATE ${\bm L} \leftarrow \mathcal{G}$
\STATE $d \leftarrow size(\bm{f})$
\WHILE {$ -\infty <\sigma < +\infty$} 
\STATE solve $\dot{\bm{x}}_{i}(t) = \bm{f}(\bm{x}_i) - \sigma \sum_{j=1}^N L_{ij} \bm{h}(\bm{x}_j)$ and $\dot{\bm{\varepsilon}}_{i}(t) = {\bm F}(\bm{s}(t)) \bm{\varepsilon}_i(t) - \sigma\sum_{j=1}^N L_{ij} {\bm H}(\bm{s}(t))\bm{\varepsilon}_j(t)$
\STATE Find $Nd$ Lyapunov exponents
\ENDWHILE
\STATE Find a region ($\mathcal{R}_{\sigma}$) in $\sigma$ all transverse Lyapunov exponents are negative.
\RETURN $\mathcal{R}_{\sigma}$
\end{algorithmic}
\end{algorithm}

\begin{algorithm}[h]
\caption{MSF($\bm{f}$, $\bm{h}$)} \label{MSF_algo}
\begin{algorithmic}
\STATE $d \leftarrow size(\bm{f})$
\WHILE {{$ -\infty <\kappa < +\infty$}}
\STATE solve $\dot{\bm{s}}(t) = \bm{f}(\bm{s}(t))$ and $\dot{\bm{\psi}}(t)=\widetilde{{\bm F}}(s(t), \kappa) \bm{\psi}(t)$
\STATE Find maximum Lyapunov exponents ($\Lambda^{max}(\kappa)$) 
\ENDWHILE
\STATE Find a region ($\mathcal{R}_{\kappa}$) where $\Lambda^{max}(\kappa) < 0$
\RETURN $\mathcal{R}_{\kappa}$
\end{algorithmic}
\end{algorithm}

\begin{algorithm}[h]
\caption{MSFclassification($\mathcal{G}(V,E)$, $\mathcal{R}_{\kappa}$)} \label{Appl_MSF_algo}
\begin{algorithmic}
\STATE ${\bm L} \leftarrow \mathcal{G}$
\STATE Find eigenvalues ($\mu_1,\mu_2, \ldots, \mu_N$) from ${\bm L}$
\IF{$R_{\kappa} = \{ \kappa | \Lambda^{max}(\kappa) < 0 \}=\O$}
\STATE Class $\mathcal{C}_0$: Synchronization state is unstable\\
\ELSIF{$R_{\kappa} = \{ (\kappa_1,\infty)| \Lambda^{max}(\kappa) < 0 \}$}
\STATE Class $\mathcal{C}_1$: Synchronization state is stable for $\sigma>\frac{\kappa_1}{\mu_2}$ 
\ELSIF{$R_{\kappa} = \{ (\kappa_1,\kappa_2)| \Lambda^{max}(\kappa) < 0 \}$}
\STATE Class $\mathcal{C}_2$: Synchronization state is stable for $\frac{\kappa_1}{\mu_2} < \sigma < \frac{\kappa_2}{\mu_N}$
\ELSIF{$R_{\kappa} = \{ (\kappa_1,\kappa_2),(\kappa_3,\infty)| \Lambda^{max}(\kappa) < 0 \}$}
\STATE 
\[
\text{Class } \mathcal{C}_3:
\begin{cases}
\text{Synchronization state is stable for } \frac{\kappa_1}{\mu_2} < \sigma < \frac{\kappa_2}{\mu_N},\\
\text{Synchronization state is stable for } \sigma > \frac{\kappa_3}{\mu_2}, \\
\text{Synchronization state is stable for } \frac{\kappa_1}{\mu_2} < \sigma < \frac{\kappa_2}{\mu_j} \text{ and } \frac{\kappa_3}{\mu_{j+1}} < \sigma \text{ where } j=3,\ldots,N-1.
\end{cases}
\]
\ELSIF{$R_{\kappa} = \{ (\kappa_1,\kappa_2),(\kappa_3,\kappa_4),\ldots | \Lambda^{max}(\kappa) < 0 \},$  $p$ \text{ finite intervals}} 
\STATE Class $\mathcal{C}_p$: Repeat $p$ times the cases of $\mathcal{C}_3$
\ENDIF
\end{algorithmic}
\end{algorithm}

\subsection{Classification of MSFs}\label{msf-class}
\begin{figure*}[tbh]
	\centering
\includegraphics[width=5.2in, height=3.2in]{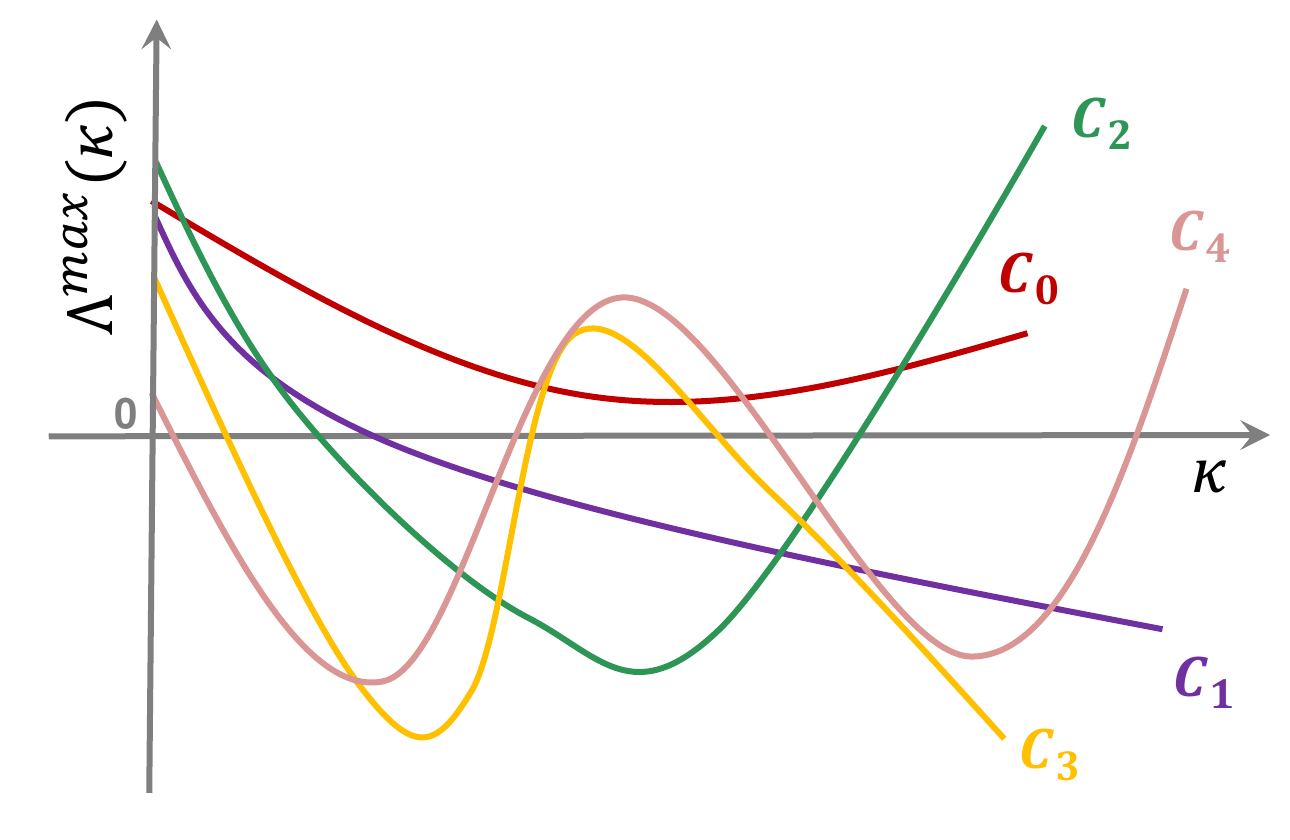}
\caption{{\bf Different classes of MSF.}  The MSF $\Lambda^{max}$ is plotted as a function of $\kappa$. The MSF can be classified into different classes based on the number of crossing points with the $x$-axis. For example, here we display five classes, $\mathcal{C}_0$ (red curve), $\mathcal{C}_1$ (purple curve), $\mathcal{C}_2$ (green curve), $\mathcal{C}_3$ (yellow curve) and $\mathcal{C}_4$ (brick red curve), and these classes have crossing points zero, one, two, three and four with the $x$-axis respectively.}
\label{MSF_different_classes}
\end{figure*}

We can classify the behaviors of the MSF from the numerical results in terms of the behaviors of $\Lambda^{max}$ crossing the $\kappa$ axis. We denote $\mathcal{C}_{p}$
be the class where $\Lambda^{max}$ has $p$ cross points with the $\kappa$ axis (Fig. \ref{MSF_different_classes}) and discussed more detail in Algorithm \ref{Appl_MSF_algo}.

\vspace{1mm}
\noindent {\bf Class $\mathcal{C}_{0}$:}  
$\Lambda^{max}$ has no finite intersection points with the $\kappa$-axis. As a result, $\Lambda^{max}$ remains strictly positive, preventing the coupled network of such oscillators from achieving a synchronized state.

\vspace{1mm}
\noindent {\bf Class $\mathcal{C}_{1}$:}
$\Lambda^{\max}$ has a single finite crossing point at $\kappa_1$. Beyond this point, for all $\kappa > \kappa_1$, $ \Lambda^{\max}$ remains negative. In this class, synchronization in the coupled network of oscillators occurs when $\sigma \mu_2 > \kappa_1$, where $\mu_2$ is the smallest nonzero eigenvalue of the Laplacian matrix  $\bm{L}$.

\vspace{1mm}
\noindent {\bf Class $\mathcal{C}_{2}$:}
$\Lambda^{\max}$ intersects the $\kappa$-axis at two finite points, $\kappa_1$ and $\kappa_2$, where $\kappa_1 < \kappa_2$. In this class, $ \Lambda^{\max}$ becomes negative at $ \kappa_1$ and remains negative within the finite interval $(\kappa_1, \kappa_2)$, before turning positive again at $\kappa_2$ and staying positive thereafter. For a coupled oscillator system in this category, synchronization is achieved if $\sigma \mu_2>\kappa_1$ and $\sigma \mu_N <\kappa_2$, where $\mu_2$ and $\mu_N$ are the smallest and largest eigenvalues of $\bm{L}$.

\vspace{1mm}
\noindent {\bf Class $\mathcal{C}_{3}$:}
$\Lambda^{\max}$ has three finite cross points $\kappa_1, \kappa_2, \text{ and } \kappa_3$. For this class, the coupled oscillator system is synchronizable if the eigenvalues of the coupling matrix satisfy that $\sigma \mu_i$, $i=2,\ldots, N$ all reside in the intervals $(\kappa_1,\kappa_2)$ and $(\kappa_3,\infty)$. More precisely, all $\sigma \mu_i$ reside in either in one interval, i.e., $\sigma \mu_2>\kappa_1$ and $\sigma \mu_N < \kappa_2$, or $\sigma \mu_2 >\kappa_3$. A general cases exists for certain $j$, satisfying the relations $\sigma \mu_2> \kappa_1$ and $\sigma \mu_j <\kappa_2$ and $\sigma \mu_{j+1} >\kappa_3$ where $j=3,\ldots,N-1$.

\vspace{1mm}
\noindent {\bf Class $\mathcal{C}_{p}$, $p=2,3,\ldots$:} $\Lambda^{\max}$ has $2p$ finite cross points, which can be grouped into $(\kappa_1,\kappa_2),(\kappa_3,\kappa_4),\ldots,(\kappa_r,\kappa_{r+1})$. For these classes, synchronization of the coupled oscillators can be more complex. Specifically, it occurs only when \( \sigma \mu_i \) for \( i = 2, \ldots, N \) all fall within the \( p \) stable intervals of the MSF. These values may either be contained within a single interval or distributed across all \( p \) stable intervals of the MSF \cite{HuangPRE2009}.

\subsection{Master Stability of coupled R\"ossler oscillators}
The R{\" o}ssler oscillator is a system of three coupled, first-order ordinary differential equations originally studied by the German biochemist Otto R{\" o}ssler in $1976$ \cite{rossler1976equation}. The R{\" o}ssler system is a classic example of a chaotic dynamical system, exhibiting a wide range of complex behaviors. We can find applications of the R{\" o}ssler oscillator in various fields due to its chaotic behavior and its ability to generate complex, non-periodic dynamics. Some applications include -- nonlinear dynamics studies, climate modeling, and finance \cite{HuangPRE2009, lee2011prediction}. 
It provides insights into the behavior of coupled oscillators and synchronization phenomena. Overall, the R{\"o}ssler oscillator is a valuable tool for studying complex, nonlinear dynamics and has applications in a wide range of fields beyond those listed above \cite{easaw2023estimation}. We consider chaotic R{\" o}ssler oscillators to model the dynamical evolution of nodes. The basic model \cite{rossler1976equation} is as follows
\begin{equation}\label{roseqn}
\begin{split}	
\dot{x} &=-y-z\\
\dot{y} &= x + a y\\
\dot{z} &= b + (x-c)z
\end{split}    
\end{equation}
where $a$, $b$, and $c$ are the R{\"o}ssler parameters, the values of these parameters are chosen such that the dynamical evolution of the oscillator will be chaotic. Let us now consider a network of $N$ coupled identical R{\"o}ssler oscillators. In the network, the dynamics of each node $i$ are modeled as follows 
\begin{equation}\label{rossler_equations}
\begin{split}
  \dot{x_i} & = - y_i-z_i+\sigma \sum_{j=1}^{N}a_{ij}(x_j-x_i)+\sigma \sum_{j=1}^{N}a_{ij}(y_j-y_i)+\sigma \sum_{j=1}^{N}a_{ij}(z_j-z_i)\\ 
  \dot{y_i} & = x_i + a y_i+\sigma \sum_{j=1}^{N}a_{ij}(x_j-x_i)+\sigma \sum_{j=1}^{N}a_{ij}(y_j-y_i)+\sigma \sum_{j=1}^{N}a_{ij}(z_j-z_i) \\ 
\dot{z_i} & = b + z_i(x_i - c)+\sigma \sum_{j=1}^{N}a_{ij}(x_j-x_i)+\sigma \sum_{j=1}^{N}a_{ij}(y_j-y_i)+\sigma \sum_{j=1}^{N}a_{ij}(z_j-z_i)
\end{split}
\end{equation} 
where $x_i$, $y_i$, and $z_i$ are the dynamical state variables of nodes in a network ranging $i=1,\ldots, N$. Here, $a_{ij}$ represents a connection between nodes $i$ and $j$ in $\mathcal{G}$, and $\sigma$ denotes the coupling strength between the connected nodes. Further, the coupling term can be simplified and expressed in the form of the Laplacian matrix $L$. For example, a diffusive coupling term in the $x$ variable in Eq. (\ref{rossler_equations}) is simplified as follow
\begin{equation}
\begin{split}
&\sigma \sum_{j=1}^{N} a_{ij}({x}_{j}-{x}_{i})=\sigma \sum_{j=1}^{N} a_{ij}{x}_{j}-\sigma \sum_{j=1}^{N} a_{ij}{x}_{i}\\
&=\sigma \sum_{j=1}^{N} a_{ij} {x}_{j}-\sigma k_i {x}_{i}\\
&=\sigma[a_{i1} {x}_{1}+a_{i2} {x}_{2}+\ldots+\ldots+a_{iN} {x}_{N}]-\sigma k_i {x}_{i}\\
&=\sigma[a_{i1} {x}_{1}+a_{i2} {x}_{2}+\ldots- k_i {x}_{i}+\ldots+a_{iN} {x}_{N}]\\
&= - \sigma[L_{i1} {x}_{1}+L_{i2} {x}_{2}+\ldots+L_{ii} {x}_{i}+\ldots+L_{iN} {x}_{N}]\\
&= - \sigma \sum_{j=1}^{N} L_{ij} {x}_{j}
\end{split} 
\end{equation}
where $k_i=\sum_{j=1}^{N} a_{ij}$, is the degree of node $i$ and $L_{ij}$ is the Laplacian matrix entry (Eq. (\ref{laplacian_matrix})). Similarly, we can rewrite all the coupling terms in the Laplacian form; hence, dynamics of node $i$ in the coupled R{\"o}ssler oscillator can be represented as follows
\begin{equation}
\begin{split}
\dot{x}_i&=-y_i-z_i - \sigma \sum_{j=1}^N L_{ij} x_j - \sigma \sum_{j=1}^N L_{ij} y_j - \sigma \sum_{j=1}^N L_{ij} z_j\\
\dot{y}_i&=-x_i+a y_i - \sigma \sum_{j=1}^N L_{ij} x_j - \sigma \sum_{j=1}^N L_{ij} y_j - \sigma \sum_{j=1}^N L_{ij} z_j\\
\dot{z}_i&=b+z_i(x_i-c) - \sigma \sum_{j=1}^N L_{ij} x_j - \sigma \sum_{j=1}^N L_{ij} y_j - \sigma \sum_{j=1}^N L_{ij} z_j \\
\end{split}
\label{rossler_withLaplacianForm}
\end{equation}

For simplicity, we define the functional form of the coupled system for each state variable and rewrite Eq. (\ref{rossler_withLaplacianForm}) as follow
\begin{equation}
\begin{split}
\dot{x}_i&=f^{1}(x_i,y_i,z_i) - \sigma \sum_{j=1}^N L_{ij} h^{1}(x_j,y_j,z_j)\\
\dot{y}_i&=f^{2}(x_i,y_i,z_i) - \sigma \sum_{j=1}^N L_{ij} h^{2}(x_j,y_j,z_j)\\
\dot{z}_i&=f^{3}(x_i,y_i,z_i) - \sigma \sum_{j=1}^N L_{ij} h^{3}(x_j,y_j,z_j) \\
\end{split}
\label{rossler_withLaplacianForm2}
\end{equation}
where $f^{1}(x_i,y_i,z_i) = -y_i-z_i$, $f^{2}(x_i,y_i,z_i) = -x_i + ay_i$, $f^{3}(x_i,y_i,z_i) = b+z_i(x_i-c)$ and the coupling functions as $h^{1}(x_j,y_j,z_j) = x_j+y_j+z_j$, $h^{2}(x_j,y_j,z_j)=x_j+y_j+z_j$, and $h^{3}(x_j,y_j,z_j)=x_j+y_j+z_j$. From Eq. (\ref{rossler_withLaplacianForm2}), we can get a synchronized oscillatory state and expressed as $x_i(t) = x^s(t)$, $y_i(t) = y^s(t)$, and $z_i(t) = z^s(t) \;\;\forall i$. Note that the solution is not a fixed point. We give a small perturbation around the solution ($x_i(i) = x^s(t) + \varepsilon^{x}_{i}(t)$, $y_i(t) = y^s(t) + \varepsilon^{y}_{i}(t)$, $z_i(i) = z^s(t) + \varepsilon^{z}_{i}(i)$) and linearize the system. After perturbation, the first equation of the Eq. (\ref{rossler_withLaplacianForm2}) is expressed as 
\begin{equation}\label{rossler_withLaplacianForm3}
\begin{split}
\dot{x}^s + \dot{\varepsilon}_{x}^{i} &= f^{1}(x^s + \varepsilon^{x}_{i},y^s + \varepsilon^{y}_{i}, z^s + \varepsilon^{z}_{i}) - \sigma \sum_{j=1}^N L_{ij} h^{1}(x^{s}+\varepsilon^{x}_{i},y^{s}+\varepsilon^{y}_{i},z^{s}+\varepsilon^{z}_{i})
\end{split}
\end{equation}

\noindent Now, after applying Taylor series expansion and ignoring the higher-order terms gives
\begin{equation*}
\begin{split}
f^1\big(x^s + \varepsilon_i^x, y^s + \varepsilon_i^y, z^s + \varepsilon_i^z\big) &\approx f^1(x^s, y^s, z^s) + \varepsilon_i^x \frac{\partial f^1}{\partial x_i}\bigg|_{(x^s, y^s, z^s)} + \varepsilon_i^y \frac{\partial f^1}{\partial y_i}\bigg|_{(x^s, y^s, z^s)} + \varepsilon_i^z \frac{\partial f^1}{\partial z_i}\bigg|_{(x^s, y^s, z^s)}\\
&= \dot{x}^s-\varepsilon^{y}_{i}-\varepsilon^{z}_{i}
\end{split}
\end{equation*}
where in synchronized state, $\dot{x}^s= f^1(x^s,y^s,z^s)$, $\dot{y}^s= f^2(x^s,y^s,z^s)$, and $\dot{z}^s= f^3(x^s,y^s,z^s)$ (Eq. (\ref{syncdyn})). On the other hand $f^1(x_i,y_i,z_i) = -y_i-z_i$, thus $\frac{\partial f^1}{\partial x_i} = 0$, $\frac{\partial f^1}{\partial y_i}=-1$, and $\frac{\partial f^1}{\partial z_i}=-1$. Furthermore, from the properties of the Laplacian matrix, we have $\sum_{j=1}^N L_{ij}=0$ (Eq. (\ref{laplacian_condition})). Finally, $\frac{\partial h^{1}(x_j,y_j,z_j)}{\partial x_j} =\frac{\partial}{\partial x_j} (x_j+y_j+z_j)=1$, $\frac{\partial h^{1}(x_j,y_j,z_j)}{\partial y_j} =\frac{\partial}{\partial y_j} (x_j+y_j+z_j)=1$, and $\frac{\partial h^{1}(x_j,y_j,z_j)}{\partial z_j} =\frac{\partial}{\partial z_j} (x_j+y_j+z_j)=1$. Hence,  
\begin{equation*}
\begin{split}
\sigma \sum_{j=1}^N L_{ij}h^1\big(x^s + \varepsilon_i^x, y^s + \varepsilon_i^y, z^s + \varepsilon_i^z\big) &\approx  \sigma \sum_{j=1}^N L_{ij} 
 h^1(x^s, y^s, z^s) \\
 &+ \sigma \sum_{j=1}^N L_{ij} \biggl[\varepsilon_i^x \frac{\partial h^1}{\partial x_i}\bigg|_{(x^s, y^s, z^s)} + \varepsilon_i^y \frac{\partial h^1}{\partial y_i}\bigg|_{(x^s, y^s, z^s)} + \varepsilon_i^z \frac{\partial h^1}{\partial z_i}\bigg|_{(x^s, y^s, z^s)}\biggr]\\
 &= \sigma \sum_{j=1}^N L_{ij} \biggl[\varepsilon^{x}_{i}+\varepsilon^{y}_{i}+\varepsilon^{z}_{i}\biggr]
\end{split}
\end{equation*}
Hence, from the above, by combining, we can write (Eq. (\ref{rossler_withLaplacianForm3})) the linearized system  as follows
\begin{equation}
\begin{split}
\dot{\varepsilon}^{x}_{i} &= -\varepsilon^{y}_{i}-\varepsilon^{z}_{i} - \sigma \sum_{j=1}^N L_{ij}\varepsilon^{x}_{j}  - \sigma \sum_{j=1}^N L_{ij}  \varepsilon^{y}_{j} - \sigma \sum_{j=1}^N L_{ij}\varepsilon^{z}_{j}
\end{split}
\end{equation}
Similarly, the linearized equation of the state variables $y_i$ and $z_i$ can be written as follows
\begin{equation}
\begin{split}
\dot{\varepsilon}^{y}_{i} &= -\varepsilon^{x}_{i} +a \varepsilon^{y}_{i} - \sigma \sum_{j=1}^N L_{ij}\varepsilon^{x}_{j} - \sigma \sum_{j=1}^N L_{ij}  \varepsilon^{y}_{j} - \sigma \sum_{j=1}^N L_{ij}\varepsilon^{z}_{j}\\
\dot{\varepsilon}^{z}_{i} &= z^{s} \varepsilon^{x}_{i}  + (x^{s}-c)\varepsilon^{z}_{i} - \sigma \sum_{j=1}^N L_{ij}\varepsilon^{x}_{j} - \sigma \sum_{j=1}^N L_{ij}  \varepsilon^{y}_{j} - \sigma \sum_{j=1}^N L_{ij}\varepsilon^{z}_{j}
\end{split}
\end{equation}
Hence, for the $ith$ node, the self-dynamic term and coupling matrix can be represented as in matrix notation (Eq. (\ref{lindynpert}))
\begin{equation}\label{ross_matrix}
    \begin{gathered}
\begin{pmatrix}
 \dot{\varepsilon^{x}_{i}} \\
 \dot{\varepsilon^{y}_{i}} \\
 \dot{\varepsilon^{z}_{i}} \\
\end{pmatrix}
=
\begin{pmatrix}
  0 & -1 & -1 \\
  1 & a & 0 \\
  z^{s}(t) & 0 & x^{s}(t)-c 
\end{pmatrix}
\begin{pmatrix}
\varepsilon^{x}_{i} \\
\varepsilon^{y}_{i} \\
\varepsilon^{z}_{i} \\
\end{pmatrix} - \sigma
\sum_{j=1}^{N} L_{ij}\begin{pmatrix}
  1 & 1 & 1 \\
  1 & 1 & 1 \\
  1 & 1 & 1 
\end{pmatrix}\begin{pmatrix}
 \varepsilon^{x}_{j} \\
 \varepsilon^{y}_{j} \\
 \varepsilon^{z}_{j} \\
\end{pmatrix}
\end{gathered}
\end{equation}
\begin{equation}
\begin{split}
\dot{\bm{\varepsilon}}_{i}(t) &= {\bm F}(\bm{s}(t))\bm{\varepsilon}_i(t) - \sigma \sum_{j=1}^N L_{ij}  {\bm H}(\bm{s}(t)) \bm{\varepsilon}_{j}(t)
\end{split}
\end{equation}
\begin{figure*}[htb]
\centering
\includegraphics[width=6.5in, height=3.8in]{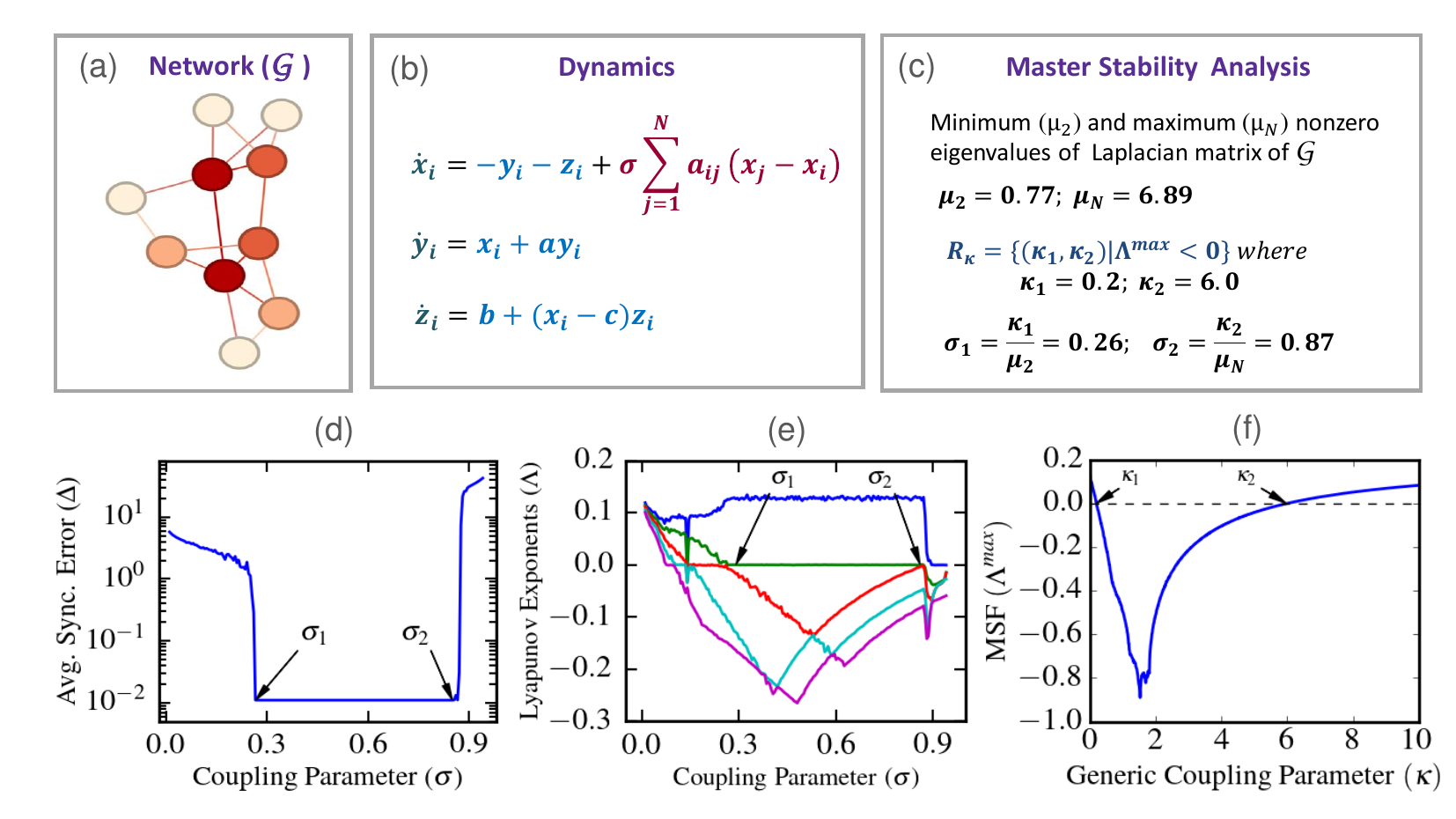}
\caption{{\bf Master Stability Analysis of coupled R\"ossler Oscillators.} To analyze master stability, we take (a) a network of $N=10$ nodes with random configuration. (b) Identical R{\"o}ssler oscillators with parameters $a = 0.15, b = 0.2, \text{ and } c = 10.0$ are placed on the nodes of this network, and these oscillators are coupled in the $x$ component. To find the range of the coupling parameter for stable synchronization, (d) we can determine the average synchronization error from Eq.~(\ref{sync_err}) or (e) the Lyapunov exponents. From (d) and (e), we can observe that the synchronization state is stable in the range $(\sigma_1\leq \sigma\leq\sigma_2)$ where $\sigma_1\approx 0.26$ and $\sigma_2\approx 6.0$. The same $\sigma$ range can be retrieved from the Master Stability Analysis as in (c). (c) The minimum and maximum nonzero eigenvalues of the Laplacian matrix of the network in (a) are $\mu_2 = 0.77, \text{ and } \mu_N = 6.89$, respectively. Further using the values of $\kappa_1$ and $\kappa_2$ from (f) we can determine the range of the coupling strength $\sigma$ where the synchronization state of the coupled R{\"o}ssler oscillators is stable. (f) The Master Stability Function for $x$ component coupled R\"ossler oscillator $(\Lambda^{max})$ is plotted as a function of generic coupling function $\kappa$. The MSF is negative in the range $\kappa_1=0.2$ and $\kappa_2=6.0$. This defines the stable region $R_{\kappa}=\{(\kappa_1,\kappa_2) \vert \Lambda^{max} < 0\}$. }
\label{MSF_results}
\end{figure*}
where $\bm{\varepsilon}_i=(\varepsilon^{x}_{i},\varepsilon^{y}_{i},\varepsilon^{z}_{i})^{T}$ and $\bm{s}(t)=({x}^{s}(t),{y}^{s}(t),{z}^{s}(t))^{T}$. Now we consider the network of $N$ number of oscillators. To find the perturbation propagation behavior around the synchronous state over the network of $N$ number of nodes, each having three oscillator variables, we introduce a matrix with $d=3$ rows for R{\"o}ssler oscillator and $N$ columns as
\begin{equation}
\begin{split}
{\bm E}(t) &= \begin{pmatrix}
\varepsilon^{x}_{1} & \varepsilon^{x}_{2}&\hdots & \varepsilon^{x}_{N} \\
\varepsilon^{y}_{1}&\varepsilon^{y}_{2}&\hdots&\varepsilon^{y}_{N}\\
\varepsilon^{z}_{1}&\varepsilon^{z}_{2}&\hdots&\varepsilon^{z}_{N} \\
\end{pmatrix}_{3 \times N}
\end{split}
\end{equation}
Now in matrix form 
\begin{equation}
\dot{{\bm E}}_{3\times N}= {\bm F}_{3\times 3}{\bm E}_{3 \times N} - \sigma {\bm H}_{3 \times 3}{\bm E}_{3\times N}{\bm L}^{T}_{N\times N} 
\end{equation}
We assume that ${\bm L}$ is a diagonalizable matrix and can be decomposed into separate equations along eigenvectors of ${\bm L}$. By multiplying the eigenvector of ${\bm L}$, we get Eq. (\ref{spectral_decom}). Finally, we get the master stability equation (Eq. (\ref{mse1})), and by solving it, we get the master stability function for the R\"ossler oscillator. For other coupled oscillators, we can easily get ${\bm F}$ (Table \ref{table_MSF}) and the MSF.

\begin{figure*}[htb]
\centering
\includegraphics[width=6.5in, height=4.6in]{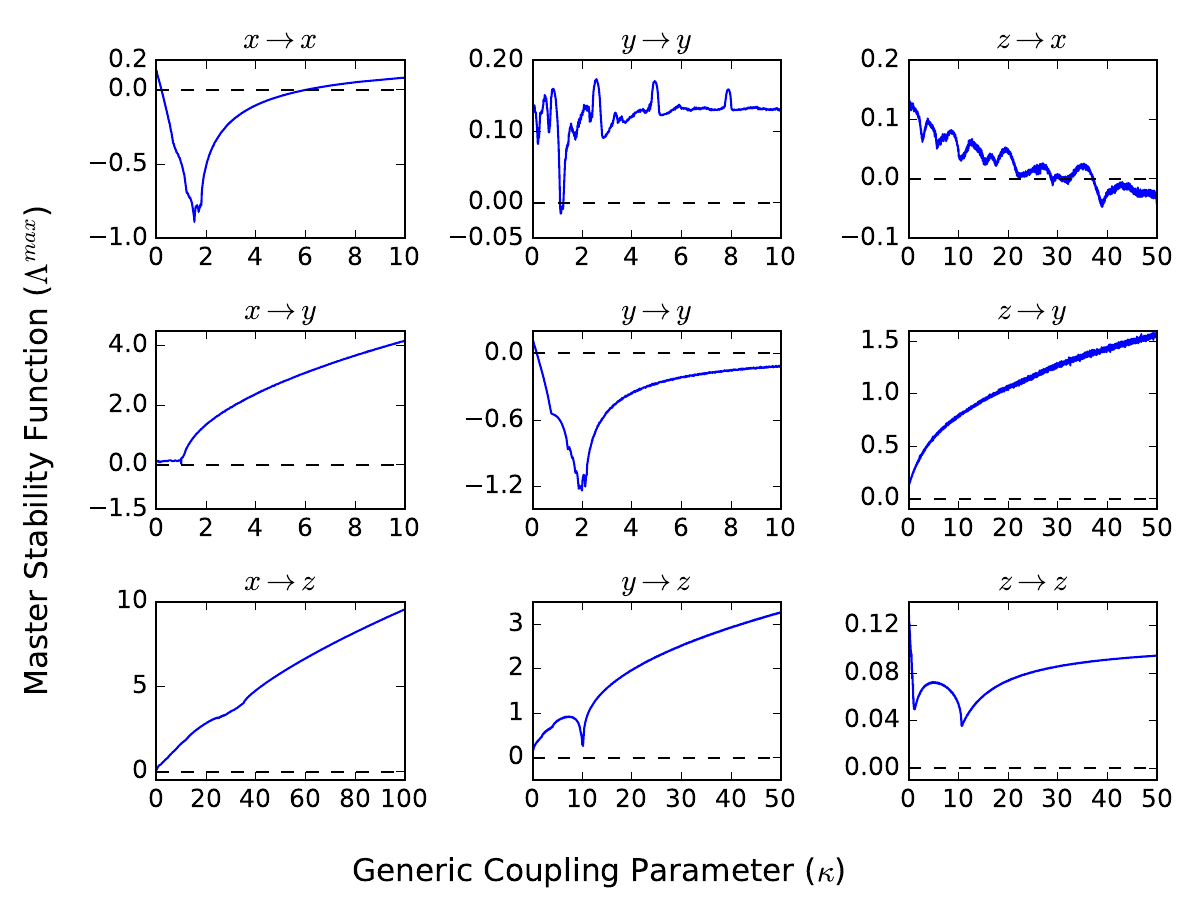}
\caption{{\bf Master Stability Function of R\"ossler oscillators underlying various coupling schemes.} We implement coupling in different variables and plot the MSF as a function of the generic parameter $\kappa$. Here the coupling $x \rightarrow x$, $y \rightarrow x$ and $z \rightarrow x$ means the $x$ variable of the $i$th oscillator is coupled through the $x$, $y$ and $z$ variable of the $j$th oscillator respectively and others follow. Based on the transitions from unsynchronized state to synchronized state and vice-versa, we can classify the different classes of the MSF as discussed in section~\ref{msf-class}.}
\label{MSF_results_classes}
\end{figure*}

For numerical analysis, we consider a network of $N=10$ identical R{\"o}ssler oscillators  (Eq. (\ref{rossler_equations}) ) connected through the $x$-component in a network structure given in Fig.~(\ref{MSF_results}(a, b)). The R{\"o}ssler parameters are $a=0.15, b=0.2, \text{ and } c =10.0$ so that all oscillators have chaotic dynamics. The dimension of a single chaotic R{\"o}ssler oscillator is $d=3$, and the number of coupled oscillators is $N=10$ in a network. Thus, the dimension of the phase space is $Nd = 30$. We solve the system (Eq.~(\ref{1stdyneqn})) numerically by varying the scalar coupling parameter $\sigma$. We compute the average synchronization error $\Delta$ (Fig.~(\ref{MSF_results}(d))) and the Lyapunov exponents $\Lambda$ (Fig. \ref{MSF_results}(e)) as a function of the coupling parameter $\sigma$. The average synchronization error ($\Delta$) is define as
\begin{equation}
    \label{sync_err}
    \Delta = \left\langle \left( \frac{1}{N(N-1)}\sum_{i, j=1}^{N}  \lVert \bm{x}_i - \bm{x}_j \rVert \right)^{\frac{1}{2}} \right \rangle_{T}
\end{equation}
here, $\langle\rangle_{T}$ denotes time average. As the scalar coupling parameter $\sigma$ increases from zero, the coupled R\"ossler oscillators synchronize at $\sigma=\sigma_1\approx 0.26$ (Fig.~\ref{MSF_results}(d, e)). The global synchronization error $\Delta$ becomes small $(\sim 10^{-2})$ (Fig. \ref{MSF_results}(d)) and all transverse Lyapunov exponents become negative (Fig. \ref{MSF_results}(e)). There is one positive and one zero Lyapunov exponent corresponding to the expanding and neutral directions on the synchronization manifold. As the coupling parameter is further increased, the coupled R\"ossler oscillators desynchronize at $\sigma=\sigma_2\approx 0.87$ (Fig. \ref{MSF_results}(d, e)) \cite{AcharyyaChaos2011}. This provides the stability region for the given network in Fig.~(\ref{MSF_results})(a) as $(\sigma_1, \sigma_2)$ by direct numerical study.

In Fig.~(\ref{MSF_results})(f), the Master Stability Function (MSF) for the $x$-coupled R\"ossler oscillator is plotted as a function of the generic coupling parameter $\kappa$. The MSF is determined as the largest Lyapunov Exponent $\Lambda^{\text{max}}$ as a function of the generic coupling parameter $\kappa$ by solving Eq.~(\ref{mse1}) and an isolated R\"ossler system (Eq.~(\ref{roseqn})) together (Algorithm \ref{MSF_algo}). From Fig. \ref{MSF_results}(c, f) we get the stability region $R_{\kappa} = \{(\kappa_1,\kappa_2)\vert \Lambda^{\text{max}} < 0\}$, where $\kappa_1 = 0.2$ and $\kappa_2 = 6.0$. We can use this information to determine the stability of the given network in Fig.~(\ref{MSF_results})(a) in terms of the scalar coupling parameter $\sigma$. The minimum and the maximum eigenvalues of the network are $\mu_2\approx 0.77$ and $\mu_N\approx 0.87$, respectively. This will provide us the minimum and maximum values of the scalar coupling parameter for synchronization as $\sigma_1=\kappa_1/\mu_2 \approx 0.26$ and $\sigma_2=\kappa_2/\mu_N \approx 0.87$. Thus, we can retrieve the stability of synchronization using the MSF. We can observe that MSF belongs to the $\mathcal{C}_2$ class. In the above we only consider $x\rightarrow x$ coupling inside the ${\bm H}$ matrix (Fig. \ref{MSF_results}(b)). However, considering the other coupling functions, we can get some of the other classes ($\mathcal{C}_0$ and $\mathcal{C}_1$) of MSF behavior (Fig. \ref{MSF_results_classes}).

\begin{table} 
\centering 
\renewcommand{\arraystretch}{1.3}
\begin{tabular}{|c| c| c|}
\hline 
\textbf{Oscillators} & \textbf{Dynamical Equations} & \textbf{Jacobian matrix} \\[0.5ex] 
\hline 
R{\"o}ssler Systems \cite{rossler1976equation}& $\begin{array}{lcl} \dot{x} = -y-z\\ \dot{y} = x+a y\\ \dot{z} = b+z(x-c) \end{array}$ &   ${\bm F}= \begin{pmatrix}
  0 & -1 & -1 \\
  1 & a & 0 \\
  z & 0 & x-c 
\end{pmatrix}$\\ 
\hline 
Lorenz Systems \cite{lorenz1963deterministic}& $\begin{array} {lcl} \dot{x} = \sigma(y-x)\\ \dot{y} = x(\rho-z)- y\\ \dot{z} = xy - \beta z \end{array}$ & 
${\bm F}= \begin{pmatrix}
  -\sigma & \sigma & 0 \\
  \rho-z & -1 & -x \\
  y & x & -\beta 
\end{pmatrix}$ \\ 
\hline 
Chens Systems \cite{chen1999yet}& $\begin{array} {lcl} 
\dot{x} = a(y-x)\\
\dot{y} = x(c-a-z)+ cy\\
\dot{z} = xy-\beta z \end{array}$ & ${\bm F}= \begin{pmatrix}
  -a & a & 0 \\
  c-a-z & c & -x \\
  y & x & -\beta 
\end{pmatrix}$ \\    \hline 
HR neuron \cite{dhamala2004transitions}& $\begin{array} {lcl} 
\dot{x} = y+3x^{2}-x^{3}-z+I\\
\dot{y} = 1-5x^{2}- y\\
\dot{z} = -rz +rs(x+1.6) \end{array}$ &${\bm F}= \begin{pmatrix}
  3x(2-x) &1  &-1  \\
   -10x&  -1&0  \\
  rs &  0&-r 
\end{pmatrix}$ \\ 
\hline 
Forced Duffing \cite{stefanski2007ragged}& $\begin{array} {lcl} 
\dot{x} = y\\
\dot{y} = -hy-x^{3}+ q\sin (\eta t)
 \end{array}$ & ${\bm F}= \begin{pmatrix}
  0 &1  \\
   -3x^2&  -h  
\end{pmatrix}$\\ 
\hline
Forced van der Pol \cite{mettin1993bifurcation}& $\begin{array} {lcl}
\dot{x} = y\\
\dot{y} = -x+d(1-x^2)y+ F \sin (\eta t)
 \end{array}$ &${\bm F}= \begin{pmatrix}
  0 &1  \\
   -1-2dx y&  d(1-x^2)  
\end{pmatrix}$\\ 
\hline
\end{tabular}
\caption{We present the dynamical equations and the corresponding Jacobian matrices of various oscillators.}
\label{table_MSF}
\end{table}

\newcommand\bigzero{\makebox(0,0){\text{\huge0}}}

\section{MSF in Identical Systems with Natural Coupling}

The dynamical systems found in nature are rarely related by linear coupling function. Instead, most of the time, the coupling function is complex in nature~\cite{StankovskiRMP2017}. So, it is essential to extend the MSF analysis for the more general form of coupling functions. In this section, we consider a general form of the coupling function, which is relaxing the linearity condition. Let the dynamical equation for the $ith$ oscillator be
\begin{equation}\label{1stdyneqn_nc}
	\begin{split}
		\dot{\bm{x}}_{i}(t) &= \bm{f}(\bm{x}_i(t)) + \sigma \sum_{j=1}^N a_{ij} \bm{h}(\bm{x}_i(t),\bm{x}_j(t)),\;\; i = 1, \ldots, N 
	\end{split}    
\end{equation}
where $\bm{x}_i(t) \in \mathbb{R}^d$, 
$\bm{f}(\bm{x}_i):\mathbb{R}^d\rightarrow \mathbb{R}^d$ is the nonlinear self dynamics, and $\bm{h}(\bm{x}_i(t),\bm{x}_j(t)): \mathbb{R}^d \rightarrow \mathbb{R}^d$ is a \emph{nonlinear} coupling function that captures the effect of the nearest neighbor vertex $j$ of the network on the vertex $i$. Here, $a_{ij}$ is the adjacency matrix term, and $\sigma$ is the scalar coupling parameter. The only condition we impose on the coupling function is that it is zero on the synchronization manifold. So, the coupling function satisfies $\bm{ h}(\bm{x},\bm{x})=0$. In the synchronization state $\bm{x}_i=\bm{s}(t),\forall\; i$. The stability of the synchronization is analyzed with the dynamics of the small perturbations, ${\bm \varepsilon}_i(t)={\bm x}_i(t) - {\bm s}(t)$
\begin{equation}\label{pertdyn_nc1}
	\dot{\bm \varepsilon}_i = \left.\frac{\partial {\bm f}}{\partial \bm{x}_i}\right\vert_{\bm{s}} \bm{\varepsilon}_i+\sigma \sum_{j=1}^{N} a_{ij} \biggl[ \frac{\partial \bm{h}(\bm{x}_i,\bm{x}_j)}{\partial \bm{x}_i}\bm{\varepsilon}_i+\frac{\partial \bm{h}(\bm{x}_i,\bm{x}_j)}{\partial \bm{x}_j} \bm{\varepsilon}_j \biggr] 
\end{equation}
We know $\bm{h}(\bm{x},\bm{x})=0$. The complete derivation at $(\bm{s},\bm{s})$ 
\begin{equation}\label{identity1_nc}
	\left.\frac{\partial \bm{h}(\bm{x}_i, \bm{x}_j)}{\partial \bm{x}_i}\right\vert_{\bm{s}}+\left.\frac{\partial \bm{h}(\bm{x}_i,\bm{x}_j)}{\partial \bm{x}_j}\right\vert_{\bm{s}}=0 
\end{equation}
Using Eq.~(\ref{identity1_nc}) in Eq.~(\ref{pertdyn_nc1}) we get
\begin{equation}
	\begin{split}
		\dot{\bm \varepsilon}_i &= {\bm F}(\bm{s}) \bm{\varepsilon}_i+\sigma \sum_{j=1}^{N} a_{ij} \biggl[ \left.\frac{\partial \bm{h}(\bm{x}_i,\bm{x}_j)}{\partial \bm{x}_i}\right\vert_{\bm{s}}\bm{\varepsilon}_i+\left.\frac{\partial \bm{h}(\bm{x}_i,\bm{x}_j)}{\partial \bm{x}_j}\right\vert_{\bm{s}}\bm{\varepsilon}_j \biggr] \\
		&= {\bm F}(\bm{s}) \bm{\varepsilon}_i- \sigma \left.\frac{\partial \bm{h}(\bm{x}_i,\bm{x}_j)}{\partial \bm{x}_j}\right\vert_{\bm{s}}  \biggl[ k_i \bm{\varepsilon}_i- \sum_{j=1}^{N} a_{ij} \bm{\varepsilon}_j \biggr] \\
		&= {\bm F}(\bm{s}) \bm{\varepsilon}_i-\sigma \left.\frac{\partial \bm{h}(\bm{x}_i,\bm{x}_j)}{\partial \bm{x}_j}\right\vert_{\bm{s}} \sum_{j=1}^{N} L_{ij} \bm{\varepsilon}_j\\
		&= {\bm F}(\bm{s}) \bm{\varepsilon}_i - \sigma {\bm H}(\bm{s},\bm{s}) \sum_{j=1}^{N} L_{ij}  \bm{\varepsilon}_j
	\end{split}
\end{equation}
Here, we denote $\left.\frac{\partial {\bm f}}{\partial \bm{x}_i}\right\vert_{\bm{s}} = \bm{F}(\bm{s})$ and $\left.\frac{\partial \bm{h}(\bm{x}_i,\bm{x}_j)}{\partial \bm{x}_j}\right\vert_{\bm{s}} = \bm{H}(\bm{s},\bm{s})$ as the Jacobian of the self-dynamics $(\bm{f})$ and the coupling function $(\bm{h})$ at the synchronization solution $(\bm{s})$ respectively. $L_{ij}$ is the Laplacian matrix elements as defined in Eq.~(\ref{laplacian_matrix}) For $N$ number of oscillators, we can write in matrix form as 
\begin{equation}	\label{pertdyn_mat_nc}
\dot{\bm{E}} = {\bm F}(\bm{s}) \bm{E} - \sigma {\bm H}(\bm{s},\bm{s}) \bm{E} {\bm L}^{T}
\end{equation}
where $\bm{E}=(\bm{\varepsilon}_1,\bm{\varepsilon}_2,\ldots,\bm{\varepsilon}_N )$ be a $d \times N$ matrix such that $\bm{\varepsilon_i}=(\varepsilon_i^{1},\varepsilon_i^2,\ldots,\varepsilon_i^d )^T$. We know ${\bm L}^{T} \bm{e}_k = \mu_k \bm{e}_k$, multiplying both sides of Eq.~(\ref{pertdyn_mat_nc}) by $\bm{e}_k$ we get  
\begin{equation} \label{pertdyn_em_1_nc}
\dot{\bm{E}}\bm{e}_k = {\bm F}(\bm{s}) \bm{E}\bm{e}_k - \sigma {\bm H}(\bm{s},\bm{s}) \bm{E} {\bm L}^{T}\bm{e}_k
\end{equation}
for $k=1,2,\ldots, N$. This above equation decouples the perturbations along the eigenmodes of the Laplacian matrix. The summation of the row elements in the Laplacian matrix is zero. This leads to a zero eigenvalue, say $\mu_1=0$ corresponding to the eigenvector $\bm{e}_1 = (1 \ 1 \ \ldots \ 1)^T$. The perturbation along the eigenmode of the zero eigenvalue $\mu_1$ will be on the synchronization manifold and thus will not affect the stability of the synchronization manifold. The perturbations along the remaining eigenmode (excluding $\mu_1=0$) will correspond to the transverse directions to the synchronization manifold. The synchronization state will be stable when the transverse perturbations are zero, so the perturbations corresponding to the eigenvalues $\mu_k;\;k=2,\ldots, N$ are zero. Let's denote $\bm{\psi}_k=\bm{E}\bm{e}_k$ and $\beta_k = \sigma \mu_k$
\begin{equation} \label{pertdyn_em_2_nc}
	\dot{\bm{\psi}_k} = \biggl[{\bm F}(s)  - \beta_k  {\bm H}(\bm{s},\bm{s})\biggr] \bm{\psi}_k;\; k = 2, \ldots, N.
\end{equation}
The synchronization state is stable when the perturbations in Eq.~(\ref{pertdyn_em_2_nc}) are zero. We construct the Master Stability Equation (MSE) from Eq.~(\ref{pertdyn_em_2_nc}) by considering a generic coupling parameter $\kappa=\beta_k$ and dropping the index $k$
\begin{equation} \label{mse_nc}
    	\dot{\bm{\psi}} = \biggl[{\bm F}(s)  - \kappa  {\bm H}(\bm{s},\bm{s})\biggr] \bm{\psi}.
\end{equation}
The maximum Lyapunov exponent $\Lambda^{\text{max}}$ computed from Eq.~(\ref{mse_nc}) as a function of the generic coupling parameter $\kappa$ is the Master Stability Function (MSF). The synchronization state is stable for $\Lambda^{max}(\kappa) < 0$ and the stability region is given as $\mathcal{R} = \{ \kappa | \Lambda^{max} < 0 \}$.

\section{MSF in Directed Graphs}

In this section, we discuss the derivation of the Master Stability Function in directed and/or weighted networks~\cite{NishikawaPRE2006,NishikawaPhysicaD2006}. Until now, we have discussed the derivation of the MSF in undirected and unweighted networks. In undirected and unweighted networks, the Laplacian matrix is symmetric and diagonalizable. And all the eigenvalues of the Laplacian matrix are real. However, real-world networks are rarely undirected and unweighted. Many real-world networks where synchronization is important are directed, weighted, and highly heterogeneous~\cite{BaratPNAS2004,YookPRL2001}. Examples includes the neural networks~\cite{FriesPNAS1997}, power-grid networks~\cite{motter2013spontaneous}, and transportation networks~\cite{BaratPNAS2004}. For directed networks, the Laplacian matrix is not symmetric, its eigenvalues may be complex, and the Laplacian matrix itself can be nondiagonalizable. Thus, it is important to extend the analysis of the MSF to directed and weighted networks.

We consider a network of $N$ identical oscillators interacting through a diffusive coupling function. The dynamics of the oscillator $i$ on the network is given by
\begin{equation}\label{dyneqn1_dirnet}
    \dot{\bm{x}}_i(t) = \bm{f}(\bm{x}_i) + \sigma\sum_{j=1}^N a_{ij} (\bm{h}(\bm{x}_j) - \bm{h}(\bm{x}_i));\; i=1,\ldots,N
\end{equation}
where, $\bm{x}_i\in \mathbb{R}^d$ is the $d$ dimensional state variable of the oscillator $i$ and $\bm{f}:\mathbb{R}^d \mapsto \mathbb{R}^d$ provides the dynamics of an isolated oscillator, $\sigma$ is overall coupling strength and $\bm{h}:\mathbb{R}^d \mapsto \mathbb{R}^d$ is the linear coupling function. The direction and the weight of interaction are encapsulated in the Adjacency matrix $\bm{A} = [a_{ij}]$, where $a_{ij} = w_{ij};i\neq j$, if there is a link from oscillator $j$ to oscillator $i$ and $w_{ij}$ is the strength of interaction, otherwise $a_{ij}=0$. The Laplacian matrix of the directed and weighted network is defined as $L_{ij} = - a_{ij};\; i\neq j$ and $L_{ii} = -\sum_{j\neq i}a_{ij}$. So the Laplacian matrix satisfies $\sum_j L_{ij} = 0$, confirming one zero eigenvalue, say $\mu_1=0$, corresponding to eigenvector $(1,1,\ldots,1)^T$.   The dynamical equation Eq.~(\ref{dyneqn1_dirnet}) can be alternatively written as
\begin{equation}\label{dyneqn2_dirnet}
    \dot{\bm{x}}_i(t) = \bm{f}(\bm{x}_i) - \sigma\sum_{j=1}^N L_{ij} \bm{h}(\bm{x}_j);\; i=1,\ldots,N.
\end{equation}

For suitable coupling strength $\sigma$ and coupling function $\bm{h}$, the coupled oscillators in Eq.~(\ref{dyneqn1_dirnet}) will converge to a global synchronization state given as $\bm{x}_i (t) = \bm{s}(t);\; \forall i$. The linear stability of the synchronization state can be analyzed by extending the linear stability analysis of Pecora and Carroll~\cite{PecoraCarrollPRL1998} (section \ref{msf-iden-diff} ) for the case where the Laplacian matrix $L$ is not necessarily diagonalizable. Let us consider $\bm{\varepsilon}_i(t) = \bm{x}_i(t) - \bm{s}(t)$ is a small perturbation to the synchronization state corresponding to the $i$th oscillator. Following the analysis in section \ref{msf-iden-diff}, the linearized dynamics of the perturbations can be written as
\begin{equation}\label{lindyn1_dirnet}
    \dot{\bm{\varepsilon}}_i = \bm{F}(\bm{s}) \bm{\varepsilon}_i - \sigma \sum_{j=1}^N L_{ij} \bm{H}(\bm{s}) \bm{\varepsilon}_j; \; i=1,\ldots,N.
\end{equation}
where $\bm{F}(\bm{s})$ and $\bm{H}(\bm{s})$ are the Jacobian matrices of the self-dynamics and the coupling function, respectively, at the synchronization solution $\bm{s}(t)$. Collecting the small perturbations in a $d\times N$ matrix $\bm{E}$ the linearized dynamics of the perturbations can be written in a single matrix equation as
\begin{equation}\label{lindyn2_dirnet}
    \dot{\bm{E}} = \bm{F}(\bm{s}) \bm{E} - \sigma \bm{H}(\bm{s}) \bm{E} \bm{L}^T
\end{equation}
where $\bm{L}^T$ is the transpose of the Laplacian matrix $\bm{L}$. Note that in Eq.~(\ref{lindyn2_dirnet}), the Laplacian matrix is not symmetric and may not be diagonalizable. Thus, the straightforward Pecora-Carroll analysis of diagonalizing the matrix $\bm{L}$ to separate the perturbations along the eigenmodes of $\bm{L}$ is not applicable. Nishikawa and Motter~\cite{NishikawaPRE2006,NishikawaPhysicaD2006} used the Jordan cannonical transformation on $\bm{L}$~\cite{HornBook}. According to Jordan's canonical transformation, every square matrix resembles an upper triangular matrix. Thus, for the $N \times N$ Laplacian $\bm{L}$ there exists an invertible matrix $\bm{V}$ of the generalized eigenvectors of $\bm{L}$ that transform $\bm{L}$ into Jordan canonical form as
\begin{equation}
    \bm{V}^{-1} \bm{L} \bm{V} = \bm{J} =
    \begin{pmatrix}
     0 & & & \\
       & B_1 & & \\
       & & \ddots & \\
       & & & B_l
    \end{pmatrix}, \text{ and } 
        \bm{B}_i(\mu) = \begin{pmatrix}
        \mu & 1 & &  & \\
          & \mu & 1 & & \\
          & & \ddots & \ddots & \\
          & & & \mu & 1 \\
          & & & & \mu \\
    \end{pmatrix}
\end{equation}
where $B_i$'s are the Jordan block and have the form of an upper triangular matrix. Here, $\mu$ is one of the (possibly complex) eigenvalues of $\bm{L}$. Now multiplying Eq.~(\ref{lindyn2_dirnet}) by $(\bm{V}^{-1})^T$ from right we get
\begin{eqnarray}\label{lindyn3_dirnet}
\begin{split}    
    \dot{\bm{E}} (\bm{V}^{-1})^T & = \bm{F}(\bm{s}) \bm{E} (\bm{V}^{-1})^T - \sigma \bm{H}(\bm{s}) \bm{E} (\bm{V}\bm{V}^{-1})^T \bm{L}^T (\bm{V}^{-1})^T \\
    & = \bm{F}(\bm{s}) \bm{E} (\bm{V}^{-1})^T - \sigma \bm{H}(\bm{s}) \bm{E} (\bm{V}^{-1})^T (\bm{V}^{-1} \bm{L} \bm{V})^T \\
    & = \bm{F}(\bm{s}) \bm{E} (\bm{V}^{-1})^T - \sigma \bm{H}(\bm{s}) \bm{E} (\bm{V}^{-1})^T \bm{J}^T \\
\end{split}
\end{eqnarray}
By denoting $\bm{\psi}(t) = {\bm{E}} (\bm{V}^{-1})^T$ Eq.~(\ref{lindyn3_dirnet}) can be written as
\begin{equation} \label{lindyn4_dirnet}
    \dot{\bm{\psi}}(t) = \bm{F}(\bm{s}) \bm{\psi}(t) - \sigma \bm{H}(\bm{s}) \bm{\psi}(t) \bm{J}^T
\end{equation}

Now, if we consider the Laplacian matrix $\bm{L}$ is diagonalizable and then the Jordan matrix $\bm{J}$ is a diagonal matrix with the eigenvalues of $\bm{L}$, $(\mu_1,\ldots,\mu_N)$ as its diagonal entries. Then the equations of perturbations from Eq.~(\ref{lindyn4_dirnet}) can be separated along the different eigenmodes of $\bm{L}$
\begin{equation}\label{lindyn5_dirnet}
    \dot{\bm{\psi}}_k(t) = \bm{F}(\bm{s}) \bm{\psi}_k(t) - \sigma\mu_k \bm{H}(\bm{s}) \bm{\psi}_k(t);\; k=1,\ldots,N,
\end{equation}
which takes the form of Eq.~(\ref{spectral_decom}) from section \ref{msf-iden-diff}. The stability analysis of synchronization will follow the same from section \ref{msf-iden-diff} except that the eigenvalues of the Laplacian matrix $\bm{L}$ can be complex. Let's say the eigenvalue $\mu_1 = 0$ and the corresponding mode of perturbation from Eq.~(\ref{lindyn5_dirnet}) is parallel to the synchronization manifold and will not affect the stability of the synchronization manifold. The remaining perturbations corresponding to the rest of the eigenvalues $\mu_2,\mu_3,\ldots,\mu_N$ will be along the transverse manifold. The synchronization manifold will be stable when the perturbations along the transverse manifold will be zero, i.e., the Lyapunov exponents $\Lambda(\sigma\mu_k);\;k=2,\ldots, N$ will be negative.

Let us now return to the non-diagonalizable Laplacian matrix $\bm{L}$. In this case, the Jordan canonical matrix $\bm{J}$ will have the form of a block diagonal matrix. Each block $\bm{B}_k$ of the matrix $\bm{J}$ will correspond to a subset of the perturbation matrix $\bm{\psi}$. The Jordan block $\bm{B}_k$ corresponding to the eigenvalue $\mu_k$ of $\bm{L}$ is $l\times l$ matrix. Then, the subset of the linearized dynamical equation for this block is
\begin{subequations}\label{lindyn6_dirnet}
\begin{align}  
    \dot{\bm{\psi}}_1 & = [\bm{F}(\bm{s}) - \sigma\mu_k \bm{H}(\bm{s})] \bm{\psi}_1  \label{lindyn7_dirnet} \\
    \dot{\bm{\psi}}_2 & = [\bm{F}(\bm{s}) - \sigma\mu_k \bm{H}(\bm{s})] \bm{\psi}_2 + \sigma \bm{H}(\bm{s}) \bm{\psi}_1  \label{lindyn8_dirnet} \\
    & \ldots\\
    \dot{\bm{\psi}}_l & = [\bm{F}(\bm{s}) - \sigma\mu_k \bm{H}(\bm{s})] \bm{\psi}_l + \sigma\bm{H}(\bm{s})\bm{\psi}_{l-1}  \label{lindyn9_dirnet}
\end{align}
\end{subequations}
here, $\bm{\psi}_1,\ldots,\bm{\psi}_l$ represents the modes of perturbations in the generalized eigenspace corresponding to the eigenvalue $\mu_k$ of $\bm{L}$. For the stability of the synchronization manifold, we need all perturbations corresponding to the nonzero eigenvalues of the Laplacian matrix $\bm{L}$ to be zero. The perturbation in Eq.~(\ref{lindyn7_dirnet}) will be zero if $\Lambda(\sigma\mu_k) < 0$. Next, we consider the stability in Eq.~(\ref{lindyn8_dirnet}). If $\bm{H}(\bm{s})$ is bounded and $\Lambda(\sigma\mu_k)<0$, then the third term in Eq.~(\ref{lindyn8_dirnet}) will be very small as time evolves, essentially converging to an equation of $\bm{\psi}_2$ equivalent to Eq.~(\ref{lindyn7_dirnet}). Thus the same stability condition $\Lambda(\sigma\mu_k)<0$ will assure that $\bm{\psi}_2 \rightarrow 0$ as $t \rightarrow \infty$. Repeating the same reasoning, it can be shown that the remaining modes of perturbations $\bm{\psi}_3,\ldots,\bm{\psi}_l$ must also converge to zero if $\Lambda(\sigma\mu_k) < 0$. Thus, the master stability equation for the non-diagonalizable networks can be obtained by introducing a generic coupling parameter $\kappa \equiv \sigma\mu_k$ in Eq.~(\ref{lindyn7_dirnet}) and removing the index
\begin{equation} \label{mse-dirnet}
    \dot{\bm{\psi}}  = [\bm{F}(\bm{s}) - \kappa \bm{H}(\bm{s})] \bm{\psi}.
\end{equation}
The largest Lyapunov exponent $\Lambda^{max}(\kappa)$ computed as a function of $\kappa$ from Eq.~(\ref{mse-dirnet}) is the Master Stability Function. The \emph{stability region} is defined by the subset of the complex plane in which the master stability function $\Lambda^{\max}(\kappa)$ is negative, i.e., $\mathcal{R} = \{\kappa \in \mathbb{C} | \Lambda^{max}(\kappa) < 0\}$. 

Although the stability conditions for diagonalizable and non-diagonalizable networks are the same, a crucial difference exists between them. In diagonalizable networks, the different modes of the perturbations are decoupled and become independent of each other. Thus, these modes will converge to zero simultaneously. In non-diagonalizable networks, this is not possible as the modes are coupled to each other. We can see from Eq.~(\ref{lindyn6_dirnet}), for $\bm{\psi}_2$ to be zero, $\bm{\psi}_1$ needed to be zero first. So $\bm{\psi}_2$ has to wait. Similarly, $\bm{\psi}_3$ may need to wait for $\bm{\psi}_2$ to be zero before it starts converging to zero. So all the succeeding modes may need to wait for their former modes to be zero. The larger the size $l$ of the Jordan block is, the longer time it will take for all modes of perturbations to be zero. Thus, we will expect longer transients in nondiagonalizable networks than their diagonalizable counterparts.

\begin{figure*}[htb]
	\centering
\includegraphics[width=6.5in, height=1.8in]{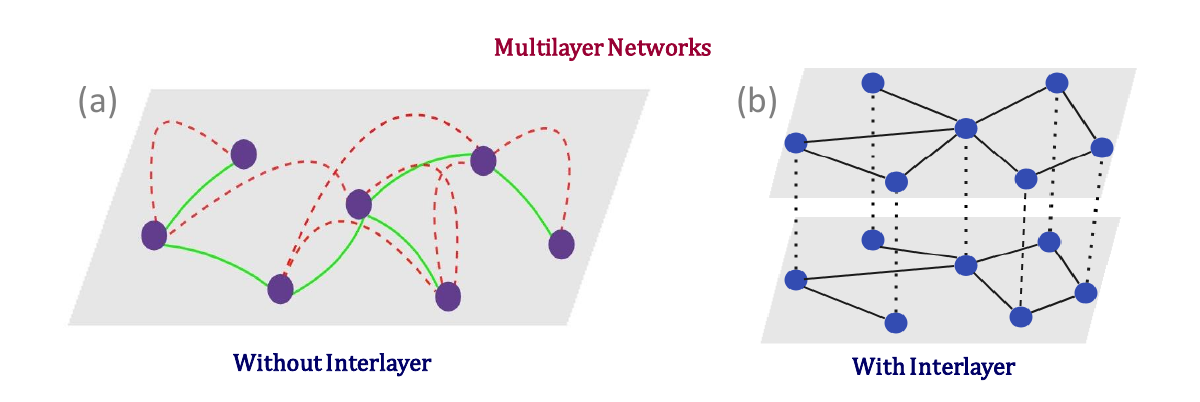}
\caption{{\bf Schematic representation of the two types of multilayer networks.} (a) A multilayer network without interlayer connections. For example, within the same layer, nodes can have two types of independent connections, such as two modes of transportation within a city, denoted by solid green and dashed red lines and represented by Laplacians $\bm{L}^{(1)}$ and $\bm{L}^{(2)}$. (b) A two-layer multilayer network includes both intra-layer and inter-layer connections. An example of this is social interactions across different platforms, such as LinkedIn and Twitter. Here, intra-layer interactions within each layer are represented by solid lines and encoded in the Laplacian matrices $\bm{L}^{(1)}$ and $\bm{L}^{(2)}$. The inter-layer interactions, which connect copies of the same node across layers, are represented by dashed lines and encoded in the Laplacian matrix $\mathcal{B}$ (Eq. (\ref{intralayer_lsup_lap_mlil})).}
\label{schematic_multilayer}
\end{figure*}
\section{Master Stability Function in Multilayer Networks}
Several applications of complex systems, such as synchronization, pattern formation, and network control, are significantly influenced by the topologies of multilayer networks \cite{bianconi2018multilayer,de2016physics,SciAdv.2.e1601679}. In a multilayer network, synchronization can manifest in different forms: global synchronization involving all nodes across the entire network, intra-layer synchronization within individual layers, or inter-layer synchronization between corresponding (replica) nodes across layers (Fig. \ref{schematic_multilayer}). These dynamics are particularly relevant in real-world applications, including international trade, communication systems, power-grid engineering, neuroscience, and beyond \cite{de2016physics, boccaletti2014structure, bassett2017network, jalan2018localization}.

A critical tool for studying the stability of global synchronization in such multilayer networks of coupled oscillators is the Master Stability Function (MSF). The MSF provides a framework for analyzing the conditions under which synchronization is achieved and sustained in multilayer networks. Its application to multiplex networks has yielded key insights, such as the conditions required for robust global synchronization and the interplay between intra and inter-layer coupling. These findings have broad implications for optimizing and controlling multilayer systems in various domains \cite{della2020symmetries, he2017multiagent, menichetti2016control}.

\subsection{Master Stability in multi-layer networks without inter-layer coupling}

We consider a network comprising $N$ nodes that interact through $M$ distinct connection layers, where each layer typically has its own set of links representing various types of interactions between the nodes \cite{bianconi2018multilayer, SciAdv.2.e1601679}. We assume that a node $i$ in one layer is identical to the same node $i$ in any other layer (Fig. \ref{schematic_multilayer}(a)). In this framework, nodes interacting within a single layer are treated the same across all layers. The connections between nodes in layer $m$ ($ m = 1, 2, \ldots, M$) are encoded in the adjacency matrix $\mathbf{A}^{(m)}$. Each node represents a $d$-dimensional dynamical system, describing its state by a vector $ \bm{x}_i$ containing $d$ components. The self-dynamics of the nodes are governed by a set of distinct equations, which define the evolution of the states over time. 
\begin{equation}
	\dot{\bm{x}} = \bm{f}(\bm{x}_i)
\end{equation}
Here, $\bm{f}$ represents an arbitrary vector field that governs the self-dynamics of individual nodes. Similarly, the interactions within layer $m$ are modeled by a continuous and differentiable vector field $\bm{h}^{(m)}$, scaled by a layer-specific coupling strength $\sigma_m$. We assume that interactions between nodes $i$ and $j$ are diffusive in nature. This implies that, for each layer where nodes $i$ and $j$ are connected, their coupling depends on the difference between $\bm{h}^{(m)}$ evaluated at the states of $ \bm{x}_j$ and $\bm{x}_i$. Under these assumptions, the dynamics of all nodes in the coupled network are governed by the following set of equations, which incorporate both the self-dynamics of the nodes and the interactions across all layers.
\begin{equation}\label{1stdyneqn_3}
	\begin{split}
		\dot{\bm{x}}_{i}(t) &= \bm{f}(\bm{x}_i(t)) + \sum_{m=1}^M \sigma_{m} \sum_{j=1}^N a_{ij}^{(m)} \bm{h}^{(m)}(\bm{x}_j(t)-\bm{x}_i(t)),\;\; i = 1, \ldots, N 
	\end{split}    
\end{equation}
where $\bm{x}_i(t)\in \mathbb{R}^d$ is the state variable and $\bm{f}(\bm{x}_i):\mathbb{R}^d\rightarrow \mathbb{R}^d$ is the self dynamics or the uncoupled dynamics of the node $i$, and the second term is the coupling term that capture the effect of the nearest neighbor vertices of vertex $i$ on it. Here, ${\bm A}^{(m)} = [a_{ij}^{(m)}]$ is the adjacency matrix of $m$ layer, and $\sigma_{m}$ is the scalar coupling parameter and $\bm{h}^{(m)}(\bm{x}_j): \mathbb{R}^d \rightarrow \mathbb{R}^d$ is the coupling function, which we consider as a linear function. Since, $\bm{h}^{(m)}(\bm{x})$ is a linear function, we can write Eq. (\ref{1stdyneqn_3}) in terms of the network Laplacian matrix as below
\begin{equation}\label{2stdyneqn}
	\begin{split}
		\dot{\bm{x}}_{i}(t) 
		&= \bm{f}(\bm{x}_i) + \sum_{m=1}^M \sigma_{m} \sum_{j=1}^N a_{ij}^{(m)} [\bm{h}^{(m)}(\bm{x}_j)-\bm{h}^{(m)}(\bm{x}_i)]\\
		&= \bm{f}(\bm{x}_i) + \sum_{m=1}^M \sigma_{k} \left[\sum_{j=1}^N a_{ij}^{(m)} \bm{h}^{(m)}(\bm{x}_j)-\bm{h}^{(m)}(\bm{x}_i)\sum_{j=1}^N a_{ij}^{(m)}\right]\\
		&= \bm{f}(\bm{x}_i) + \sum_{m=1}^M \sigma_{m} \left[\sum_{j=1}^N a_{ij}^{(m)} \bm{h}^{(m)}(\bm{x}_j)-\bm{h}^{(m)}(\bm{x}_i) k_i^{(m)}\right]\\
		&= \bm{f}(\bm{x}_i) - \sum_{m=1}^M \sigma_{m} \sum_{j=1}^N L_{ij}^{(m)} \bm{h}^{(m)}(\bm{x}_j)
	\end{split}    
\end{equation}
where, 
\begin{equation}\label{laplacian_matrix_multilayer}
	L_{ij}^{(m)}=
	\begin{cases}
		- a_{ij}^{(m)}       & \quad \text{if} \ i \neq j \\
		k_i^{(m)}  & \quad i=j\\
	\end{cases}
\end{equation}
${\bm L}^{(m)}=[L_{ij}^{(m)}]={\bm D}^{(m)} - {\bm A}^{(m)}$ is the network Laplacian matrix and ${\bm D}^{(m)}=\text{diag}(k_1^{(m)}, k_2^{(m)}, \ldots, k_N^{(m)})$ is a diagonal matrix with $i$th diagonal entry as the degree of node $i$ of $m$ layer. For a suitable coupling function $\bm{h}^{(m)}(\bm {x})$ and scalar coupling parameter $\sigma_{m}$, the coupled systems will undergo synchronization. In the synchronization, we can write $\bm{x}_1 = \bm{x}_2 = \cdots = \bm{x}_N = \bm{s}(t)$, where $\bm{s}(t)$ is the solution of uncoupled system
\begin{equation}
	\dot{\bm{s}}(t) = \bm{f}(\bm{s}(t))
	\label{syncdyn_M}
\end{equation}
where in synchronized state $\sigma_{m} \sum_{j=1}^N L_{ij}^{(m)} \bm{h}^{(m)}(\bm{s}(t))=\sigma_{m}\bm{h}^{(m)}(\bm{s}(t)) \sum_{j=1}^N L_{ij}^{(m)}=0$, as for Laplacian matrix $\sum_{j=1}^N L_{ij}^{(m)}=0$. In the synchronized state, the dynamic evolution of the coupled oscillators would converge to a $d$-dimensional hyperplane in the $d\times N$ phase space.

The linear stability of the synchronization state $\bm{s}(t)$ is determined by deriving the dynamics of small perturbations to the synchronization state. We consider  $\bm{x}_i(t) 
= \bm{s}(t) + \bm{\varepsilon}_i(t)$ is the small perturbation to the synchronization state of the oscillator $i$ where $\bm{x}_i(t)=(x_1(t),x_2(t),\ldots,x_d(t))^{T}$, $\bm{s}(t)=(s_1(t),s_2(t),\ldots,s_d(t))^{T}$ and $\bm{\varepsilon}_i(t)=(\varepsilon_1(t),\varepsilon_2(t),\ldots,\varepsilon_d(t))^{T}$. Now from Eq.~(\ref{1stdyneqn_3}) we can write
\begin{equation}
	\dot{\bm{s}}(t) +  \dot{\bm{\varepsilon}}_i(t) = \bm{f}(\bm{s}(t) + \bm{\varepsilon}_i(t)) - \sum_{m=1}^M \sigma_{m} \sum_{j=1}^N L_{ij}^{(m)} \bm{h}^{(m)}(\bm{s}(t) + \bm{\varepsilon}_j(t))
	\label{pertdyn_1_M}
\end{equation}
Expanding $\bm{f}(\bm{s}(t) + \bm{\varepsilon}_i(t))$ and $\bm{h}^{(m)}(\bm{s}(t) + \bm{\varepsilon}_j(t))$ in Taylor's series and ignoring the higher order terms we can write,
\begin{subequations}
	\label{tse_fh_M}
	\begin{align}
		\bm{f}(\bm{s}(t) + \bm{\varepsilon}_i(t)) = \bm{f}(\bm{s}(t)) +  {\bm F}(\bm{s}(t))\bm{\varepsilon}_i(t)  \label{tse_f_M}
		\\
		\bm{h}^{(m)}(\bm{s}(t) + \bm{\varepsilon}_j(t)) = \bm{h}^{(m)}(\bm{s}(t)) +  {\bm H}(\bm{s}(t)) \bm{\varepsilon}_j(t)  \label{tse_h_M}
	\end{align}
\end{subequations}
where, ${\bm F}(\bm{s}(t))=\frac{\partial f^i}{\partial x^j}$ and ${\bm H}^{(m)}(\bm{s}(t))=\frac{\partial h^i}{\partial x^j}$ are the $d\times d$ Jacobian matrices of $\bm{f}$ and $\bm{h}$ respectively at the synchronized solution $\bm{s}(t)$. Now replacing Eqs.~(\ref{tse_f_M}) and ~(\ref{tse_h_M}) in Eq.~(\ref{pertdyn_1_M}) we get
\begin{equation}
	\dot{\bm{s}}(t) +  \dot{\bm{\varepsilon}}_i(t) = \bm{f}(\bm{s}(t)) + {\bm F}(\bm{s}(t))\bm{\varepsilon}_i(t) -  \sum_{m=1}^M \sigma_{m} \sum_{j=1}^N L_{ij}^{(m)}\left[ \bm{h}^{(m)}(\bm{s}(t)) + {\bm H}^{(m)}(\bm{s}(t))\bm{\varepsilon}_j(t)  \right]
    \label{dynamics_multilayer}
\end{equation}

In the above equation the first term, in the L.H.S. ($\frac{d \bm{s}(t)}{dt}$) is equal to the first term in the R.H.S. ($\bm{f}(\bm{s}(t))$) from Eq. (\ref{syncdyn_M}). The second interaction term of the Eq. (\ref{dynamics_multilayer}) can be simplify as

\begin{eqnarray}
	\sum_{m=1}^M \sigma_{m}\sum_{j=1}^N L_{ij}^{(m)}\left[ \bm{h}^{(m)}(\bm{s}(t)) + {\bm H}^{(m)}(\bm{s}(t))\bm{\varepsilon}_j  \right] 
	& = & \sum_{m=1}^M \sigma_{m}\sum_{j=1}^N L_{ij}^{(m)} \bm{h}^{(m)}(\bm{s}(t)) + \sum_{m=1}^M \sigma_{m}\sum_{j=1}^N L_{ij}^{(m)}\left[ {\bm H}^{(m)}(\bm{s}(t))\bm{\varepsilon}_j \right] \nonumber \\
	& = & \sum_{m=1}^M \sigma_{m} \bm{h}^{(m)}(\bm{s}(t)) \sum_{j=1}^N L_{ij}^{(m)} + \sum_{m=1}^M \sigma_{m} {\bm H}^{(m)}(\bm{s}(t)) \sum_{j=1}^N L_{ij}^{(m)}\bm{\varepsilon}_j  \nonumber \\
	& = & \sum_{m=1}^M \sigma_{m}\sum_{j=1}^N  L_{ij}^{(m)} {\bm H}^{(m)}(\bm{s}(t)) \bm{\varepsilon}_j
\end{eqnarray}
since $\sum_{j=1}^{N} L_{ij}^{(m)} =0$. Thus, the linearized dynamics of the small perturbation to a node $i$ can be written as
\begin{equation}\label{lindynpert_M}
	\dot{\bm{\varepsilon}}_i(t) = {\bm F}(\bm{s}(t))\bm{\varepsilon}_i(t) - \sum_{m=1}^M \sigma_{m} \sum_{j=1}^N L_{ij}^{(m)} {\bm H}^{(m)}(\bm{s}(t))\bm{\varepsilon}_j(t),\;\; i=1,\ldots,N
\end{equation}
For $N$ coupled nodes the coupling matrix ${\bm L}^{(m)}=[L_{ij}^{(m)}]$ will be defined as 

\begin{equation}
{\bm L}^{(m)}= \begin{pmatrix}
		k_{1}^{(m)} & -a_{12}^{(m)}&\hdots & -a_{1N}^{(m)} \\
		-a_{21}^{(m)}&k_{2}^{(m)}&\hdots&-a_{2N}^{(m)}\\
		\vdots&\vdots&\hdots&\vdots \\
		-a_{N1}^{(m)}&-a_{N2}^{(m)}&\hdots&k_{N} ^{(m)}\\
	\end{pmatrix}_{N \times N}
\end{equation}
with $\sum_{j=1}^{N} L_{ij}^{(m)}=0$. Hence, the perturbation associated with each node $i$ ranging from $i=1$ to $N$ in a network will evolve as 
\begin{equation}
\begin{gathered}
			\begin{pmatrix}
					\dot{\bm{\varepsilon}_{1}} \\
					\dot{\bm{\varepsilon}_{2}} \\
					\vdots\\
					\dot{\bm{\varepsilon}_{N}}
				\end{pmatrix}
			=
	\begin{pmatrix}
		{\bm F} & &\bigzero \\
		& \ddots  & \\
		\bigzero &  &{\bm F}
	\end{pmatrix}
			\begin{pmatrix}
					\bm{\varepsilon}_{1} \\
					\bm{\varepsilon}_{2} \\
					\vdots\\
					\bm{\varepsilon}_{N}
				\end{pmatrix} - \sum_{m=1}^{M}\sigma_{m} %
			\begin{pmatrix}
					L_{11}^{(m)}{\bm H}^{(m)} & L_{12}^{(m)}{\bm H}^{(m)} & \cdots & L_{1N}^{(m)}{\bm H}^{(m)} \\
					L_{21}^{(m)}{\bm H}^{(m)} & L_{22}^{(m)}{\bm H}^{(m)} & \cdots & L_{2N}^{m}{\bm H}^{(m)} \\ 
					\vdots & \vdots & \vdots & \vdots\\ 
					L_{N1}^{(m)}{\bm H}^{(m)} & L_{N2}^{(m)}{\bm H}^{(m)} & \cdots & L_{NN}^{(m)}{\bm H}^{(m)} 
				\end{pmatrix} 
			\begin{pmatrix}
					\bm{\varepsilon}_{1} \\
					\bm{\varepsilon}_{2} \\
					\vdots\\
					\bm{\varepsilon}_{N}
				\end{pmatrix}
	\end{gathered}
\end{equation}
Let, $\bm{\varepsilon} = (\bm{\varepsilon}_1,\bm{\varepsilon}_2,\ldots,\bm{\varepsilon}_N)^{T} \in \mathbb{R}^{1 \times Nd}$, $\bm{\varepsilon}_i=(\varepsilon_i^{1},\varepsilon_i^{2},\ldots,\varepsilon_i^{d})^{T}\in \mathbb{R}^{1 \times d}$ and we get
\begin{equation}\label{multi_MSE}
\dot{\bm{\varepsilon}}=\biggl(\mathbb{I}_N \otimes {\bm F}(\bm{s}) - \sum_{m=1}^{M}\sigma_{m}{\bm L}^{(m)} \otimes {\bm H}^{(m)}(\bm{s})\biggr) \bm{\varepsilon}  
\end{equation}

The Laplacian matrix in each layer $m$ is diagonalizable by a matrix ${\bm V}^{(m)}$ consists of the eigenvectors of the respective Laplacian matrix, so that ${\bm L}^{(m)}={\bm V}^{(m)}{\bm B}^{(m)}{\bm V}^{(m)^{T}}$, and ${\bm V}^{(m)^{T}}{\bm L}^{(m)}={\bm B}^{(m)}{\bm V}^{(m)^T}$, where ${\bm B}^{(m)}$ is the diagonal matrix of the eigenvalues of layer $m$. We spectrally decompose $\bm{\varepsilon}$ in Eq.~(\ref{multi_MSE}) and project onto the Laplacian eigenvectors of a layer \cite{SciAdv.2.e1601679}. We choose the first layer for the projection as Laplacian eigenvectors are always a basis of $\mathbb{R}^{N}$. We represent $\mathbb{I}_d$ as a d-dimensional identity matrix and multiply Eq.~(\ref{multi_MSE}) on the left by $({\bm V}^{(1)^{T}} \otimes \mathbb{I}_d)$. Here, ${\bm V}^{(1)}$ is the matrix of eigenvectors of the Laplacian of layer 1. We use the relation $({\bm A}_1 \otimes {\bm A}_2)({\bm A}_3 \otimes {\bm A}_4) = ({\bm A}_1{\bm A}_3)\otimes({\bm A}_2{\bm A}_4)$ \cite{SciAdv.2.e1601679} and get
\begin{equation}\nonumber
\begin{split}
	({\bm V}^{(1)^{T}} \otimes \mathbb{I}_d)\dot{\bm{\varepsilon}}&=\biggl(({\bm V}^{(1)^{T}} \otimes \mathbb{I}_d)(\mathbb{I}_N \otimes {\bm F}(\bm{s})) - \sum_{m=1}^{M}\sigma_{m}({\bm V}^{(1)^{T}} \otimes \mathbb{I}_d)({\bm L}^{(m)} \otimes {\bm H}^{(m)}(\bm{s}))\biggr) \bm{\varepsilon}\\
    &=\biggl(({\bm V}^{(1)^{T}} \otimes \mathbb{I}_d)(\mathbb{I}_N \otimes {\bm F}(\bm{s})) - \sigma_{1}({\bm V}^{(1)^{T}} \otimes \mathbb{I}_d)({\bm L}^{(1)} \otimes {\bm H}^{(1)}(\bm{s}))\biggr) \bm{\varepsilon}\\
    &- \sum_{m=2}^{M}\sigma_{m}({\bm V}^{(1)^{T}} \otimes \mathbb{I}_d)({\bm L}^{(m)} \otimes {\bm H}^{(m)}(\bm{s}))\biggr) \bm{\varepsilon}\\
    &=\biggl(({\bm V}^{(1)^{T}}\mathbb{I}_N) \otimes (\mathbb{I}_d{\bm F}(\bm{s})) - \sigma_{1}({\bm V}^{(1)^{T}} \otimes \mathbb{I}_d)({\bm L}^{(1)} \otimes {\bm H}^{(1)}(\bm{s}))\biggr) \bm{\varepsilon}\\
    &- \sum_{m=2}^{M}\sigma_{m}({\bm V}^{(1)^{T}} \otimes \mathbb{I}_d)({\bm L}^{(m)} \otimes {\bm H}^{(m)}(\bm{s}))\biggr) \bm{\varepsilon}\\
    &=\biggl(({\bm V}^{(1)^{T}} \otimes {\bm F}(\bm{s})) - \sigma_{1}({\bm V}^{(1)^{T}}{\bm L}^{(1)} \otimes \mathbb{I}_d{\bm H}^{(1)}(\bm{s}))\biggr) \bm{\varepsilon} - \biggl( \sum_{m=2}^{M}\sigma_{m}({\bm V}^{(1)^{T}}{\bm L}^{(m)}) \otimes \mathbb{I}_d {\bm H}^{(m)}(\bm{s}))\biggr) \bm{\varepsilon}\\
    &=\biggl(({\bm V}^{(1)^{T}} \otimes {\bm F}(\bm{s})) - \sigma_{1}({\bm B}^{(1)}{\bm V}^{(1)^{T}} \otimes {\bm H}^{(1)}(\bm{s}))\biggr) \bm{\varepsilon} - \biggl(\sum_{m=2}^{M}\sigma_{m}({\bm V}^{(1)^{T}}{\bm L}^{(m)}) \otimes {\bm H}^{(m)}(\bm{s}))\biggr) \bm{\varepsilon}
    \end{split}
\end{equation}
where ${\bm V}^{(1)^{T}}{\bm L}^{(1)}={\bm B}^{(1)}{\bm V}^{(1)^T}$. In the above, we left-multiply the first occurrence of ${\bm V}^{(1)^{T}}$ in the right-hand side by a $N \times N$ identity matrix $\mathbb{I}_N$, and right-multiply ${\bm F}$ and ${\bm H}^{(1)}$ by a $d \times d$ identity matrix $\mathbb{I}_d$. Finally, we use the relation $({\bm A}_1{\bm A}_3)\otimes({\bm A}_2{\bm A}_4)=({\bm A}_1 \otimes {\bm A}_2)({\bm A}_3 \otimes {\bm A}_4)$ and get	
\begin{equation}\nonumber
\begin{split}
    ({\bm V}^{(1)^{T}} \otimes \mathbb{I}_d)\dot{\bm{\varepsilon}}
    &=\biggl((\mathbb{I}_N{\bm V}^{(1)^{T}} \otimes {\bm F}(\bm{s})\mathbb{I}_d) - \sigma_{1}{\bm B}^{(1)}{\bm V}^{(1)^{T}} \otimes {\bm H}^{(1)}(\bm{s})\mathbb{I}_d\biggr) \bm{\varepsilon} - \biggl(\sum_{m=2}^{M}\sigma_{m}({\bm V}^{(1)^{T}}{\bm L}^{(m)}) \otimes {\bm H}^{(m)}(\bm{s}))\biggr) \bm{\varepsilon}\\
    &=\biggl((\mathbb{I}_N\otimes {\bm F}(\bm{s}))( {\bm V}^{(1)^{T}} \otimes \mathbb{I}_d) - ( \sigma_{1}{\bm B}^{(1)} \otimes {\bm H}^{(1)}(\bm{s})) ({\bm V}^{(1)^{T}} \otimes \mathbb{I}_d)\biggr) \bm{\varepsilon} - \biggl(\sum_{m=2}^{M}\sigma_{m}({\bm V}^{(1)^{T}}{\bm L}^{(m)}) \otimes {\bm H}^{(m)}(\bm{s}))\biggr) \bm{\varepsilon}\\
    &=\biggl((\mathbb{I}_N\otimes {\bm F}(\bm{s}))( {\bm V}^{(1)^{T}} \otimes \mathbb{I}_d) - ( \sigma_{1}{\bm B}^{(1)} \otimes {\bm H}^{(1)}(\bm{s})) ({\bm V}^{(1)^{T}} \otimes \mathbb{I}_d)\biggr) \bm{\varepsilon}\\
    & - \sum_{m=2}^{M}\sigma_{m}{\bm V}^{(1)^{T}}{\bm L}^{(m)} \otimes {\bm H}^{m}(\bm{s})) ({\bm V}^{(1)} \otimes \mathbb{I}_d)({\bm V}^{(1)^{T}} \otimes \mathbb{I}_d)\bm{\varepsilon}
        \end{split}
\end{equation}
Now we take $\bm{\psi}=({\bm V}^{(1)^{T}} \otimes \mathbb{I}_d)\bm{\varepsilon}$ and get
\begin{equation}\nonumber
\begin{split}
    \dot{\bm{\psi}} &= \biggl((\mathbb{I}_N\otimes {\bm F}(\bm{s})) - ( \sigma_{1}{\bm B}^{(1)} \otimes {\bm H}^{(1)}(\bm{s}))  \biggr) \bm{\psi} -  \sum_{m=2}^{M}\sigma_{m}{\bm V}^{(1)^{T}}{\bm L}^{(m)} \otimes {\bm H}^{m}(\bm{s})) ({\bm V}^{(1)} \times \mathbb{I}_d)\bm{\psi}\\
    &= \biggl((\mathbb{I}_N\otimes {\bm F}(\bm{s})) - ( \sigma_{1}{\bm B}^{(1)} \otimes {\bm H}^{(1)}(\bm{s}))  \biggr) \bm{\psi} - \sum_{m=2}^{M}\sigma_{m}{\bm V}^{(1)^{T}}{\bm V}^{(m)}{\bm V}^{(m)^{T}}{\bm L}^{m} \otimes {\bm H}^{m}(\bm{s})) ({\bm V}^{(1)} \times \mathbb{I}_d)\bm{\psi} \\
    &= \biggl((\mathbb{I}_N\otimes {\bm F}(\bm{s})) - ( \sigma_{1}{\bm B}^{(1)} \otimes {\bm H}^{(1)}(\bm{s}))  \biggr) \bm{\psi} - \sum_{m=2}^{M} (\sigma_{m} {\bm V}^{(1)^{T}} {\bm V}^{(m)} {\bm B}^{(m)} {\bm V}^{(m)^{T}} \otimes {\bm H}^{(m)}(\bm{s})) ({\bm V}^{(1)} \times \mathbb{I}_d)\bm{\psi}\\ 
    &= \biggl((\mathbb{I}_N\otimes {\bm F}(\bm{s})) - ( \sigma_{1}{\bm B}^{(1)} \otimes {\bm H}^{(1)}(\bm{s}))  \biggr) \bm{\psi} - \sum_{m=2}^{M} (\sigma_{m} {\bm V}^{(1)^{T}} {\bm V}^{(m)} {\bm B}^{(m)} {\bm V}^{(m)^{T}} {\bm V}^{(1)} \otimes {\bm H}^{(m)}(\bm{s})) \bm{\psi}\\  
    &= \biggl((\mathbb{I}_N\otimes {\bm F}(\bm{s})) - ( \sigma_{1}{\bm B}^{(1)} \otimes {\bm H}^{(1)}(\bm{s}))  \biggr) \bm{\psi} - \sum_{m=2}^{M} (\sigma_{m} {\bm \Gamma}^{(m)^{T}} {\bm B}^{(m)} {\bm \Gamma}^{(m)} \otimes {\bm H}^{(m)}(\bm{s})) \bm{\psi}    
\end{split}
\end{equation}
where \( \bm{\psi} = (\bm{\psi}_1, \bm{\psi}_2, \dots, \bm{\psi}_N)^T\) is an \( N d \)-dimensional column vector representing the perturbed state of the system, $\bm{\psi}_k \in \mathbb{R}^{d}$ and \( \bm{B}^{(m)} = \text{diag}(\mu_1^{(m)}, \mu_2^{(m)}, \dots, \mu_N^{(m)})\) is a diagonal \( N \times N \) matrix that holds the eigenvalues of layer $m$. Finally, we define \( \bm{\Gamma}^{(m)} \equiv \bm{V}^{(m)^{T}}\bm{V}^{(1)} \), where \( \bm{V}^{(m)} \) represents the matrix of eigenvectors of the Laplacian of layer \( m \) which transforms the eigenmodes from layer $1$ to layer $m$. We can rewrite the above equation explicitly using the Kronecker product property for an oscillator \cite{SciAdv.2.e1601679} as 
\begin{equation}\label{mult_without_interllayer}
\dot{\bm{\psi}}_k=({\bm F}(s)-\sigma_1 \mu_k^{(1)}{\bm H}^{(1)}(s))\bm{\psi}_k- \sum_{m=2}^{M} \sigma_m \sum_{i=2}^{N}\sum_{j=2}^{N} \mu_j^{(m)} \Gamma_{ji}^{(m)} \Gamma_{jk}^{(m)}{\bm H}^{(m)}(s) \bm{\psi}_i\;\; \text{for } k=2,\ldots,N    
\end{equation}
Here, $\mu_j^{(m)}$ is the $j$th eigenvalue of the Laplacian matrix of layer $m$, and $\bm{\psi}_k$ is the vector coefficient of the eigen-mode decomposition of $\bm{\varepsilon}$. The sums over $i$ and $j$ start from $2$ because the first eigenvalue ($\mu_1$) of the Laplacian matrix is always $0$ for all layers. Since all Laplacians are real symmetric matrices, their eigenvectors in each layer are ensured to be orthonormal. Moreover, the first eigenvector, to which all others are orthogonal, is common to all layers.
Now, if the Laplacians commute \cite{SciAdv.2.e1601679}, they can be simultaneously diagonalized using a common eigenvector basis. In this case, \( \bm{V}^{(m)^{T}} = \bm{V}^{(1)} \equiv \bm{V} \) for all \( m \). Consequently, this leads to \( \bm{\Gamma}^{(m)} = 1 \) for all \( m \), reducing Eq. (\ref{mult_without_interllayer}) to its simplified form.
\begin{equation}\nonumber
\dot{\bm{\psi}}_k=\biggl({\bm F}(s)-\sigma_1 \mu_k^{(1)}{\bm H}^{(1)}(s)\biggr)\bm{\psi}_k - \sum_{m=2}^{M} \sigma_m \sum_{i=2}^{N}\sum_{j=2}^{N} \mu_j^{(m)} \delta_{ji}^{(m)} \delta_{jk}^{(m)}{\bm H}^{(m)}(s) \bm{\psi}_i
\end{equation}
Hence,
\begin{equation}\label{multilayer_final_MSF}
\dot{\bm{\psi}}_k=\biggl({\bm F}(s) - \sum_{m=1}^{M} \sigma_m  \mu_k^{(m)} {\bm H}^{(m)}(s)\biggr) \bm{\psi}_k
\end{equation}
Let us consider $\beta^{(m)}_k=\sigma_m  \mu_k^{(m)}$, $k=2,\ldots,N$ and $m=1,\ldots,M$. Now to form the master stability equation from Eq. (\ref{multilayer_final_MSF}), we consider a generic parameter $\kappa^{(m)}\equiv \beta^{(m)}_k$
\begin{equation}\label{Multi_MSE}
\dot{\bm{\psi}}=\biggl({\bm F}(s) - \sum_{m=1}^{M} \kappa^{(m)} {\bm H}^{(m)}(s)\biggr) \bm{\psi}    
\end{equation}
The stability of the synchronized state is completely specified by the maximum Lyapunov exponent $\Lambda^{max}$ from Eq. (\ref{Multi_MSE}) as a function of $(\kappa^{(1)},\ldots,\kappa^{(M)})$. The fully synchronized state will be stable against small perturbations only if $\Lambda^{max}<0$. The stability region is then given by $\mathcal{R}=\{\kappa^{(1)},\ldots,\kappa^{(M)}|\Lambda^{max}<0\}$.

\subsection{Master Stability in multi-layer networks with inter-layer coupling}

In this section, we discuss the derivation of the master stability function in multilayer networks with inter-layer coupling \cite{PRE.99.012304.2019}. We consider a multilayer network consisting of $M$ layers and $N$ identical nodes in each layer (Fig. \ref{schematic_multilayer}(b)). The time dependent state variable of the $i$th node in the $m$th layer is specified by a $d$-dimensional column vector $\bm{x}^{(m)}_i(t)=\left(x^{(m)}_{i1}, x^{(m)}_{i2},\ldots,x^{(m)}_{id}\right)\in \mathbb{R}^{d}$. We consider the dynamical evolution of the state variable of the $i$th node in the $m$th layer is given by
\begin{equation}\label{1stdyneqn_mlil}
    \dot{\bm{x}}^{(m)}_i(t) = \bm{f}(\bm{x}^{(m)}_i) - \sigma^{(I)}\sum_{j=1}^N L^{(m)}_{ij}\bm{h}(\bm{x}^{(m)}_j) - \sigma^{(L)}\sum_{l=1}^M B_{ml} \bm{g}(\bm{x}^{(l)}_i);\; i=1,\ldots,N, \text{ and } m=1,\ldots,M,
\end{equation}
where, $\bm{f}: \mathbb{R}^d \mapsto \mathbb{R}^d$ is the self dynamics of each node, $\bm{h}:\mathbb{R}^d \mapsto \mathbb{R}^d$ and $\bm{g}:\mathbb{R}^d \mapsto \mathbb{R}^d$ are the intra-layer and inter-layer coupling functions respectively, $\sigma^{(I)}$ and $\sigma^{(L)}$ are the intra-layer and inter-layer scalar coupling parameters respectively. Here, $L^{(m)} = [L^{(m)}_{ij}]$ is the $N \times N$ Laplacian matrix within layer $m$ with entries as following

\begin{equation}
    L^{(m)}_{ij} = 
    \begin{cases}
        -1, \;\; \text{if the nodes $i$ and $j$ in the $m$th layer are connected}\\
        -\sum_{j\neq i}^N L^{(m)}_{ij}, \;\; \text{if $i=j$}\\
        0, \;\; \text{Otherwise}
    \end{cases}
\end{equation}
And ${\bm B} = [B_{ml}]$ is the $M \times M$ inter-layer Laplacian matrix with entries as following
\begin{equation}
    B_{ml} = 
    \begin{cases}
        -1, \;\; \text{if a node in layer $m$ connected to its replica in layer $l$}\\
        -\sum_{l\neq m}^M B_{ml}, \;\; \text{if $m=l$}\\
        0, \;\; \text{Otherwise}.
    \end{cases}
\end{equation}

For suitable coupling functions $\bm{h}(\bm{x}^{(m)}_j)$ and $\bm{g}(\bm{x}^{(l)}_i)$ and coupling parameters $\sigma^{(I)}$ and $\sigma^{(L)}$ the coupled systems in all layers will converge to a complete or identical synchronization state given by 
\begin{equation}\label{syncstate_mlil}
\bm{x}^{(m)}_i(t) = \bm{s}(t);\; \forall i    
\end{equation}
where, $\dot{\bm{s}}(t) = \bm{f}(\bm{s}(t))$ is the dynamics of isolated system. The stability of the synchronized state can be analyzed using linear stability analysis. Let $\bm{\varepsilon}^{(m)}_i(t) = \left(\bm{x}^{(m)}_i(t) - \bm{s}(t)\right)\in \mathbb{R}^d$ is a small perturbation to the synchronized state corresponding to the oscillator $i$ in the $m$th layer. From Eq. (\ref{1stdyneqn_mlil}), we can write the linear dynamics of the perturbations as
\begin{equation}\label{lindyn1_mlil}
    \dot{\bm{\varepsilon}}^{(m)}_i(t) = \bm{F}(\bm{s}(t)) \bm{\varepsilon}^{(m)}_i - \sigma^{(I)}\sum_{j=1}^N L^{(m)}_{ij} \bm{H}(\bm{s}(t)) \bm{\varepsilon}^{(m)}_j - \sigma^{(L)}\sum_{l=1}^M B_{ml} \bm{G}(\bm{s}(t)) \bm{\varepsilon}^{(l)}_i;\; i=1,\ldots,N ~\text{and}~ m=1,\ldots,M,
\end{equation}
where $\bm{F}(\bm{s}(t))$, $\bm{H}(\bm{s}(t))$ and $\bm{G}(\bm{s}(t))$ are the Jacobian matrices of the self-dynamics, intra-layer coupling function, and inter-layer coupling functions respectively, determined at the time-dependent synchronization solution $\bm{s}(t)$. We consider the following $Nd \times 1$ column matrix $\bm{E}^{(m)}$ combining the perturbations to the synchronization state in the layer $m$,
\begin{equation}
    \bm{E}^{(m)} = \left[ \bm{\varepsilon}^{(m)}_1, \bm{\varepsilon}^{(m)}_2, \ldots, \bm{\varepsilon}^{(m)}_N  \right]^T;\; m = 1, \ldots, M.
\end{equation}

The following equation gives the dynamics of perturbations on the $m$th layer.
\begin{equation}
    \dot{\bm{E}}^{(m)} = \Bigl( \mathbb{I}_N \otimes \bm{F}(\bm{s}(t)) \Bigr) \bm{E}^{(m)} - \sigma^{(I)} \Bigl(L^{(k)} \otimes \bm{H}(\bm{s}(t)) \Bigr) - \sigma^{(L)} \sum_{l=1}^M B_{ml} \Bigl( \mathbb{I}_N \otimes \bm{G}(\bm{s}(t)) \Bigr) \bm{E}^{(l)}
\end{equation}
where, $\mathbb{I}_N$ is an $N\times N$ identity matrix. Next, we consider a column matrix of dimension $MNd \times 1$, combining the perturbations of all layers
\begin{equation}
\bm{E} =  \left[ \bm{E}^{(1)}, \bm{E}^{(2)}, \ldots, \bm{E}^{(M)}  \right]^T
\end{equation}
and these two following intralayer and interlayer supra Laplacian matrices
\begin{equation}\label{intralayer_lsup_lap_mlil}
    \bm {\mathcal{L}} = \bigoplus_{m=1}^M {\bm L}^{(m)} = 
    \begin{pmatrix}
     {\bm L}^{(1)} & & & \\
     & {\bm L}^{(2)} & & \\
     & & \ddots & \\
     & & & {\bm L}^{(M)}
    \end{pmatrix},\;\; 
   \bm{\mathcal{B}} = {\bm B} \otimes \mathbb{I}_N = 
       \begin{pmatrix}
     {\bm B} & & & \\
     & {\bm B} & & \\
     & & \ddots & \\
     & & & {\bm B}
    \end{pmatrix} 
\end{equation}

With these above considerations, we can write the dynamics of the perturbations of all layers in the following single equation.
\begin{equation}\label{full_pertdyn_mlil}
    \dot{\bm{E}} = \biggl[ \Bigl( \mathbb{I}_M \otimes \mathbb{I}_N \otimes \bm{F}(\bm{s}) \Bigr) - \sigma^{(I)} \Bigl(\bm{\mathcal{L}} \otimes \bm{H}(\bm{s}) \Bigr) - \sigma^{(L)} \Bigl( \bm{\mathcal{B}} \otimes \bm{G}(\bm{s}) \Bigr) \biggr] \bm{E},
\end{equation}
The above equation captures the linearized dynamics of all perturbations. Next, our objective is to decouple these linearized dynamics along the eigenmodes of the interlayer and intralayer supra Laplacian matrices. Here, we note that the intra-layer supra Laplacian matrix $\bm{\mathcal{L}}$ and the inter-layer supra Laplacian matrix $\bm{\mathcal{B}}$ both are diagonalizable by considering a matrix $P$ of dimension $MN\times MN$, which is composed of the eigenvectors of the supra Laplacian matrices. So we can write 
\begin{subequations}\label{supmat_diag_mlil}
\begin{align}
    {\bm P}^{-1} \bm{\mathcal{L}} {\bm P} &= 
    \text{diag}\{\mu_1,\mu_2,\ldots,\mu_M,\mu_{M+1},\ldots,\mu_{MN}\} \label{intra_supmat_diag}\\
    {\bm P}^{-1} \bm{\mathcal{B}} {\bm P} &=
    \text{diag}\{\lambda_1,\lambda_2,\ldots,\lambda_M,\lambda_{M+1},\ldots,\lambda_{MN}\},
    \label{inter_supmat_diag}
\end{align}
\end{subequations}
here, $0 = \mu_1 = \mu_2 = \ldots = \mu_M \leqslant \mu_{M+1} \leqslant \ldots \leqslant \mu_{MN}$ are the eigenvalues of the intra-layer supra Laplacian matrix $\bm{\mathcal{L}}$ and $\lambda_m \geqslant 0;\; m= 1, \ldots, M$ are the eigenvalues of the inter-layer supra Laplacian matrix $\bm{\mathcal{B}}$.
Now, we rewrite the dynamics of the perturbations from Eq.~(\ref{full_pertdyn_mlil}) as below.
\begin{equation}\label{full_pertdyn_mlil_2}
    \begin{split}
        \dot{\bm{E}}  = \biggl[ \Bigl( \mathbb{I}_{MN} \otimes \bm{F}(\bm{s}) \Bigr)\Bigl({\bm P}\otimes \mathbb{I}_{d}\Bigr)\Bigl({\bm P}\otimes \mathbb{I}_{d}\Bigr)^{-1} & - \sigma^{(I)} \Bigl(\bm{\mathcal{L}} \otimes \bm{H}(\bm{s}) \Bigr)\Bigl({\bm P}\otimes \mathbb{I}_{d}\Bigr)\Bigl(\bm{P}\otimes \mathbb{I}_{d}\Bigr)^{-1} \biggr.\\
        & \biggl. - \sigma^{(L)} \Bigl( \bm{\mathcal{B}} \otimes \bm{G}(\bm{s}) \Bigr)\Bigl(\bm{P}\otimes \mathbb{I}_{d}\Bigr)\Bigl(\bm{P}\otimes \mathbb{I}_{d}\Bigr)^{-1} \biggr] \bm{E}.
    \end{split}
\end{equation}
Multiplying Eq.~(\ref{full_pertdyn_mlil_2}) by $(\bm{P}\otimes \mathbb{I}_d)^{-1}$ from left,
\begin{eqnarray}\label{full_pertdyn_mlil_21}
\begin{split}
    (\bm{P}\otimes \mathbb{I}_d)^{-1}\dot{\bm{E}}  &= 
    \biggl[ \Bigl(\bm{P}\otimes\mathbb{I}_d\Bigr)^{-1}\Bigl( \mathbb{I}_{MN} \otimes \bm{F}(\bm{s}) \Bigr)\Bigl(\bm{P}\otimes \mathbb{I}_{d}\Bigr)\Bigl(\bm{P}\otimes \mathbb{I}_{d}\Bigr)^{-1} \biggr.\\
    & \biggl.- \sigma^{(I)} \Bigl(\bm{P}\otimes\mathbb{I}_d\Bigr)^{-1} \Bigl(\bm{\mathcal{L}} \otimes \bm{H}(\bm{s}) \Bigr)\Bigl(\bm{P}\otimes \mathbb{I}_{d}\Bigr)\Bigl(\bm{P}\otimes \mathbb{I}_{d}\Bigr)^{-1} - \sigma^{(L)} \Bigl(\bm{P}\otimes\mathbb{I}_d\Bigr)^{-1} \Bigl( \bm{\mathcal{B}} \otimes \bm{G}(\bm{s}) \Bigr)\Bigl(\bm{P}\otimes \mathbb{I}_{d}\Bigr)\Bigl(\bm{P}\otimes \mathbb{I}_{d}\Bigr)^{-1} \biggr] \bm{E}.\\
    & =
    \biggl[ \Bigl(\bm{P}\otimes\mathbb{I}_d\Bigr)^{-1}\Bigl( \mathbb{I}_{MN} \otimes \bm{F}(\bm{s}) \Bigr)\Bigl(\bm{P}\otimes \mathbb{I}_{d}\Bigr) \biggr.\\
    & \biggl. - \sigma^{(I)} \Bigl(\bm{P}\otimes\mathbb{I}_d\Bigr)^{-1}\Bigl(\mathcal{L} \otimes \bm{H}(\bm{s}) \Bigr)\Bigl(\bm{P}\otimes \mathbb{I}_{d}\Bigr) - \sigma^{(L)} \Bigl(\bm{P}\otimes\mathbb{I}_d\Bigr)^{-1}\Bigl( \bm{\mathcal{B}} \otimes \bm{G}(\bm{s}) \Bigr)\Bigl(\bm{P}\otimes \mathbb{I}_{d}\Bigr) \biggr] \Bigl(\bm{P}\otimes \mathbb{I}_{d}\Bigr)^{-1} \bm{E}. &&
\end{split}
\end{eqnarray}
Let us now define a column vector $\Bigl(\bm{P}\otimes \mathbb{I}_{d}\Bigr)^{-1} \bm{E} = \bm{\Psi} =[\bm{\psi}_1^T,\bm{\psi}_2^T,\ldots,\bm{\psi}_{MN}^T]^T$ and we use the identities from Eq.~(\ref{supmat_diag_mlil}) in Eq.~(\ref{full_pertdyn_mlil_21}). Thus, we get the dynamics of all perturbations as
\begin{equation}\label{full_pertdyn_mlil_3}
\begin{split}
    \dot{\bm{\Psi}} &= \biggl[(\mathbb{I}_{MN}\otimes \bm{F}(\bm{s})) - \sigma^{(I)} (\text{diag}\{\mu_1,\mu_2,\ldots,\mu_M,\mu_{M+1},\ldots,\mu_{MN}\})\otimes \bm{H}(\bm{s}) \biggr.\\
    & \biggl. - \sigma^{(L)} (\text{diag}\{\lambda_1,\ldots,\lambda_M,\lambda_{M+1},\ldots,\lambda_{MN}\})\otimes \bm{G}(\bm{s})\biggl] \bm{\Psi}    
\end{split}
\end{equation}
Now, we can decouple all perturbations from Eq.~(\ref{full_pertdyn_mlil_3}) along the eigenmodes of the supra Laplacian matrices.
\begin{equation}\label{kth_perdyn_mlil}
    \dot{\bm{\psi}}_k = \biggl[ \bm{F}(\bm{s}) - \sigma^{(I)} \mu_k \bm{H}(\bm{s}) - \sigma^{(L)} \lambda_k \bm{G}(\bm{s})\biggl] \bm{\psi}_k;\; k=1,\ldots,MN,
\end{equation}
Here, $\bm{\psi}_k$ is the perturbation associated with the eigenvalues $\mu_k$ and $\lambda_k$ of the supra Laplacian matrices $\mathcal{L}$ and $\mathcal{B}$ respectively. The synchronization is stable when all transverse Lyapunov exponents are negative. As noted before, the perturbations corresponding to $\mu_k=0$ and $\lambda_k=0$ will not affect the stability of the synchronization manifold. We construct the master stability equation for multilayer networks with intra-layer and inter-layer coupling from Eq.~(\ref{kth_perdyn_mlil}) by removing the index dependence $k$ and considering the generic coupling parameters $\kappa^{(I)} = \kappa^{(I)}_k = \sigma^{(I)}\mu_k$ and $\kappa^{(L)} = \kappa^{(L)}_k = \sigma^{(L)}\lambda_k$,
\begin{equation}\label{mse_mlil}
    \dot{\bm{\psi}} = \biggl[ \bm{F}(\bm{s}) - \kappa^{(I)} \bm{H}(\bm{s}) - \kappa^{(L)} \bm{G}(\bm{s})\biggl] \bm{\psi}. 
\end{equation}
The maximum Lyapunov exponent $\Lambda^{max}(\kappa^{(I)},\kappa^{(L)})$ calculated from Eq.~(\ref{mse_mlil}) as a function $\kappa^{(I)}$ and $\kappa^{(L)}$ is the master stability function. The satble synchronization region is $\mathcal{R}_{\kappa^{(I)},\kappa^{(L)}} = \{ (\kappa^{(I)},\kappa^{(L)})\vert \Lambda^{max}(\kappa^{(I)},\kappa^{(L)})<0\}$.

\section{Master Stability Function in Higher Order Networks with natural coupling functions}

For the past three decades, networks with pairwise interactions have been effectively used to model various systems in natural sciences, engineering, and social sciences~\cite{Newman2018}. However, recent studies have pointed out that this traditional approach is not sufficient, asserting that modeling real-world systems using networks with pairwise connections is effective only in specific scenarios and provides a limited description of the systems where elements interact in groups~\cite{BianoniBookHOI, PhysRep.874.1}. For instance,  many biological~\cite{JCompNeuroSci.41.1}, social~\cite{BianoniBookHOI, Science.353.163}, physical~\cite{Roitz2014} and ecological~\cite{Nature.546.56, Nature.548.210} systems, as well as semantic networks \cite{semantic_networks} exhibit group interactions. 
In such complex networks, decomposing the group interactions into pairwise connections fails to capture the true nature of the system. Examples include functional and structural brain networks ~\cite{PetriJRSI2014, Lee2012},  protein interaction networks~\cite{JTheoBio.438.46}, coauthorship networks~\cite{EPJDataSci.6.18}, and ecological networks~\cite{Mayfield2017, Bairey2017}, where interactions inherently involve multiple elements and can not be meaningfully reduced to pairwise relationships~\cite{PhysRep.874.1}.

Hypergraphs \cite{BergeRevuzBook} and simplicial complexes \cite{BianoniBookHOI} are the mathematical objects that encode higher-order (group) interactions in complex systems \cite{SIAMRev.65.686}. Hypergraphs are the collections of hyperedges.
 The hyperedge describes the interaction of a group of nodes, and this hyperedge is characterized by its order $m$. For instance, the hyperedge of order $m=1$ represents pairwise interactions, the hyperedge of order $m=2$ represents the group interaction of $3$ nodes, and so on.
A simplicial complex is a collection of simplices, objects that generalize links and can be of different dimensions. A $d$-simplex, or simplex of dimension $d$, is a collection of $(d+1)$ nodes. In this way, a $0$-simplex is a node, a $1$-simplex is a link, a $2$-simplex is a (full) triangle, and so on. A \emph{simplicial complex} $\cal{S}$ is on a given set of nodes is a collection of $M$ simplices, $\mathcal{S}= \{\Sigma_1, \Sigma_2, \ldots, \Sigma_M\}$ with the extra requirement that for any simplice $\Sigma\in\mathcal{S}$, all subsimplices $\Sigma'\in\Sigma$ are also contained in $\mathcal{S}$. On the other hand, do not need to satisfy the above requirement in hypergraphs.
For instance, if three people are writing an article, then the requirement for the simplicial complex is that every pair of the triangle written a paper $\{A, B, C\}$, $Nodes=\{A, B, C\}$, $simplex=\{\{A\}, \{B\},\{C\},\{A, B, C\},\{A, B\}, \{B, C\}, \{C, A\} \} $. However, in hypergraphs, the writing of an article by three authors can be encoded only by a hyperedge $\{A, B, C\}$ of order $2$.

This section discusses the derivation of master stability function (MSF) in higher-order networks (Fig. \ref{schematic_MSF}), such as simplicial complexes with natural or general coupling functions \cite{NatComm.12.1255}. We consider $N$ coupled identical oscillators on a general $D$-dimensional simplicial complex. The following equation gives the dynamical equation of motion of the $i$th oscillator.
\begin{equation}\label{1stdyneq_hoi}
\begin{split}
    \dot{\bm{x}}_i & = \bm{f}(\bm{x}_i) + \sigma_1 \sum_{{j_1}=1}^N a_{i{j_1}}^{(1)} \bm{h}^{(1)}(\bm{x}_i,\bm{x}_{j_1}) + \sigma_2\sum_{{j_1}=1}^N\sum_{{j_2}=1}^N a_{i{j_1}{j_2}}^{(2)} \bm{h}^{(2)}(\bm{x}_i,\bm{x}_{j_1},\bm{x}_{j_2}) +  \\
    & \ldots + \sigma_D\sum_{{j_1}=1}^N \ldots \sum_{{j_D}=1}^N a_{i{j_1}\ldots{j_D}}^{(D)}\bm{h}^{(D)}(\bm{x}_i,\bm{x}_{j_1},\ldots,\bm{x}_{j_D});\; i=1,\ldots,N
\end{split}
\end{equation}
where, $\bm{x}_i\in \mathbb{R}^{d}$ is the $d$-dimensional state vector of oscillator $i$, $\bm{f}(\bm{x}_i):\mathbb{R}^{d}\mapsto \mathbb{R}^{d}$ is the identical self dynamics, $\sigma_1,\sigma_2,\ldots,\sigma_D$ are the scalar coupling parameters, $a^{(m)}_{i{j_1}{j_2}\ldots{j_m}}$ are the entries of the adjacency tensor ${\bm A}^{(m)};\; m=1,\ldots,D$. These tensors describe the structural patterns of interaction of any order that takes place in the higher-order networks.
Coupling functions of order $m$, denoted as
$\bm{h}^{(m)}(\bm{x}_i,\bm{x}_{j_1},\ldots,\bm{x}_{j_m}):\mathbb{R}^{(m+1)d}\mapsto\mathbb{R}^{d}; (m=1,\ldots,D)$, satisfies 
\begin{equation}\label{cfcond}
\bm{h}^{(m)}(\bm{x},\bm{x},\ldots,\bm{x})\equiv 0;\forall m.   \end{equation}
at the global synchronization state. For suitable coupling parameter values and coupling functions, the identical coupled oscillators converge to a global synchronization state provided by $\bm{x}_i(t)=\bm{s}(t);\;\forall i$, where $\dot{\bm{s}}(t) = \bm{f}(\bm{s}(t))$. The linear stability of the synchronization state can be studied by introducing small perturbations to the synchronization state, let $\bm{\varepsilon}_i(t) = \bm{x}_i(t) - \bm{s}(t)$ be such a perturbation to the synchronization solution of the $i$th oscillator. The linearized dynamical equation of the perturbation can be written as 
\begin{equation}\label{lineq_hoi}
\begin{split}
\dot{\bm{\varepsilon}_i}(t) & = \bm{F}(\bm{s})\bm{\varepsilon}_i + \sigma_1 \sum_{j_{1}=1}^{N} a_{ij_{1}}^{(1)} \left[ \left.\frac{\partial \bm{h}^{(1)}(\bm{x}_i,\bm{x}_{j_1})}{\partial\bm{x}_i}\right \vert_{(\bm{s},\bm{s})}\bm{\varepsilon}_i + \left.\frac{\partial\bm{h}^{(1)}(\bm{x}_i,\bm{x}_{j_1})}{\partial\bm{x}_{j_1}}\right\vert_{(\bm{s},\bm{s})}\bm{\varepsilon}_{j_1}\right] \\ & + \sigma_2 \sum_{j_{1}=1}^{N} \sum_{j_{2}=1}^{N} a_{ij_{1}j_{2}}^{(2)} \left[ \left.\frac{\partial \bm{h}^{(2)}(\bm{x}_i,\bm{x}_{j_{1}},\bm{x}_{j_{2}})}{\partial \bm{x}_i}\right\vert_{(\bm{s},\bm{s},\bm{s})}\bm{\varepsilon}_i + \left.\frac{\partial \bm{h}^{(2)}(\bm{x}_i,\bm{x}_{j_{1}},\bm{x}_{j_{2}})}{\partial \bm{x}_{j_1}} \right\vert_{(\bm{s},\bm{s},\bm{s})}\bm{\varepsilon}_{j_1} \right. 
\\ & \left. + \left.\frac{\partial\bm{h}^{(2)}(\bm{x}_i,\bm{x}_{j_1},\bm{x}_{j_2})}{\partial\bm{x}_{j_2}}\right \vert_{(\bm{s},\bm{s},\bm{s})}\bm{\varepsilon}_{j_2} \right] + \ldots + \sigma_D \sum_{{j_1}=1}^{N}\ldots\sum_{{j_D}=1}^{N} a_{i{j_1}\ldots{j_D}}^{(D)}\left[ \left.\frac{\partial\bm{h}^{(D)}(\bm{x}_i,\bm{x}_{j_1},\ldots,\bm{x}_{j_D})}{\partial\bm{x}_i}\right\vert_{(\bm{s},\bm{s},\ldots,\bm{s})}\bm{\varepsilon}_i \right. \\ &
     \left. + \left.\frac{\partial\bm{h}^{(D)}(\bm{x}_i,\bm{x}_{j_1},\ldots,\bm{x}_{j_D})}{\partial\bm{x}_{j_1}}\right\vert_{(\bm{s},\bm{s},\ldots,\bm{s})}\bm{\varepsilon}_{j_1}\right. + \ldots + \left.\left.\frac{\partial\bm{h}^{(D)}(\bm{x}_i,\bm{x}_{j_1},\ldots,\bm{x}_{j_D})}{\partial\bm{x}_{j_D}}\right\vert_{(\bm{s},\bm{s},\ldots,\bm{s})}\bm{\varepsilon}_{j_D} \right]
\end{split}
\end{equation}
where, $\bm{F}(\bm{s})$ is the $d\times d$ Jacobian of the self dynamics $\bm{f}(\bm{x})$ at the synchronization solution $\bm{s}(t)$. Now due to the conditions on the coupling function $\bm{h}^{(m)}(\ldots)$ from Eq.~(\ref{cfcond}), the total derivatives of the coupling function vanishes on the synchronization solution $\bm{s}(t)$
\begin{subequations}
\label{totderiv}
    \begin{equation}
        \label{td_1}
        \left.\frac{\partial\bm{h}^{(1)}(\bm{x}_i,\bm{x}_{j_1})}{\partial\bm{x}_i}\right\vert_{(\bm{s},\bm{s})} + \left.\frac{\partial\bm{h}^{(1)}(\bm{x}_i,\bm{x}_{j_1})}{\partial\bm{x}_{j_1}}\right\vert_{(\bm{s},\bm{s})} = 0 
    \end{equation}
    \begin{equation}
        \label{td_2}
        \left.\frac{\partial\bm{h}^{(2)}(\bm{x}_i,\bm{x}_{j_1},\bm{x}_{j_2})}{\partial\bm{x}_i}\right\vert_{(\bm{s},\bm{s},\bm{s})} + \left.\frac{\partial\bm{h}^{(2)}(\bm{x}_i,\bm{x}_{j_1},\bm{x}_{j_2})}{\partial\bm{x}_{j_1}}\right\vert_{(\bm{s},\bm{s},\bm{s})} + \left.\frac{\partial\bm{h}^{(2)}(\bm{x}_i,\bm{x}_{j_1},\bm{x}_{j_2})}{\partial\bm{x}_{j_2}}\right\vert_{(\bm{s},\bm{s},\bm{s})}= 0 
    \end{equation}
    \begin{equation}
        \label{td_D}
        \left.\frac{\partial\bm{h}^{(D)}(\bm{x}_i,\bm{x}_{j_1},\ldots,\bm{x}_{j_D})}{\partial\bm{x}_i}\right\vert_{(\bm{s},\bm{s},\ldots,\bm{s})} + \left.\frac{\partial\bm{h}^{(D)}(\bm{x}_i,\bm{x}_{j_1},\ldots,\bm{x}_{j_D})}{\partial\bm{x}_{j_1}}\right\vert_{(\bm{s},\bm{s},\ldots,\bm{s})} +\ldots+ \left.\frac{\partial\bm{h}^{(D)}(\bm{x}_i,\bm{x}_{j_1},\ldots,\bm{x}_{j_D})}{\partial\bm{x}_{j_D}}\right\vert_{(\bm{s},\bm{s},\ldots,\bm{s})}= 0.
    \end{equation}
\end{subequations}

Now form Eq.~(\ref{td_1}) replacing $\left.\frac{\partial\bm{h}^{(1)}(\bm{x}_i,\bm{x}_{j_1})}{\partial\bm{x}_{j_1}}\right\vert_{\bm{s},\bm{s}}$ in the second term of R.H.S. of Eq.~(\ref{lineq_hoi}) we get
\begin{eqnarray}\label{2ndterm1}
\begin{split}
        \sigma_1\sum_{{j_1}=1}^{N} a_{i{j_1}}^{(1)}\left[ \left.\frac{\partial\bm{h}^{(1)}(\bm{x}_i,\bm{x}_{j_1})}{\partial\bm{x}_i}\right\vert_{(\bm{s},\bm{s})}\bm{\varepsilon}_i - \left.\frac{\partial\bm{h}^{(1)}(\bm{x}_i,\bm{x}_{j_1})}{\partial\bm{x}_{i}}\right\vert_{(\bm{s},\bm{s})}\bm{\varepsilon}_{j_1}\right] & = \sigma_1 \left.\frac{\partial\bm{h}^{(1)}(\bm{x}_i,\bm{x}_{j_1})}{\partial\bm{x}_{i}}\right\vert_{(\bm{s},\bm{s})} \left[ k_{i}^{(1)}\bm{\varepsilon}_i - \sum_{{j_1}=1}^{N} a_{i{j_1}}^{(1)}\bm{\varepsilon}_{j_1}\right]
\end{split}
\end{eqnarray}
where, $\sum_{i_{j_1}}^N a_{i_{j_1}}^{(1)} = k_i^{(1)}$ is the degree of the $i$th node in the network. Similarly, replacing $\left.\frac{\partial\bm{h}^{(2)}(\bm{x}_i,\bm{x}_{j_1},\bm{x}_{j_2})}{\partial\bm{x}_i}\right\vert_{(\bm{s},\bm{s},\bm{s})}$ from Eq.~(\ref{td_2}) in the third term in the R.H.S. of the linearized equation Eq.~(\ref{lineq_hoi}) we get
\begin{equation}\label{3rdterm1}
\begin{split}
    \sigma_2 \sum_{{j_1}=1}^{N} \sum_{{j_2}=1}^{N} a_{i{j_1}{j_2}}^{(2)}\left[ \left.\frac{\partial\bm{h}^{(2)}(\bm{x}_i,\bm{x}_{j_1},\bm{x}_{j_2})}{\partial\bm{x}_i}\right\vert_{(\bm{s},\bm{s},\bm{s})}\bm{\varepsilon}_i + \left.\frac{\partial\bm{h}^{(2)}(\bm{x}_i,\bm{x}_{j_1},\bm{x}_{j_2})}{\partial\bm{x}_{j_1}}\right\vert_{\bm{s},\bm{s},\bm{s}}\bm{\varepsilon}_{j_1} + \left.\frac{\partial\bm{h}^{(2)}(\bm{x}_i,\bm{x}_{j_1},\bm{x}_{j_2})}{\partial\bm{x}_{j_2}}\right\vert_{(\bm{s},\bm{s},\bm{s})}\bm{\varepsilon}_{j_2} \right] \\
    = -\sigma_2\left.\frac{\partial\bm{h}^{(2)}(\bm{x}_i,\bm{x}_{j_1},\bm{x}_{j_2})}{\partial\bm{x}_{j_1}}\right\vert_{(\bm{s},\bm{s},\bm{s})}\left[2k_{i}^{(2)}\bm{\varepsilon}_i - \sum_{{j_1}=1}^N k_{i{j_1}}^{(2)}\bm{\varepsilon}_{j_1} \right]  -\sigma_2\left.\frac{\partial\bm{h}^{(2)}(\bm{x}_i,\bm{x}_{j_1},\bm{x}_{j_2})}{\partial\bm{x}_{j_2}}\right\vert_{(\bm{s},\bm{s},\bm{s})}\left[2k_{i}^{(2)} - \sum_{{j_2}=1}^N k_{i{j_2}}^{(2)}\bm{\varepsilon}_{j_2}\right] 
\end{split}
\end{equation}
where $k_{i}^{(2)}=\frac{1}{2}\sum_{j}\sum_{l}a_{i{j}{l}}^{(2)}$ is the generalized 2-degree of node $i$ and $k_{ij}^{(2)}=\sum_{l}a_{ijl}^{(2)}$ is the generalized 2-degree of edge $(i,j)$. A similar analysis on the last term of Eq.~(\ref{lineq_hoi}) will provide the following equation
\begin{equation}\label{lastterm1}
    \begin{split}
        -\sigma_D\left.\frac{\partial\bm{h}^{(D)}(\bm{x}_i,\bm{x}_{j_1},\ldots,\bm{x}_{j_D})}{\partial\bm{x}_{j_1}}\right\vert_{(\bm{s},\ldots,\bm{s})}\left[D!k_{i}^{(D)}\bm{\varepsilon}_{i} - \sum_{{j_1}=1}^{N} (D-1)!k_{i{j_1}}^{(D)}\bm{\varepsilon}_{j_1}\right]\\
        -\sigma_D\left.\frac{\partial\bm{h}^{(D)}(\bm{x}_i,\bm{x}_{j_1},\ldots,\bm{x}_{j_D})}{\partial\bm{x}_{j_2}}\right\vert_{(\bm{s},\ldots,\bm{s})}\left[D!k_{i}^{(D)}\bm{\varepsilon}_{i} - \sum_{{j_2}=1}^{N} (D-1)!k_{i{j_2}}^{(D)}\bm{\varepsilon}_{j_2}\right]\\
        - \ldots -\sigma_D\left.\frac{\partial\bm{h}^{(D)}(\bm{x}_i,\bm{x}_{j_1},\ldots,\bm{x}_{j_D})}{\partial\bm{x}_{j_D}}\right\vert_{(\bm{s},\ldots,\bm{s})}\left[D!k_{i}^{(D)}\bm{\varepsilon}_{i} - \sum_{{j_D}=1}^{N} (D-1)!k_{i{j_D}}^{(D)}\bm{\varepsilon}_{j_D}\right]
    \end{split}
\end{equation}
where, $k_i^{(D)}=\frac{1}{D!}\sum_{j_1}\ldots\sum_{j_D}a_{i{j_1}\ldots{j_D}}^{(D)}$ and $k_{ij}^{(D)}=\frac{1}{(D-1)!}\sum_{{l_2}}\ldots\sum_{l_D}a_{ij{l_2}\ldots{l_D}}^{(D)}$ are the generalized $D$-degree of the node $i$ and link $(i,j)$ respectively. Let us now define a tensor ${\bm T}^{(m)}$ with elements
\begin{equation}\label{tau_def}
    \tau_{i{j_1}{j_2}\ldots{j_m}}^{(m)} = m!k_i^{(m)}\delta_{i{j_1}{j_2}\ldots{j_m}} - a_{i{j_1}{j_2}\ldots{j_m}}^{(m)};\; i,j_1,j_2,\ldots,j_m = 1,\ldots,N;\; m=2,\ldots,D.
\end{equation}
where $\delta_{i{j_1}\ldots{j_m}}=1$ if $i=j_1=j_2=\ldots=j_m$ and zero otherwise. To see the importance of the above definition of $\tau$ in Eq.~(\ref{tau_def}) let us consider the case for $m=2$ and the following two summations $\sum_{{j_1},{j_2}}\tau_{i{j_1}{j_2}}^{(2)}\bm{\varepsilon}_{j_1}$ and $\sum_{{j_1},{j_2}}\tau_{i{j_1}{j_2}}^{(2)}\bm{\varepsilon}_{j_2}$ below

\begin{equation}
\begin{split}
    \sum_{{j_1},{j_2}}\tau_{i{j_1}{j_2}}^{(2)}\bm{\varepsilon}_{j_1} 
    & = \sum_{{j_1},{j_2}}\left(2k_i^{(2)}\delta_{i{j_1}{j_2}} - a_{i{j_1}{j_2}}^{(2)}\right)\bm{\varepsilon}_{j_1}\\
    & = 2k_i^{(2)}\sum_{{j_1},{j_2}}\delta_{i{j_1}{j_2}}\bm{\varepsilon}_{j_1} - \sum_{j_1}\bm{\varepsilon}_{j_1}\sum_{j_2}a_{i{j_1}{j_2}}^{(2)}\\
    & = 2k_i^{(2)}\bm{\varepsilon}_{i} - \sum_{j_1}k_{i{j_1}}^{(2)}\bm{\varepsilon}_{j_1}.     
\end{split}
\end{equation}

Similarly, we can obtain,
\begin{equation}
    \sum_{{j_1},{j_2}}\tau_{i{j_1}{j_2}}^{(2)}\bm{\varepsilon}_{j_2} = 2k_i^{(2)}\bm{\varepsilon}_{i} - \sum_{j_2}k_{i{j_2}}^{(2)}\bm{\varepsilon}_{j_2}.
\end{equation}

And for the generalized dimension $m=D$, we will have,
\begin{eqnarray}
    \sum_{{j_1},\ldots,{j_D}}\tau_{i{j_1}\ldots{j_D}}^{(D)}\bm{\varepsilon}_{j_1} & = & D!k_i^{(D)}\bm{\varepsilon}_{i} - \sum_{j_1}(D-1)!k_{i{j_1}}^{(D)}\bm{\varepsilon}_{j_1}\\
    \sum_{{j_1},\ldots,{j_D}}\tau_{i{j_1}\ldots{j_D}}^{(D)}\bm{\varepsilon}_{j_2} & = & D!k_i^{(D)}\bm{\varepsilon}_{i} - \sum_{j_2}(D-1)!k_{i{j_2}}^{(D)}\bm{\varepsilon}_{j_2}\\
    \sum_{{j_1},\ldots,{j_D}}\tau_{i{j_1}\ldots{j_D}}^{(D)}\bm{\varepsilon}_{j_D} & = & D!k_i^{(D)}\bm{\varepsilon}_{i} - \sum_{j_D}(D-1)!k_{i{j_D}}^{(D)}\bm{\varepsilon}_{j_D}.
\end{eqnarray}

Let us also use the following compact notations for the derivatives of the coupling function
\begin{eqnarray}
    \left.\frac{\partial\bm{h}^{(1)}(\bm{x}_i,\bm{x}_{j_1})}{\partial\bm{x}_{j_1}}\right\vert_{(\bm{s},\bm{s})} & = & \bm{H}^{(1)}_{1}(\bm{s},\bm{s}) \\
    \left.\frac{\partial\bm{h}^{(2)}(\bm{x}_i,\bm{x}_{j_1},\bm{x}_{j_2})}{\partial\bm{x}_{j_1}}\right\vert_{(\bm{s},\bm{s},\bm{s})} & = & \bm{H}^{(2)}_{1}(\bm{s},\bm{s},\bm{s})\\
    \left.\frac{\partial\bm{h}^{(2)}(\bm{x}_i,\bm{x}_{j_1},\bm{x}_{j_2})}{\partial\bm{x}_{j_2}}\right\vert_{(\bm{s},\bm{s},\bm{s})}  & = & \bm{H}^{(2)}_{2}(\bm{s},\bm{s},\bm{s})\\
    \left.\frac{\partial\bm{h}^{(D)}(\bm{x}_i,\bm{x}_{j_1},\ldots,\bm{x}_{j_D})}{\partial\bm{x}_{j_1}}\right\vert_{(\bm{s},\ldots,\bm{s})} & = & \bm{H}^{(D)}_{1}(\bm{s,\ldots,\bm{s}})\\
    \left.\frac{\partial\bm{h}^{(D)}(\bm{x}_i,\bm{x}_{j_1},\ldots,\bm{x}_{j_D})}{\partial\bm{x}_{j_2}}\right\vert_{(\bm{s},\ldots,\bm{s})} & = & \bm{H}^{(D)}_{2}(\bm{s,\ldots,\bm{s}})\\
    \left.\frac{\partial\bm{h}^{(D)}(\bm{x}_i,\bm{x}_{j_1},\ldots,\bm{x}_{j_D})}{\partial\bm{x}_{j_D}}\right\vert_{(\bm{s},\ldots,\bm{s})} & = & \bm{H}^{(D)}_{D}(\bm{s,\ldots,\bm{s}})
\end{eqnarray}
Incorporating the above calculations, we can rewrite the linearized dynamical equation (Eq.~(\ref{lineq_hoi})) for the small perturbation as
\begin{equation}\label{pertdynhoi_1}
    \begin{split}
        \dot{\bm{\varepsilon}}_i 
        & = \bm{F}(\bm{s})\bm{\varepsilon}_{i} - \sigma_1 \bm{H}^{(1)}_{1}(\bm{s},\bm{s}) \left[k_i^{(1)}\bm{\varepsilon}_{i} - \sum_{j_1}a_{i{j_1}}^{(1)}\bm{\varepsilon_{j_1}}\right]  -\sigma_2\left[ \bm{H}^{(2)}_{1}(\bm{s},\bm{s})\sum_{{j_1}}\sum_{j_2}\tau_{i{j_1}{j_2}}^{(2)}\bm{\varepsilon}_{j_1} + \bm{H}^{(2)}_{2}(\bm{s},\bm{s})\sum_{{j_1}}\sum_{j_2}\tau_{i{j_1}{j_2}}^{(2)}\bm{\varepsilon}_{j_2}\right] - \ldots\\
        &-\sigma_{D}\left[ \bm{H}^{(D)}_{1}(\bm{s},\ldots,\bm{s})\sum_{{j_1}\ldots{j_D}}\tau_{i{j_1}\ldots{j_D}}^{(D)}\bm{\varepsilon}_{j_1} + \bm{H}^{(D)}_{2}(\bm{s},\ldots,\bm{s})\sum_{{j_1}\ldots{j_D}}\tau_{i{j_1}\ldots{j_D}}^{(D)}\bm{\varepsilon}_{j_2} + \ldots + \bm{H}^{(D)}_{D}(\bm{s},\ldots,\bm{s})\sum_{{j_1}\ldots{j_D}}\tau_{i{j_1}\ldots{j_D}}^{(D)}\bm{\varepsilon}_{j_D}\right]\\
        & = {\bm F}(\bm{s})\bm{\varepsilon}_{i} - \sigma_1 \bm{H}^{(1)}(\bm{s},\bm{s}) \left[k_i^{(1)}\bm{\varepsilon}_{i} - \sum_{j_1}a_{i{j_1}}^{(1)}\varepsilon_{j_1}\right]  -\sigma_2\left[\bm{H}^{(2)}_{1}(\bm{s},\bm{s})\sum_{{j_1}}\bm{\varepsilon}_{j_1}\sum_{j_2}\tau_{i{j_1}{j_2}}^{(2)} + \bm{H}^{(2)}_{2}(\bm{s},\bm{s})\sum_{j_2}\bm{\varepsilon}_{j_2}\sum_{{j_1}}\tau_{i{j_1}{j_2}}^{(2)}\right] - \ldots \\
        & -\sigma_D\left[\bm{H}^{(D)}_{1}(\bm{s},\ldots,\bm{s})\sum_{j_1}\bm{\varepsilon}_{j_1} \sum_{{j_2}\ldots{j_D}}\tau_{i{j_1}\ldots{j_D}}^{(D)} + \bm{H}^{(D)}_{2}(\bm{s},\ldots,\bm{s})\sum_{j_2} \bm{\varepsilon}_{j_2} \sum_{{j_1}{j_3}\ldots{j_D}}\tau_{i{j_1}\ldots{j_D}}^{(D)} + \ldots\right. \\
        & \left. + \bm{H}^{(D)}_{D}(\bm{s},\ldots,\bm{s})\sum_{j_D} \bm{\varepsilon}_{j_D} \sum_{{j_1}\ldots{j_{(D-1)}}}\tau_{i{j_1}\ldots{j_D}}^{(D)} \right]
    \end{split}
\end{equation}
From Eq.~(\ref{tau_def}) we will find the below symmetry properties
\begin{subequations}
    \begin{equation}
        \sum_{j_2}\tau_{i{j_1}{j_2}}^{(2)} = \sum_{j_2}\tau_{i{j_2}{j_1}}^{(2)} 
    \end{equation}
    \begin{equation}
        \sum_{{j_1}{j_3}\ldots{j_D}}\tau_{i{j_1}{j_2}\ldots{j_D}}^{(D)} = \sum_{{j_1}{j_3}\ldots{j_D}}\tau_{i{j_2}{j_1}\ldots{j_D}}^{(D)}
    \end{equation}
    \begin{equation}
        \sum_{{j_1}\ldots{j_{(D-1)}}}\tau_{i{j_1}\ldots{j_D}}^{(D)} = \sum_{{j_1}\ldots{j_{(D-1)}}}\tau_{i{j_D}{j_2}\ldots{j_{D-1}}{j_1}}^{(D)}
    \end{equation}
\end{subequations}

The elements of the generalized Laplacian matrix of order $m=1,2,\ldots,D$ are as follows
\begin{equation}
    \mathcal{L}_{ij}^{(m)} = \begin{cases}
        0, \text{ for $i\neq j$ and $a_{ij}^{(1)}=0$}\\
        -(m-1)!k_{ij}^{(m)}, \text{ for $i\neq j$ and $a_{ij}^{(1)}=1$}\\
        m!k_i^{(m)}, \text{ for $i=j$}
    \end{cases}
\end{equation}

Incorporating the these symmetry properties of $\tau$ and the first order Laplacian matrix $\mathcal{L}^{(1)}$ Eq.~(\ref{pertdynhoi_1}) can be rewritten as
\begin{equation}\label{pertdynhoi_2}
    \begin{split}
        \dot{\bm{\varepsilon}}_i 
        & = \bm{F}(\bm{s})\bm{\varepsilon}_i - \sigma_1 \bm{H}^{(1)}_{1}(\bm{s},\bm{s})\sum_{j_1}\mathcal{L}_{i{j_1}}^{(1)}\bm{\varepsilon_{j_1}}  - \sigma_2\left[ \bm{H}^{(2)}_{1}(\bm{s},\bm{s})\sum_{{j_1}}\bm{\varepsilon}_{j_1}\sum_{j_2}\tau_{i{j_1}{j_2}}^{(2)} + \bm{H}^{(2)}_{2}(\bm{s},\bm{s})\sum_{j_2}\bm{\varepsilon}_{j_2}\sum_{{j_1}}\tau_{i{j_2}{j_1}}^{(2)}\right] - \ldots \\
        &  -\sigma_D\left[\bm{H}^{(D)}_{1}(\bm{s},\ldots,\bm{s})\sum_{j_1} \bm{\varepsilon}_{j_1} \sum_{{j_2}\ldots{j_D}}\tau_{i{j_1}\ldots{j_D}}^{(D)} + \bm{H}^{(D)}_{2}(\bm{s},\ldots,\bm{s})\sum_{j_2} \bm{\varepsilon}_{j_2} \sum_{{j_1}{j_3}\ldots{j_D}}\tau_{i{j_2}{j_1}\ldots{j_D}}^{(D)} + \ldots\right. \\
        & \left. + \bm{H}^{(D)}_{D}(\bm{s},\ldots,\bm{s})\sum_{j_D} \bm{\varepsilon}_{j_D} \sum_{{j_1}\ldots{j_{(D-1)}}}\tau_{i{j_D}{j_2}\ldots{j_{(D-1)}}{j_1}}^{(D)} \right]
    \end{split}
\end{equation}

We note here that the indices $j_1,j_2,\ldots,j_D$ are  dummy indices in Eq.~(\ref{pertdynhoi_2}). By suitably swapping these indices in the Eq.~(\ref{pertdynhoi_2}) we get
\begin{equation}\label{pertdynhoi_3}
    \begin{split}
        \dot{\bm{\varepsilon}}_i 
        & = \bm{F}(\bm{s})\bm{\varepsilon}_i - \sigma_1 D\bm{h}^{(1)}(\bm{s},\bm{s})\sum_{j_1}\mathcal{L}_{i{j_1}}^{(1)}\bm{\varepsilon_{j_1}}  - \sigma_2\left[ \bm{H}^{(2)}_1(\bm{s},\bm{s})\sum_{{j_1}}\bm{\varepsilon}_{j_1}\sum_{j_2}\tau_{i{j_1}{j_2}}^{(2)} + \bm{H}^{(2)}_{2}(\bm{s},\bm{s})\sum_{j_1}\bm{\varepsilon}_{j_1}\sum_{{j_2}}\tau_{i{j_1}{j_2}}^{(2)}\right] - \ldots \\
        &  -\sigma_D\left[\bm{H}^{(D)}_{1}(\bm{s},\ldots,\bm{s})\sum_{j_1} \bm{\varepsilon}_{j_1} \sum_{{j_2}\ldots{j_D}}\tau_{i{j_1}\ldots{j_D}}^{(D)} + \bm{H}^{(D)}_{2}(\bm{s},\ldots,\bm{s})\sum_{j_1} \bm{\varepsilon}_{j_1} \sum_{{j_2}{j_3}\ldots{j_D}}\tau_{i{j_1}{j_2}\ldots{j_D}}^{(D)} + \ldots\right. \\
        & \left. + \bm{H}^{(D)}_{D}(\bm{s},\ldots,\bm{s})\sum_{j_1} \bm{\varepsilon}_{j_1} \sum_{{j_2}\ldots{j_{D}}}\tau_{i{j_1}{j_2}\ldots{j_{(D-1)}}{j_D}}^{(D)} \right]
    \end{split}
\end{equation}

Now, from Eq.~(\ref{tau_def}) we will find that
\begin{equation}
    \sum_{{j_2}\ldots{j_m}}\tau_{i{j_1}{j_2}\ldots{j_m}}^{(m)} = \sum_{{j_2}\ldots{j_m}}m!k_i^{(m)}\delta_{i{j_1}{j_2}\ldots{j_m}} - \sum_{{j_2}\ldots{j_m}}a_{i{j_1}\ldots{j_m}}^{(m)} =  \begin{cases}
        m!k_{i}^{(m)}, \text{ for $i={j_1}$}\\
        0, \text{ for $i\neq {j_1}$ and $a_{i{j_1}}^{(1)}=0$}\\
        -(m-1)!k_{i{j_1}}^{(m)}, \text{ for $i\neq {j_1}$ and $a_{i{j_1}}^{(1)}=1$}
    \end{cases}
\end{equation}
which is the generalized $m$-order Laplacian $\mathcal{L}_{i{j_1}}^{(m)}$. With this we can rewrite Eq.~(\ref{pertdynhoi_3}) as
\begin{equation}\label{pertdynhoi_4}
    \begin{split}
        \dot{\bm{\varepsilon}}_i  & = \bm{F}(\bm{s})\bm{\varepsilon}_i - \sigma_1 \bm{H}^{(1)}_{1}(\bm{s},\bm{s})\sum_{j_1}\mathcal{L}_{i{j_1}}^{(1)}\bm{\varepsilon}_{j_1}  - \sigma_2\left[ \bm{H}^{(2)}_{1}(\bm{s},\bm{s}) + \bm{H}^{(2)}_{2}(\bm{s},\bm{s})\right]\sum_{{j_1}}\mathcal{L}_{i{j_1}}^{(2)}\bm{\varepsilon}_{j_1} - \ldots \\
        &  -\sigma_D\left[\bm{H}^{(D)}_{1}(\bm{s},\ldots,\bm{s}) + \bm{H}^{(D)}_{2}(\bm{s},\ldots,\bm{s}) + \ldots + \bm{H}^{(D)}_{D}(\bm{s},\ldots,\bm{s}) \right]\sum_{j_1}\mathcal{L}_{i{j_1}}^{(D)} \bm{\varepsilon}_{j_1}
    \end{split}
\end{equation}

By denoting $\bm{F}(\bm{s})= \bm{\mathcal{F}}$, $\bm{H}^{(1)}_{1}(\bm{s},\bm{s})=\bm{\mathcal{H}}^{(1)}$,
$\left[\bm{H}^{(2)}_{1}(\bm{s},\bm{s}) + \bm{H}^{(2)}_{2}(\bm{s},\bm{s})\right] = \bm{\mathcal{H}}^{(2)}$
and,
$$\left[\bm{H}^{(D)}_{1}(\bm{s},\ldots,\bm{s}) + \bm{H}^{(D)}_{2}(\bm{s},\ldots,\bm{s}) + \ldots + \bm{H}^{(D)}_{D}(\bm{s},\ldots,\bm{s}) \right] = \bm{\mathcal{H}}^{(D)}$$
 in  Eq.~(\ref{pertdynhoi_4}) we get
\begin{equation}\label{pertdynhoi_5}
    \dot{\bm{\varepsilon}}_i = \bm{\mathcal{F}}\bm{\varepsilon}_i - \sigma_1\bm{\mathcal{H}}^{(1)} \sum_{j_1}\mathcal{L}_{i{j_1}}^{(1)}\bm{\varepsilon_{j_1}}  - \sigma_2 \bm{\mathcal{H}}^{(2)}\sum_{{j_1}}\mathcal{L}_{i{j_1}}^{(2)}\bm{\varepsilon}_{j_1}  - \ldots -\sigma_D \bm{\mathcal{H}}^{(D)}\sum_{j_1}\mathcal{L}_{i{j_1}}^{(D)} \bm{\varepsilon}_{j_1} ;\; i=1,\ldots,N
\end{equation}

Let us consider the following $d \times N$ matrix $\bm{\mathcal{E}}=(\bm{\varepsilon}_1 \bm{\varepsilon}_2 \ldots \bm{\varepsilon}_N)$. Eq.~(\ref{pertdynhoi_5}) can be written in the matrix form
\begin{equation}\label{pertdynhoi_6}
	\begin{split}
		\dot{\bm{\mathcal{E}}} = \bm{\mathcal{F}}\bm{\mathcal{E}} - \sigma_1 \bm{\mathcal{H}}^{(1)}\bm{\mathcal{E}}\left(\bm{\mathcal{L}}^{(1)}\right)^T - \sigma_2 \bm{\mathcal{H}}^{(2)}\bm{\mathcal{E}}\left(\bm{\mathcal{L}}^{(2)}\right)^T - \ldots - \sigma_D \bm{\mathcal{H}}^{(D)}\bm{\mathcal{E}}\left(\bm{\mathcal{L}}^{(D)}\right)^T
	\end{split}
\end{equation}
where ${\bm A}^{T}$ denotes the transpose of matrix ${\bm A}$. Eq.~(\ref{pertdynhoi_6}) encloses the dynamics of all perturbations in a single matrix equation. We now decouple these dynamics along the eigenmodes of the classical Laplacian matrix $\bm{\mathcal{L}}^{(1)}$ of the network with pairwise connections only. Let $\bm{v}_k^{(1)}$ be the eigenvector of $\left(\bm{\mathcal{L}}^{(1)}\right)^T$  corresponding to eigenvalue $\mu_k^{(1)}$
\begin{equation}
\left(\bm{\mathcal{L}}^{(1)}\right)^T \bm{v}_k^{(1)} = \mu_k^{(1)} \bm{v}_k^{(1)};\; k=1,\ldots,N,
\end{equation}
and $\bm{V}^{(1)}=(\bm{v}_1^{(1)} \bm{v}_2^{(1)} \ldots \bm{v}_N^{(1)})$ is $N \times N$ matrix that diagonalizes $\left(\bm{\mathcal{L}}^{(1)}\right)^T$.

\begin{equation}	\left(\bm{V}^{(1)}\right)^{-1}\left(\bm{\mathcal{L}}^{(1)}\right)^T\bm{V}^{(1)} = \bm{\Gamma}^{(1)},
\end{equation}
where ${\bm \Gamma}^{(1)} = \text{diag}(\mu_1^{(1)} \mu_2^{(1)} \ldots \mu_N^{(1)})$ are diagonal matrices with diagonal entries are the eigenvalues of the transpose of the classical Laplacian matrix. Multiplying Eq.~(\ref{pertdynhoi_6}) by $\bm{V}^{(1)}$ from right
\begin{equation}\label{pertdynhoi_7}
	\begin{split}
	\dot{\bm{\mathcal{E}}} \bm{V}^{(1)} = \bm{\mathcal{F}}\bm{\mathcal{E}} \bm{V}^{(1)} - \sigma_1 \bm{\mathcal{H}}^{(1)}\bm{\mathcal{E}}\left(\bm{\mathcal{L}}^{(1)}\right)^T \bm{V}^{(1)} - \sigma_2 \bm{\mathcal{H}}^{(2)}\bm{\mathcal{E}}\left(\bm{\mathcal{L}}^{(2)}\right)^T \bm{V}^{(1)} -
		 \ldots - \sigma_D \bm{\mathcal{H}}^{(D)}\bm{\mathcal{E}}\left(\bm{\mathcal{L}}^{(D)}\right)^T \bm{V}^{(1)}
	\end{split}
\end{equation}
We define $\bm{\mathcal{E}}\bm{V}^{(1)}=\bm{\Psi} = [\bm{\psi}_1 \bm{\psi}_2 \ldots \bm{\psi}_N]$ and $\left(\bm{V}^{(1)}\right)^{-1}\left(\bm{\mathcal{L}}^{(m)}\right)^T\bm{V}^{(1)} = \tilde{\bm{\mathcal{L}}}^{(m)};\; m=2,\ldots,D$ are the matrices obtained by transforming $\bm{\mathcal{L}}^{(m)}$ by the matrix $\bm{V}^{(1)}$. With these definitions Eq.~(\ref{pertdynhoi_7}) takes the below form
\begin{equation}\label{projdynhoi_1}
 \begin{split}
 	\dot{\bm{\Psi}} = \bm{\mathcal{F}}\bm{\Psi} - \sigma_1 \bm{\mathcal{H}}^{(1)}\bm{\Psi}\Gamma^{(1)} - \sigma_2 \bm{\mathcal{H}}^{(2)} \bm{\Psi}\tilde{\bm{\mathcal{L}}}^{(2)} - \ldots - \sigma_D \bm{\mathcal{H}}^{(D)}\bm{\Psi}\tilde{\bm{\mathcal{L}}}^{(D)}.
 	\end{split}
 \end{equation}

The eigenvalues of the classical Laplacian matrix $\bm{\mathcal{L}}^{(1)}$ are $0=\mu^{(1)}_1\leq\mu^{(1)}_2\leqslant\ldots\leqslant\mu^{(1)}_N$ and the row sum of the Laplacian matrices $\bm{\mathcal{L}}^{(m)};\;m=1,\ldots,D$ are zero we can decouple Eq.~(\ref{projdynhoi_1}) as,
\begin{subequations}\label{projdynhoi_decoupled}
\begin{align}
    \dot{\bm{\psi}}_1 & = \bm{\mathcal{F}}\bm{\psi}_1 \label{projdynhoi_decoupled_1}\\
    \dot{\bm{\psi}}_i & = \bm{\mathcal{F}}\bm{\psi}_i - \sigma_1 \mu^{(1)}_i \bm{\mathcal{H}}^{(1)}\bm{\psi}_i - \sigma_2 \bm{\mathcal{H}}^{(2)} \sum_{j=2}^N\tilde{\mathcal{L}}^{(2)}_{ij}\bm{\psi}_j - \ldots - \sigma_D \bm{\mathcal{H}}^{(D)}\sum_{j=2}^N \tilde{\mathcal{L}}^{(D)}_{ij} \bm{\psi}_j.
    \label{projdynhoi_decoupled_i}
\end{align}
\end{subequations}
The stability of the synchronization state in higher-order networks can be determined by solving the above set of Eq.~(\ref{projdynhoi_decoupled}). We note here that the first Eq.~(\ref{projdynhoi_decoupled_1}) provides the dynamics of perturbations on the synchronization manifold and the remaining Eqs.~(\ref{projdynhoi_decoupled_i}) provide the dynamics of perturbations transverse to the synchronization manifold and these equations are coupled using the transformed Laplacian matrices $\tilde{\bm{\mathcal{L}}}^{(m)};m=2,\ldots, D$. The largest transverse Lyapunov exponent $\Lambda^{max}$ calculated as function of $\sigma_1,\ldots,\sigma_D$ and $\tilde{\bm{\mathcal{L}}}^{(2)},\ldots,\tilde{\bm{\mathcal{L}}}^{(D)}$ is defined as the master stability function. The synchronization state is stable if $\Lambda^{max}<0$ and the stability region is given as $\mathcal{R} = \{\sigma_1,\ldots,\sigma_D,\tilde{\bm{\mathcal{L}}}^{(2)},\ldots,\tilde{\bm{\mathcal{L}}}^{(D)} | \Lambda^{max}<0\}$.
Similarly to the classical master stability function analysis, here, the perturbations on the synchronization manifold are also separated from the transverse perturbations.
Although due to higher complexity structures, the transverse perturbations remain coupled, i.e., they are not further separable along the eigenmodes of the classical Laplacian matrix. If we neglect the coupled terms in Eq.~(\ref{projdynhoi_decoupled_i}) due to the higher order structures in the networks, then the transverse perturbations are decoupled along the eigenmodes of the classical Laplacian matrix. In this sense, we can consider the MSF analysis for networks with pairwise connections only as a special case of higher-order networks. 

\section{Conclusion and Outlook}\label{conclusion}

The Master Stability Function (MSF) is a powerful mathematical tool to determine the dynamic stability of synchronized states in complex systems. The general dynamical framework of such systems intertwines two fundamental components: the network structure and the dynamics. These two microscopic elements of dynamical networks are key to understanding the macroscopic behavior of complex systems. In this article, we systematically demonstrate how the MSF decouples the network structure from the dynamical network framework, offering a universal criterion for synchronization stability based on the eigenvalues of the Laplacian matrix.

We first derive the MSF in a simple and concise manner to assess the stability of networks where elements interact linearly through diffusive coupling. Subsequently, we generalize this framework to incorporate nonlinear interaction mechanisms and derive the mathematical expression for the MSF to predict the stability of such dynamical networks. Furthermore, we extend the MSF framework to multilayer network structures, where nodes are connected through interlayer and intralayer couplings. We show that the MSF effectively evaluates the stability of synchronization across layers by accounting for these couplings. This approach is essential for understanding the complexity of interaction mechanisms and network structures in systems such as social networks and transportation systems. Specifically, for multilayer networks, the MSF facilitates a comprehensive analysis of synchronization stability, providing deeper insights into interconnected systems.

Additionally, the MSF framework is expanded to analyze the stability of complex systems where elements interact through higher-order interactions, such as simplicial complexes and hypergraphs. These interactions are characteristic of systems like brain networks, where neurons engage in group interactions, and ecological systems, where species cooperate collectively for survival. These advancements enhance the MSF's capability to analyze realistic systems and benefit real-world applications, such as ensuring the stability of information processing in social networks. The versatility of the MSF is evident in its applications across diverse domains, including brain dynamics, power grids, and ecological and social networks \cite{motter2013spontaneous,PhysRevE.97.032307}.

Despite being one of the most elegant and powerful tools in analyzing synchronization, the application of the master stability function has been largely limited to specific complex systems. In the past 25 years, after its development, the MSF has been largely used in the study of synchronization of coupled identical dynamical systems~\cite{arenas2008synchronization}. Recently, a few studies have extended the MSF framework to coupled nonidentical systems~\cite{SunEPL2009,SorrentinoEPL2011,AcharyyaEPL2012,AcharyyaPRE2015,SugitaniPRL2021,PanahiPRE2021,NazerianEPL2023, NazerianCommPhys2023}. However, in these studies, nonidentity typically refers to variations in tunable (non-intrinsic) parameters rather than intrinsic parameters of the system. Furthermore, most of these studies have considered only small parameter variations, resulting in minor differences among coupled oscillators, while Ref.~\cite{NazerianCommPhys2023} explored the MSF with large parameter variations. Therefore, there is a need to develop a formal analytical framework for MSF-based stability analysis in coupled systems that exhibit fundamental differences in intrinsic parameters that manifest as distinct dynamical equations for each node. 

Further, in many natural and experimental systems, diverse elements interact through different coupling mechanisms. To accurately model such systems, it is essential to account for both nonidentical dynamics of uncoupled nodes and various types of interaction functions. Hence, future research should focus on developing an analytical framework that incorporates mixed dynamics on individual nodes as well as diverse forms of coupling schemes~\cite{meena2023emergent}. Such a framework would offer a universal approach for assessing the stability of a wide range of complex systems. In the above-mentioned examples, MSF analysis is done for the networks where all nodes reside in a single layer and interact through pairwise connections. Further, MSF analysis extended to multilayer networks 
~\cite{SciAdv.2.e1601679, PhysREvE.99.012304} and higher-order networks~\cite{NatComm.12.1255}. In all these cases, MSF analysis has primarily been applied to abstract, mathematical network representations where nodes and links do not occupy physical space. 
Synchronization is important for many systems where nodes and links occupy physical space, for example, brain networks, where nodes represent neurons with non-trivial dendritic shapes, and links are synapses~\cite{bullmore2009complex}.
Recent studies focus on modeling such physical networks where practical constraints are imposed on nodes and links, treating them as physical objects~\cite{pete2024physical}. Developing and implementing an MSF-based approach for studying the stability of synchronization in physical systems such as molecular networks~\cite{panditrao2022emerging}, wood-wide web~\cite{steidinger2019climatic}, fiber materials~\cite{picu2022network} would be a significant milestone, establishing MSF as a powerful tool for analyzing synchronization stability in real-world systems. In another direction, coupled phase oscillators have been extensively studied to model various complex systems (e.g., Josephson junction~\cite{rodrigues2016kuramoto}, arrays of lasers ~\cite{jiang1993numerical}, networks of pacemaker cells in the heart~\cite{peskin1975mathematical} ). The order parameter is commonly used to determine the synchronization of such coupled phase oscillators. However, the Master Stability Function (MSF) framework to analyze the stability of synchronization in coupled phase oscillators is still lacking. Furthermore, MSF has been estimated using reservoir computing technique, given only the time series of a single, uncoupled oscillator \cite{timeseries_ML}. 

In conclusion, the Master Stability Function is a versatile and universal tool for exploring synchronization stability in complex networks. By extending its applicability to various complex networks and integrating it with machine learning techniques, the MSF framework has the potential to unlock new insights and innovations in diverse fields, from computational neuroscience and artificial intelligence to engineering. 

\vspace{4mm}
\sloppy \noindent \textcolor{blue}{Code and Data Availability:} The code and data required to reproduce and analyze the Master Stability Function (MSF) on a single-layer network are available in the GitHub repository: \url{https://github.com/sumanacharyya/Master-Stability-Function.git}.

\section*{Acknowledgment} 
We are thankful to Adilson E. Motter (Northwestern University, USA) for valuable discussions on the MSF in networks. SA acknowledges various discussions with B. Barzel, R. Cohen (Bar-Ilan University, Israel), and R. E. Amritkar (PRL, Ahmedabad, India). PP acknowledges the Anusandhan National Research Foundation (ANRF) grant TAR/2022/000657, Govt. of India. CM acknowledges support from the Anusandhan National Research Foundation (ANRF) India (Grants Numbers SRG/2023/001846 and EEQ/2023/001080). 

\bibliography{references_ms.bib}

\begin{thebibliography}{100}
\expandafter\ifx\csname url\endcsname\relax
  \def\url#1{\texttt{#1}}\fi
\expandafter\ifx\csname urlprefix\endcsname\relax\def\urlprefix{URL }\fi
\expandafter\ifx\csname href\endcsname\relax
  \def\href#1#2{#2} \def\path#1{#1}\fi

\bibitem{BarabasiBook}
M.~P{\'o}sfai, A.-L. Barab{\'a}si, Network science, Vol.~3, Cambridge University Press, 2016.

\bibitem{RevModPhys.74.47}
R.~Albert, A.-L. Barab{\'a}si, Statistical mechanics of complex networks, Reviews of modern physics 74~(1) (2002) 47.

\bibitem{Newman2018}
M.~Newman, Networks, Oxford university press, 2018.

\bibitem{Winfree1980}
A.~T. Winfree, The geometry of biological time, Vol.~2, Springer, 1980.

\bibitem{Kuramoto1984}
Y.~Kuramoto, Chemical turbulence, Springer, 1984.

\bibitem{StrogatzSyncBook}
S.~Strogatz, Sync: The emerging science of spontaneous order (2004).

\bibitem{PhysRep.469.93}
A.~Arenas, A.~D{\'\i}az-Guilera, J.~Kurths, Y.~Moreno, C.~Zhou, Synchronization in complex networks, Physics reports 469~(3) (2008) 93--153.

\bibitem{meena2017threshold}
C.~Meena, P.~D. Rungta, S.~Sinha, Threshold-activated transport stabilizes chaotic populations to steady states, Plos one 12~(8) (2017) e0183251.

\bibitem{ducrot2022differential}
A.~Ducrot, Q.~Griette, Z.~Liu, P.~Magal, Differential equations and population dynamics i, Differential Equations and Population Dynamics I (2022).

\bibitem{TurcotteBook}
D.~L. Turcotte, Fractals and chaos in geology and geophysics, Cambridge university press, 1997.

\bibitem{RMMayBook}
R.~M. May, Stability and complexity in model ecosystems, Princeton university press, 2019.

\bibitem{Science.275.334}
S.~A. Levin, B.~Grenfell, A.~Hastings, A.~S. Perelson, Mathematical and computational challenges in population biology and ecosystems science, Science 275~(5298) (1997) 334--343.

\bibitem{NatRevNeuroSci.10.186}
E.~Bullmore, O.~Sporns, Complex brain networks: graph theoretical analysis of structural and functional systems, Nature reviews neuroscience 10~(3) (2009) 186--198.

\bibitem{PhysRevD.58.073002}
J.~Pantaleone, Stability of incoherence in an isotropic gas of oscillating neutrinos, Physical Review D 58~(7) (1998) 073002.

\bibitem{PhysRevE.57.1563}
K.~Wiesenfeld, P.~Colet, S.~H. Strogatz, Frequency locking in josephson arrays: Connection with the kuramoto model, Physical Review E 57~(2) (1998) 1563.

\bibitem{BarratBook}
A.~Barrat, M.~Barthelemy, A.~Vespignani, Dynamical processes on complex networks, Cambridge university press, 2008.

\bibitem{PikovskyBook}
A.~Pikovsky, M.~Rosenblum, J.~Kurths, Synchronization a universal concept in nonlinear sciences, Cambridge university press 12 (2001).

\bibitem{PhysRep.366.1}
S.~Boccaletti, J.~Kurths, G.~Osipov, D.~Valladares, C.~Zhou, The synchronization of chaotic systems, Physics reports 366~(1-2) (2002) 1--101.

\bibitem{rungta2017network}
P.~D. Rungta, A.~Choudhary, C.~Meena, S.~Sinha, Are network properties consistent indicators of synchronization?, Europhysics Letters 117~(2) (2017) 20003.

\bibitem{ValdezJCM2020}
L.~D. Valdez, L.~Shekhtman, C.~E. La~Rocca, X.~Zhang, S.~V. Buldyrev, P.~A. Trunfio, L.~A. Braunstein, S.~Havlin, Cascading failures in complex networks, Journal of Complex Networks 8~(2) (2020) cnaa013.

\bibitem{PhysRep.424.175}
S.~Boccaletti, V.~Latora, Y.~Moreno, M.~Chavez, D.-U. Hwang, Complex networks: Structure and dynamics, Physics reports 424~(4-5) (2006) 175--308.

\bibitem{ArenasPRL2006}
A.~Arenas, A.~D{\'\i}az-Guilera, C.~J. P{\'e}rez-Vicente, Synchronization reveals topological scales in complex networks, Physical review letters 96~(11) (2006) 114102.

\bibitem{Huygens1665}
C.~Huygens, H.~Oscillatorium, Apud f, Muguet, Parisiis, France 1673.

\bibitem{buck1968mechanism}
J.~Buck, E.~Buck, Mechanism of rhythmic synchronous flashing of fireflies: fireflies of southeast asia may use anticipatory time-measuring in synchronizing their flashing, Science 159~(3821) (1968) 1319--1327.

\bibitem{rungta2018}
P.~D. Rungta, C.~Meena, S.~Sinha, Identifying nodal properties that are crucial for the dynamical robustness of multistable networks, Physical Review E 98~(2) (2018) 022314.

\bibitem{meena2020resilience}
C.~Meena, P.~D. Rungta, S.~Sinha, Resilience of networks of multi-stable chaotic systems to targetted attacks, The European Physical Journal B 93 (2020) 1--9.

\bibitem{DevaneyBook}
R.~Devaney, An introduction to chaotic dynamical systems, CRC press, 2018.

\bibitem{PNAS.95.7092}
J.~Sarnthein, H.~Petsche, P.~Rappelsberger, G.~L. Shaw, A.~von Stein, Synchronization between prefrontal and posterior association cortex during human working memory, Proceedings of the National Academy of Sciences 95~(12) (1998) 7092--7096.

\bibitem{PhysRevE.96.258102}
R.~Amritkar, G.~Rangarajan, Spatially synchronous extinction of species under external forcing, Physical review letters 96~(25) (2006) 258102.

\bibitem{AlbertPRE2004}
R.~Albert, I.~Albert, G.~L. Nakarado, Structural vulnerability of the north american power grid, Physical review E 69~(2) (2004) 025103.

\bibitem{LamportACM1978}
C.~Time, the ordering of events in a distributed system, L. Lamport 7 (1978) 558--565.

\bibitem{quantum_sync_2010}
M.~Lohe, Quantum synchronization over quantum networks, Journal of Physics A: Mathematical and Theoretical 43~(46) (2010) 465301.

\bibitem{GlendinningBook}
P.~Glendinning, Stability, instability and chaos: an introduction to the theory of nonlinear differential equations, Cambridge university press, 1994.

\bibitem{FujisakaYamada1983}
H.~Fujisaka, T.~Yamada, Stability theory of synchronized motion in coupled-oscillator systems, Progress of theoretical physics 69~(1) (1983) 32--47.

\bibitem{YamadaFujisaka1983}
T.~Yamada, H.~Fujisaka, Stability theory of synchronized motion in coupled-oscillator systems. ii: The mapping approach, Progress of Theoretical Physics 70~(5) (1983) 1240--1248.

\bibitem{PecoraCarrollPRL1990}
L.~M. Pecora, T.~L. Carroll, Synchronization in chaotic systems, Physical review letters 64~(8) (1990) 821.

\bibitem{BarreiraBook}
L.~Barreira, Y.~B. Pesin, Lyapunov exponents and smooth ergodic theory, Vol.~23, American Mathematical Soc., 2002.

\bibitem{PecoraCarrollPRL1998}
L.~M. Pecora, T.~L. Carroll, Master stability functions for synchronized coupled systems, Physical review letters 80~(10) (1998) 2109.

\bibitem{BarahonaPRL2002}
M.~Barahona, L.~M. Pecora, Synchronization in small-world systems, Physical review letters 89~(5) (2002) 054101.

\bibitem{HuangPRE2009}
L.~Huang, Q.~Chen, Y.-C. Lai, L.~M. Pecora, Generic behavior of master-stability functions in coupled nonlinear dynamical systems, Physical Review E—Statistical, Nonlinear, and Soft Matter Physics 80~(3) (2009) 036204.

\bibitem{MotterPRE2005}
A.~E. Motter, C.~Zhou, J.~Kurths, Network synchronization, diffusion, and the paradox of heterogeneity, Physical Review E—Statistical, Nonlinear, and Soft Matter Physics 71~(1) (2005) 016116.

\bibitem{motter2013spontaneous}
A.~E. Motter, S.~A. Myers, M.~Anghel, T.~Nishikawa, Spontaneous synchrony in power-grid networks, Nature Physics 9~(3) (2013) 191--197.

\bibitem{NishikawaPNAS2010}
T.~Nishikawa, A.~E. Motter, Network synchronization landscape reveals compensatory structures, quantization, and the positive effect of negative interactions, Proceedings of the National Academy of Sciences 107~(23) (2010) 10342--10347.

\bibitem{PhysRevE.97.032307}
A.~Brechtel, P.~Gramlich, D.~Ritterskamp, B.~Drossel, T.~Gross, Master stability functions reveal diffusion-driven pattern formation in networks, Physical Review E 97~(3) (2018) 032307.

\bibitem{PhysRevE.105.044310}
A.~Krau{\ss}, T.~Gross, B.~Drossel, Master stability functions for metacommunities with two types of habitats, Physical Review E 105~(4) (2022) 044310.

\bibitem{SunEPL2009}
J.~Sun, E.~M. Bollt, T.~Nishikawa, Master stability functions for coupled nearly identical dynamical systems, Europhysics Letters 85~(6) (2009) 60011.

\bibitem{SorrentinoEPL2011}
F.~Sorrentino, M.~Porfiri, Analysis of parameter mismatches in the master stability function for network synchronization, Europhysics Letters 93~(5) (2011) 50002.

\bibitem{AcharyyaEPL2012}
S.~Acharyya, R.~Amritkar, Synchronization of coupled nonidentical dynamical systems, Europhysics Letters 99~(4) (2012) 40005.

\bibitem{AcharyyaPRE2015}
S.~Acharyya, R.~Amritkar, Synchronization of nearly identical dynamical systems: Size instability, Physical Review E 92~(5) (2015) 052902.

\bibitem{NazerianEPL2023}
A.~Nazerian, S.~Panahi, F.~Sorrentino, Synchronization in networks of coupled oscillators with mismatches, Europhysics Letters 143~(1) (2023) 11001.

\bibitem{NazerianCommPhys2023}
A.~Nazerian, S.~Panahi, F.~Sorrentino, Synchronization in networked systems with large parameter heterogeneity, Communications Physics 6~(1) (2023) 253.

\bibitem{NishikawaPRE2006}
T.~Nishikawa, A.~E. Motter, Synchronization is optimal in nondiagonalizable networks, Physical Review E—Statistical, Nonlinear, and Soft Matter Physics 73~(6) (2006) 065106.

\bibitem{NishikawaPhysicaD2006}
T.~Nishikawa, A.~E. Motter, Maximum performance at minimum cost in network synchronization, Physica D: Nonlinear Phenomena 224~(1-2) (2006) 77--89.

\bibitem{BianconiBookML}
G.~Bianconi, Multilayer networks: structure and function, Oxford university press, 2018.

\bibitem{PhysREvE.99.012304}
L.~Tang, X.~Wu, J.~L{\"u}, J.-a. Lu, R.~M. D'Souza, Master stability functions for complete, intralayer, and interlayer synchronization in multiplex networks of coupled r{\"o}ssler oscillators, Physical Review E 99~(1) (2019) 012304.

\bibitem{SciAdv.2.e1601679}
C.~I. Del~Genio, J.~G{\'o}mez-Garde{\~n}es, I.~Bonamassa, S.~Boccaletti, Synchronization in networks with multiple interaction layers, Science Advances 2~(11) (2016) e1601679.

\bibitem{MulasPRE2020}
R.~Mulas, C.~Kuehn, J.~Jost, Coupled dynamics on hypergraphs: Master stability of steady states and synchronization, Physical Review E 101~(6) (2020) 062313.

\bibitem{NatComm.12.1255}
L.~V. Gambuzza, F.~Di~Patti, L.~Gallo, S.~Lepri, M.~Romance, R.~Criado, M.~Frasca, V.~Latora, S.~Boccaletti, Stability of synchronization in simplicial complexes, Nature communications 12~(1) (2021) 1255.

\bibitem{Gambuzza2022}
L.~V. Gambuzza, F.~Di~Patti, L.~Gallo, S.~Lepri, M.~Romance, R.~Criado, M.~Frasca, V.~Latora, S.~Boccaletti, The master stability function for synchronization in simplicial complexes, in: Higher-Order Systems, Springer, 2022, pp. 249--267.

\bibitem{lyapunov_1985}
A.~Wolf, J.~B. Swift, H.~L. Swinney, J.~A. Vastano, Determining lyapunov exponents from a time series, Physica D: nonlinear phenomena 16~(3) (1985) 285--317.

\bibitem{rossler1976equation}
O.~E. R{\"o}ssler, An equation for continuous chaos, Physics Letters A 57~(5) (1976) 397--398.

\bibitem{lee2011prediction}
T.~Lee, T.~B. Ouarda, Prediction of climate nonstationary oscillation processes with empirical mode decomposition, Journal of Geophysical Research: Atmospheres 116~(D6) (2011).

\bibitem{easaw2023estimation}
N.~Easaw, W.~S. Lee, P.~S. Lohiya, S.~Jalan, P.~Pradhan, Estimation of correlation matrices from limited time series data using machine learning, Journal of Computational Science 71 (2023) 102053.

\bibitem{AcharyyaChaos2011}
S.~Acharyya, R.~Amritkar, Desynchronization bifurcation of coupled nonlinear dynamical systems, Chaos: An Interdisciplinary Journal of Nonlinear Science 21~(2) (2011).

\bibitem{lorenz1963deterministic}
E.~N. Lorenz, Deterministic nonperiodic flow, Journal of atmospheric sciences 20~(2) (1963) 130--141.

\bibitem{chen1999yet}
G.~Chen, T.~Ueta, Yet another chaotic attractor, International Journal of Bifurcation and chaos 9~(07) (1999) 1465--1466.

\bibitem{dhamala2004transitions}
M.~Dhamala, V.~K. Jirsa, M.~Ding, Transitions to synchrony in coupled bursting neurons, Physical Review Letters 92~(2) (2004) 028101.

\bibitem{stefanski2007ragged}
A.~Stefa{\'n}ski, P.~Perlikowski, T.~Kapitaniak, Ragged synchronizability of coupled oscillators, Physical Review E—Statistical, Nonlinear, and Soft Matter Physics 75~(1) (2007) 016210.

\bibitem{mettin1993bifurcation}
R.~Mettin, U.~Parlitz, W.~Lauterborn, Bifurcation structure of the driven van der pol oscillator, International Journal of Bifurcation and Chaos 3~(06) (1993) 1529--1555.

\bibitem{StankovskiRMP2017}
T.~Stankovski, T.~Pereira, P.~V. McClintock, A.~Stefanovska, Coupling functions: universal insights into dynamical interaction mechanisms, Reviews of Modern Physics 89~(4) (2017) 045001.

\bibitem{BaratPNAS2004}
A.~Barrat, M.~Barthelemy, R.~Pastor-Satorras, A.~Vespignani, The architecture of complex weighted networks, Proceedings of the national academy of sciences 101~(11) (2004) 3747--3752.

\bibitem{YookPRL2001}
S.-H. Yook, H.~Jeong, A.-L. Barab{\'a}si, Y.~Tu, Weighted evolving networks, Physical review letters 86~(25) (2001) 5835.

\bibitem{FriesPNAS1997}
P.~Fries, P.~R. Roelfsema, A.~K. Engel, P.~K{\"o}nig, W.~Singer, Synchronization of oscillatory responses in visual cortex correlates with perception in interocular rivalry, Proceedings of the National Academy of Sciences 94~(23) (1997) 12699--12704.

\bibitem{HornBook}
R.~A. Horn, C.~R. Johnson, Matrix analysis, Cambridge university press, 2012.

\bibitem{bianconi2018multilayer}
G.~Bianconi, Multilayer networks: structure and function, Oxford university press, 2018.

\bibitem{de2016physics}
M.~De~Domenico, C.~Granell, M.~A. Porter, A.~Arenas, The physics of spreading processes in multilayer networks, Nature Physics 12~(10) (2016) 901--906.

\bibitem{boccaletti2014structure}
S.~Boccaletti, G.~Bianconi, R.~Criado, C.~I. Del~Genio, J.~G{\'o}mez-Gardenes, M.~Romance, I.~Sendina-Nadal, Z.~Wang, M.~Zanin, The structure and dynamics of multilayer networks, Physics reports 544~(1) (2014) 1--122.

\bibitem{bassett2017network}
D.~S. Bassett, O.~Sporns, Network neuroscience, Nature neuroscience 20~(3) (2017) 353--364.

\bibitem{jalan2018localization}
S.~Jalan, P.~Pradhan, Localization of multilayer networks by optimized single-layer rewiring, Physical Review E 97~(4) (2018) 042314.

\bibitem{della2020symmetries}
F.~Della~Rossa, L.~Pecora, K.~Blaha, A.~Shirin, I.~Klickstein, F.~Sorrentino, Symmetries and cluster synchronization in multilayer networks, Nature communications 11~(1) (2020) 3179.

\bibitem{he2017multiagent}
W.~He, G.~Chen, Q.-L. Han, W.~Du, J.~Cao, F.~Qian, Multiagent systems on multilayer networks: Synchronization analysis and network design, IEEE Transactions on Systems, Man, and Cybernetics: Systems 47~(7) (2017) 1655--1667.

\bibitem{menichetti2016control}
G.~Menichetti, L.~Dall’Asta, G.~Bianconi, Control of multilayer networks, Scientific reports 6~(1) (2016) 20706.

\bibitem{PRE.99.012304.2019}
L.~Tang, X.~Wu, J.~L{\"u}, J.-a. Lu, R.~M. D'Souza, Master stability functions for complete, intralayer, and interlayer synchronization in multiplex networks of coupled r{\"o}ssler oscillators, Physical Review E 99~(1) (2019) 012304.

\bibitem{BianoniBookHOI}
G.~Bianconi, Higher-order networks, Cambridge University Press, 2021.

\bibitem{PhysRep.874.1}
F.~Battiston, G.~Cencetti, I.~Iacopini, V.~Latora, M.~Lucas, A.~Patania, J.-G. Young, G.~Petri, Networks beyond pairwise interactions: Structure and dynamics, Physics reports 874 (2020) 1--92.

\bibitem{JCompNeuroSci.41.1}
C.~Giusti, R.~Ghrist, D.~S. Bassett, Two’s company, three (or more) is a simplex: Algebraic-topological tools for understanding higher-order structure in neural data, Journal of computational neuroscience 41 (2016) 1--14.

\bibitem{Science.353.163}
A.~R. Benson, D.~F. Gleich, J.~Leskovec, Higher-order organization of complex networks, Science 353~(6295) (2016) 163--166.

\bibitem{Roitz2014}
A.~Ritz, A.~N. Tegge, H.~Kim, C.~L. Poirel, T.~Murali, Signaling hypergraphs, Trends in biotechnology 32~(7) (2014) 356--362.

\bibitem{Nature.546.56}
J.~M. Levine, J.~Bascompte, P.~B. Adler, S.~Allesina, Beyond pairwise mechanisms of species coexistence in complex communities, Nature 546~(7656) (2017) 56--64.

\bibitem{Nature.548.210}
J.~Grilli, G.~Barab{\'a}s, M.~J. Michalska-Smith, S.~Allesina, Higher-order interactions stabilize dynamics in competitive network models, Nature 548~(7666) (2017) 210--213.

\bibitem{semantic_networks}
J.~Borge-Holthoefer, A.~Arenas, Semantic networks: Structure and dynamics, Entropy 12~(5) (2010) 1264--1302.

\bibitem{PetriJRSI2014}
G.~Petri, P.~Expert, F.~Turkheimer, R.~Carhart-Harris, D.~Nutt, P.~J. Hellyer, F.~Vaccarino, Homological scaffolds of brain functional networks, Journal of The Royal Society Interface 11~(101) (2014) 20140873.

\bibitem{Lee2012}
H.~Lee, H.~Kang, M.~K. Chung, B.-N. Kim, D.~S. Lee, Persistent brain network homology from the perspective of dendrogram, IEEE transactions on medical imaging 31~(12) (2012) 2267--2277.

\bibitem{JTheoBio.438.46}
E.~Estrada, G.~J. Ross, Centralities in simplicial complexes. applications to protein interaction networks, Journal of theoretical biology 438 (2018) 46--60.

\bibitem{EPJDataSci.6.18}
A.~Patania, G.~Petri, F.~Vaccarino, The shape of collaborations, EPJ Data Science 6 (2017) 1--16.

\bibitem{Mayfield2017}
M.~M. Mayfield, D.~B. Stouffer, Higher-order interactions capture unexplained complexity in diverse communities, Nature ecology \& evolution 1~(3) (2017) 0062.

\bibitem{Bairey2017}
E.~Bairey, E.~D. Kelsic, R.~Kishony, High-order species interactions shape ecosystem diversity, Nature communications 7~(1) (2016) 12285.

\bibitem{BergeRevuzBook}
C.~Berge, Hypergraphs: combinatorics of finite sets, Vol.~45, Elsevier, 1984.

\bibitem{SIAMRev.65.686}
C.~Bick, E.~Gross, H.~A. Harrington, M.~T. Schaub, What are higher-order networks?, SIAM review 65~(3) (2023) 686--731.

\bibitem{arenas2008synchronization}
A.~Arenas, A.~D{\'\i}az-Guilera, J.~Kurths, Y.~Moreno, C.~Zhou, Synchronization in complex networks, Physics reports 469~(3) (2008) 93--153.

\bibitem{SugitaniPRL2021}
Y.~Sugitani, Y.~Zhang, A.~E. Motter, Synchronizing chaos with imperfections, Physical review letters 126~(16) (2021) 164101.

\bibitem{PanahiPRE2021}
S.~Panahi, F.~Sorrentino, Group synchrony, parameter mismatches, and intragroup connections, Physical Review E 104~(5) (2021) 054314.

\bibitem{meena2023emergent}
C.~Meena, C.~Hens, S.~Acharyya, S.~Haber, S.~Boccaletti, B.~Barzel, Emergent stability in complex network dynamics, Nature Physics 19~(7) (2023) 1033--1042.

\bibitem{bullmore2009complex}
E.~Bullmore, O.~Sporns, Complex brain networks: graph theoretical analysis of structural and functional systems, Nature reviews neuroscience 10~(3) (2009) 186--198.

\bibitem{pete2024physical}
G.~Pete, {\'A}.~Tim{\'a}r, S.~{\"O}. Stef{\'a}nsson, I.~Bonamassa, M.~P{\'o}sfai, Physical networks as network-of-networks, nature communications 15~(1) (2024) 4882.

\bibitem{panditrao2022emerging}
G.~Panditrao, R.~Bhowmick, C.~Meena, R.~R. Sarkar, Emerging landscape of molecular interaction networks: Opportunities, challenges and prospects, Journal of Biosciences 47~(2) (2022) 24.

\bibitem{steidinger2019climatic}
B.~S. Steidinger, T.~W. Crowther, J.~Liang, M.~E. Van~Nuland, G.~D. Werner, P.~B. Reich, G.-J. Nabuurs, S.~De-Miguel, M.~Zhou, N.~Picard, et~al., Climatic controls of decomposition drive the global biogeography of forest-tree symbioses, Nature 569~(7756) (2019) 404--408.

\bibitem{picu2022network}
C.~R. Picu, Network materials: structure and properties (2022).

\bibitem{rodrigues2016kuramoto}
F.~A. Rodrigues, T.~K.~D. Peron, P.~Ji, J.~Kurths, The kuramoto model in complex networks, Physics Reports 610 (2016) 1--98.

\bibitem{jiang1993numerical}
Z.~Jiang, M.~McCall, Numerical simulation of a large number of coupled lasers, Journal of the Optical Society of America B 10~(1) (1993) 155--163.

\bibitem{peskin1975mathematical}
C.~S. Peskin, Mathematical aspects of heart physiology, Courant Inst. Math (1975).

\bibitem{timeseries_ML}
J.~D. Hart, Estimating the master stability function from the time series of one oscillator via reservoir computing, Physical Review E 108~(3) (2023) L032201.

\end{thebibliography}
%

\end{document}